\documentclass[gmd, manuscript]{copernicus}

\begin{document}
\nolinenumbers

\title{LatENSO v1.0: a learned latent state of the tropical Pacific for seasonal ENSO operators and forecasting}

\Author[1][miladpnh@outlook.com]{Milad}{Panahi}
\Author[1]{Annalisa}{Bracco}

\affil[1]{Fondazione Centro Euro-Mediterraneo sui Cambiamenti Climatici (CMCC), Bologna, Italy}

\runningtitle{LatENSO v1.0: a learned latent state for seasonal ENSO operators}

\runningauthor{M. Panahi and A. Bracco}

\received{}
\pubdiscuss{}
\revised{}
\accepted{}
\published{}

\firstpage{1}

\maketitle

\begin{abstract}
Statistical forecasts of the El Ni\~no--Southern Oscillation (ENSO) either map a modeler-chosen, simple but physically motivated description of the current climate state (such as leading sea surface temperature patterns and time series of ocean heat content) to future ENSO conditions,  or use deep networks that learn their own representation from full gridded fields, leaving a learned state that is difficult to inspect. Here we present LatENSO v1.0, which sits between these two approaches: it is a machine-learning forecast system for ENSO built around a learned, 32-dimensional state of the tropical Pacific.  A Swin-transformer autoencoder compresses thirteen monthly-averaged ocean and atmosphere fields
into 32 numbers, trained to reconstruct the fields, with no forecast target. The 32 numbers rebuild the fields as
well as O(100) empirical orthogonal functions (EOFs). Two seasonal operators use this state. The first is the extended nonlinear recharge oscillator
(XRO) with two directions of the learned state among its variables; we call it XRO-lat. The second is a Floquet
modal operator (FMO) of three damped oscillators, which forecasts a correction that is added with a fixed weight to
a least-squares base forecast made from four observed predictors. We evaluate both against the recharge oscillator,
XRO and a deep echo-state network (DESN), all refitted on identical data, cross-validation folds and verification
windows (1981--2024). Over lead times of 1--18 months, XRO-lat reaches an anomaly correlation coefficient (ACC) of
0.673, the highest of all tropical-Pacific-only (in-domain) systems and statistically indistinguishable from XRO and
DESN with their eight remote-basin indices (0.685 and 0.667). Its two directions, selected inside each training fold,
raise the ACC of the in-domain XRO in every fold, by 0.019 (95\,\% confidence interval). The gain of XRO-lat grows with lead, and from
lead 15 to lead 24 it is the most skillful of all systems. Over leads 1--18, the FMO forecast (0.619) is
statistically indistinguishable from every in-domain reference model (0.631--0.654); it detects El Ni\~no events
above 1\,$^{\circ}$C better than every reference model at every lead from 12 to 18 (area under the receiver
operating characteristic curve 0.819 over leads 10--18, against 0.752--0.767), although the 95\,\% interval of its
advantage over each reference model includes zero. What the FMO adds to its base forecast, 0.054 in ACC, is carried
by its Floquet propagation, whose generator consistently holds a prominent quasi-biennial oscillator. At leads 3--9 the FMO trails
XRO, DESN and the in-domain XRO, a deficit partly attributable to its base. 
The novelty of LatENSO lies in combining competitive skill with trustworthiness: a compact state whose physical content can be inspected, an account of what the emulator produces and why, explicit documentation of its limitations, and fully open access to code, weights and data. The proposed framework can be easily extended to the whole tropical belt, to account for remote teleconnections by directly including the information carried by the other basins.

\end{abstract}

\introduction

The El Ni\~no--Southern Oscillation (ENSO) is the dominant mode of interannual climate variability and the main
source of skill in seasonal climate forecasts \citep{McPhaden2006,Timmermann2018}. 

Shifts in the ENSO  cycle alter rainfall, storm tracks, and temperatures across much of the globe, making it a key target of seasonal to multiseasonal forecasting efforts over the past four decades \citep{Latif1998,Ehsan2024}. Forecasters use two primary method categories to track its evolution: dynamical, coupled climate model forecasts and statistical forecasts, whose skill has been summarized over the past two decades by the International Research Institute (IRI) at Columbia University, now embedded in the Center for Climate Systems Research/IRI \citep{Ehsan2024}. 

Statistical ENSO forecasting approaches, in turn, can be broadly grouped into two families. The first relies on
low-order dynamical models fitted to a small number of climate indices. Notable examples include the recharge
oscillator \citep{Jin1997}, linear inverse models built on empirical orthogonal functions (EOFs)
\citep{PenlandSardeshmukh1995}, and, more recently, the extended nonlinear recharge oscillator (XRO) of
\citet{Zhao2024}, which couples ENSO to seasonally modulated modes of the other tropical basins and achieves
skillful forecasts at leads of 16--18 months. The second family is based on deep networks, including convolutional
networks trained on global climate fields \citep{HamEtAl2019}, transformers applied to three-dimensional anomalies
from atmospheric and oceanic fields \citep{ZhouZhang2023}, and reservoir computers built on the XRO indices
\citep{Zhang2026}. Importantly, the two families differ less in skill than in their explainability. In index-based
models, the state is chosen by the modeler and remains physically legible: for example, it is straightforward to
show that much of the advantage of XRO over the recharge oscillator originates from indices outside the tropical
Pacific \citep{Zhao2024}. In deep networks, however, the state is implicit, learned jointly with the forecast, and
difficult to inspect.

In this work, we ask whether a state learned from the coupled fields of the tropical Pacific alone, compressed to a few tens of numbers and trained without any forecasting objective, can serve as a physically grounded state for machine-learning (ML) seasonal ENSO operators; we call this framework LatENSO. The potential value of a latent-space framework for ENSO was first suggested by \citet{FalascaBracco2022}. 

Two strands of work motivate the design of LatENSO. The first concerns the representation of the state and the architecture used to learn it. Autoencoders learn low-dimensional representations from which a field can be reconstructed \citep{HintonSalakhutdinov2006}. Like EOFs, these representations provide a compact description of variability, but they are learned through a nonlinear mapping. Transformer architectures, on the other hand, represent interactions among different regions through attention mechanisms \citep{Vaswani2017}. Vision transformers extend this idea to gridded data \citep{Dosovitskiy2021}, while the Swin transformer restricts attention to local, non-overlapping windows that are shifted between successive layers, which improves computational efficiency and scalability \citep{LiuEtAl2021Swin}. Variants of these architectures now form the basis of several data-driven weather forecasting systems \citep[e.g.][]{Bi2023,Chen2023FuXi}. In nearly all such applications, attention windows are square, reflecting conventions inherited from computer vision rather than the structure of the climate system. Tropical Pacific dynamics, however, are strongly anisotropic: the zonally extended Walker circulation and the meridional Hadley circulation organize variability along distinct directions. Here, we treat the geometry of the attention window as a physical modeling choice and design it to align with these dominant circulation structures.

The second strand concerns the operator that evolves this state. A longstanding goal in the study of nonlinear systems is to identify a state space in which the evolution is approximately linear \citep{Mezic2005}. In climate science, linear inverse models pursue this objective using EOF-based representations and stochastic dynamics \citep{PenlandSardeshmukh1995}, while recent machine-learning approaches use autoencoders to learn the underlying state directly from data \citep{Lusch2018}. For ENSO, however, the dynamics are not only nonlinear but strongly modulated by the seasonal cycle, with growth and decay phase-locked to the annual march of the climate system \citep{Stein2014}. This suggests adopting a time-periodic linear operator, with parameters that evolve through the calendar year, as described by Floquet theory. XRO indeed follows this approach by representing seasonality through annual and semi-annual harmonics of the operator \citep{Zhao2024}. Here we apply two seasonally varying operators to the learned state. The first, a Floquet modal operator (FMO), represents the dynamics as a small set of damped oscillatory and decaying modes whose periods and damping rates are directly interpretable. The operator is used as a correction to a four-predictor least-squares forecast. The second is the XRO formulation itself, applied to the learned state rather than to indices derived from other ocean basins.

This paper has two objectives. The first is to introduce a trustworthy new framework for ML-based statistical ENSO prediction. The proposed framework combines a physically informed latent representation, learned with a Swin autoencoder, with seasonally varying dynamical operators. The learned state is sufficiently compact that 32 latent variables reconstruct thirteen oceanic and atmospheric fields with a fidelity comparable to that obtained using 89--172 EOFs, while retaining the ability to distinguish eastern- and central-Pacific El Ni\~no variability \citep{Takahashi2011,RenJin2011,Capotondi2015}. We introduce two seasonally varying operators for evolving this state and assess their predictive skill against state-of-the-art statistical (XRO, the extension of the recharge oscillator by \citealp{Zhao2024}), and deep-learning architectures (the deep echo-state network, DESN, of \citealp{Zhang2026}). All models are trained and evaluated using identical data, cross-validation folds, and verification periods, with the strength of evidence graded using a paired year-block bootstrap, and with the acceptance criteria of the primary comparisons fixed before each comparison was made. The second objective is to bridge the communities of climate scientists and ML researchers by demonstrating that emulators can be more than high-skill forecasting tools: they can provide physically meaningful diagnostics that help explain why forecasts improve, identify the processes limiting predictability, and point to where advances in climate science are most likely to translate into better predictions and better understanding with the contribution of new tools from artificial intelligence.

The remainder of the manuscript describes the data and methodology (Sects.~\ref{sec:data}--\ref{sec:model}), the evaluation framework (Sect.~\ref{sec:evaluation}), the forecast results (Sect.~\ref{sec:results}), and their physical interpretation and implications (Sects.~\ref{sec:discussion}--\ref{sec:conclusions}).

\section{Data}
\label{sec:data}

\subsection{Climate fields}
\label{sec:fields}

The model is trained and evaluated on monthly fields from January 1981 to June 2024 (522 months), assembled
from two reanalyses. Atmospheric and surface variables are from ERA5 \citep{Hersbach2020}. We use the monthly
averaged reanalysis (monthly means of daily means) on the native $0.25^{\circ}$ grid for the following variables:
sea surface temperature (SST), the 10\,m zonal and meridional
wind components (U10, V10), mean sea-level pressure (MSL), total column water vapor (TCWV), total cloud cover
(TCC), top-of-atmosphere net long-wave radiation, and the surface latent heat, sensible heat and total
precipitation fluxes. Subsurface ocean variables are from SODA4 (version 4.15.2) \citep{Chepurin2025}, an
assimilation of ocean observations into the MOM5 ocean model forced by ERA5. We use its regridded monthly product,
the monthly means of the five-day output on a regular $0.1^{\circ}$ grid with 24 levels in the upper 1017\,m, and
consider sea surface height (SSH), the mixed-layer depth (MLD) based on a temperature criterion (the SODA4 variable
\texttt{mlt}, a $0.2\,^{\circ}$C temperature change from the surface), and the depth of the $20\,^{\circ}$C isotherm (Z20, the thermocline proxy). Z20 is computed by linear
interpolation in depth between the two levels that bracket $20\,^{\circ}$C and set to zero where the surface is colder than $20\,^{\circ}$C. The SODA fields are bilinearly interpolated onto the ERA5 grid, and every field is masked where the SODA surface temperature is missing. On that grid, a small number of isotherm depths outside 0--1100\,m are treated as missing. Gaps in Z20, from these values and over shallow water where the sea floor lies above the isotherm, are then filled by linear interpolation along longitude over at most five grid points. The thirteen fields, which we call channels, span the coupled system: the surface temperature that the forecast targets, the subsurface structure that
carries its memory, and the atmospheric fields through which the coupling acts. The SST is the ERA5 field rather
than the SODA surface temperature, so that the forecast target is defined consistently with the
atmospheric state.

The fields are cropped to the tropical Pacific, $20^{\circ}$\,S--$20^{\circ}$\,N,
$120^{\circ}$\,E--$70^{\circ}$\,W, giving $161\times681$ points per channel. Each channel is standardized by a
single scalar mean and standard deviation computed over all months and grid points, so that the thirteen fields enter
the reconstruction loss on comparable scales. Land points, about 9\,\% of the domain, are missing in every channel
of the assembled tensor (the atmospheric fields are kept over the ocean only). They are set to zero after standardization, that is, to the channel mean, and remain in the loss, so the network reproduces the coastline as
well as the fields. The same is done for the few further missing points, 0.2\,\% of the domain in SST and about
2\,\% in Z20. The resulting tensor of
$522\times13\times161\times681$ values (2.98\,GB in single precision) is the only input of the autoencoder.
The autoencoder works on complete fields, including the seasonal cycle and the spatial mean state. Anomalies are formed later, in the latent space (the space of the state that the autoencoder learns) and per cross-validation fold (Sect.~\ref{sec:cv}).

\subsection{Indices}
\label{sec:indices}

Two indices are forecast and scored: the Ni\~no 3.4 index, the main measure of skill, and the warm-water volume (WWV). Ni\~no 3.4 is the SST anomaly (in $^{\circ}$C) averaged over
$5^{\circ}$\,S--$5^{\circ}$\,N, $170^{\circ}$--$120^{\circ}$\,W. WWV is the Z20 anomaly (in m) averaged over $5^{\circ}$\,S--$5^{\circ}$\,N, $120^{\circ}$\,E--$80^{\circ}$\,W \citep{MeinenMcPhaden2000}, with
the 4\,\% of the box values that are missing (land and shallow water) held at the domain-mean depth, as in the autoencoder input. In the recharge-oscillator theory WWV leads SST and is the subsurface memory that allows a forecast to extrapolate beyond the current SST anomaly \citep{Jin1997}. Both indices are computed from the same fields the autoencoder sees. In the forecasts, their anomalies are departures from the monthly climatology of the training months of each fold (Sect.~\ref{sec:cv}).

\begin{figure*}[tp]
\includegraphics[width=\textwidth]{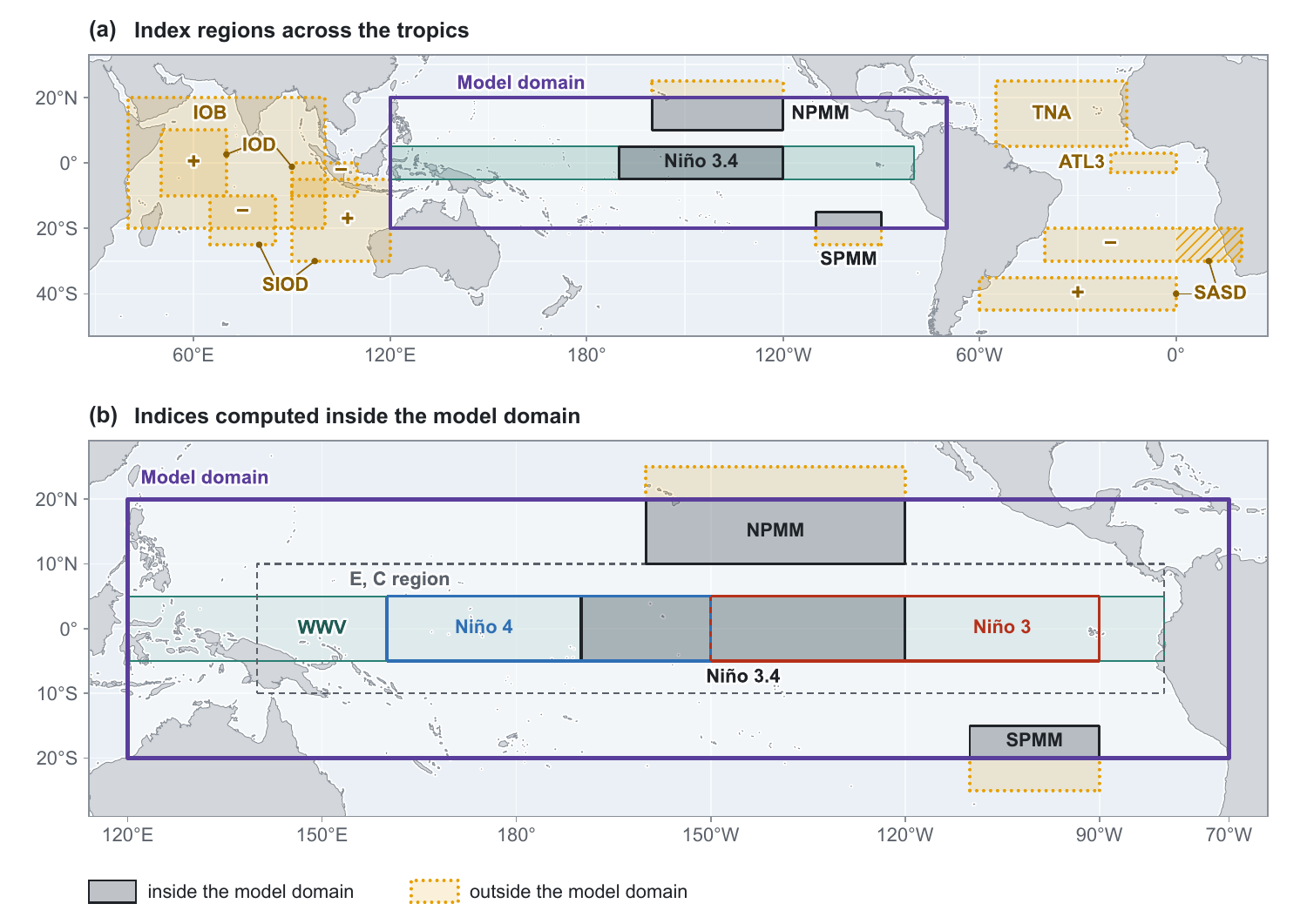}
\caption{The model domain and the areas where the indices are calculated. (a) The entire tropics. The purple frame is the model domain
($20^{\circ}$\,S--$20^{\circ}$\,N, $120^{\circ}$\,E--$70^{\circ}$\,W). Filled gray boxes with solid outlines are
SST indices computed inside the domain and used by the in-domain systems: Ni\~no 3.4 and the in-domain parts of
the two meridional modes. The teal band is the warm-water volume box (Z20). Amber regions with dotted outlines lie
outside the domain and are used only by XRO, DESN and the XRO-remote control (Sect.~\ref{sec:xrolat}): the parts of the full NPMM and SPMM
boxes beyond $20^{\circ}$, and the six indices of the other basins as defined in the XRO code \citep{Zhao2024}: IOB,
IOD (west minus east pole), SIOD (east minus west pole), TNA, ATL3 and SASD (southwest minus northeast pole);
$+$ and $-$ mark the poles. The northeast pole of SASD is coded as $40^{\circ}$\,W--$20^{\circ}$\,E; the hatched
part east of $0^{\circ}$ is selected only if the longitude grid extends beyond $360^{\circ}$\,E. (b) The Pacific
sector. The in-domain parts of NPMM ($10^{\circ}$--$20^{\circ}$\,N) and SPMM ($20^{\circ}$--$15^{\circ}$\,S) cover
68 and 51\,\% of their box areas. Ni\~no 3 (red), Ni\~no 4 (blue), which overlap Ni\~no 3.4 (gray), and the SST
region of the $E$ and $C$ indices (dashed) are also shown; $E$ and $C$ are used only to characterize ENSO diversity. All regions are drawn at
their exact bounds; coastlines from Natural Earth.}
\label{fig:indexmap}
\end{figure*}

The reference models of Sect.~\ref{sec:references} other than nRO, against which we validate our framework, and XRO-lat
(Sect.~\ref{sec:xrolat}) also use the indices representing two Pacific meridional modes. The North Pacific meridional mode (NPMM) was
introduced as the leading mode of a maximum covariance analysis of SST and surface winds over the subtropical
eastern Pacific \citep{ChiangVimont2004}, and the South Pacific meridional mode (SPMM) as its southern counterpart
\citep{Zhang2014SPMM}. XRO represents both by simpler SST indices instead, averaging SST anomalies over
$10^{\circ}$--$25^{\circ}$\,N, $160^{\circ}$--$120^{\circ}$\,W for NPMM and $25^{\circ}$--$15^{\circ}$\,S,
$110^{\circ}$--$90^{\circ}$\,W for SPMM \citep{Zhao2024}, and we follow these definitions. Both boxes extend beyond
our domain, so we compute the two indices from the same ERA5 SST over the parts of the boxes that lie inside it:
$10^{\circ}$--$20^{\circ}$\,N, $160^{\circ}$--$120^{\circ}$\,W for NPMM and $20^{\circ}$--$15^{\circ}$\,S,
$110^{\circ}$--$90^{\circ}$\,W for SPMM. These in-domain indices correlate at 0.93 with the full-box indices described
below, and they are the meridional modes used by XRO-dom and DESN-dom (in-domain reference models, Sect.~\ref{sec:references}) and the XRO-lat designs that include them. XRO
and DESN with remote basins use instead the full-box NPMM and SPMM, together with six indices of the other basins: the Indian Ocean basin mode (IOB), the Indian Ocean dipole (IOD), the subtropical Indian Ocean dipole (SIOD), the tropical North Atlantic (TNA), the Atlantic Ni\~no (ATL3) and the South Atlantic subtropical dipole (SASD). All are taken from the ORAS5-based index file of the XRO index set (1958--2025) distributed with the RC-ENSO code
\citep{Zuo2019,Zhao2024,Zhang2026}. These eight indices, which we call the remote-basin indices (or remote indices), are used only by the reference models that are allowed
information from outside the domain, and by the control experiment of Sect.~\ref{sec:xrolat}.
Figure~\ref{fig:indexmap} shows the regions of all these indices, including the parts of the two meridional-mode
boxes that lie inside the domain.

Furthermore, we use two pairs of indices to characterize ENSO diversity \citep{Capotondi2015}, that is, the fact that an El Ni\~no's peak warming can sit in the eastern Pacific (the cold tongue) or in the central Pacific (the warm pool).
No model takes these indices as input. Section~\ref{sec:res-coords} uses them, together with Ni\~no~3
and Ni\~no~4 themselves, to identify which latent coordinates carry each type of event, with all anomalies taken
with respect to the monthly climatology of the full record.

The first pair is the $E$ (canonical, eastern-Pacific or cold-tongue) and $C$ (central-Pacific or warm-pool)
indices of \citet{Takahashi2011}. They are built from the two leading principal components, PC1 and PC2, of the SST
anomalies over \mbox{$10^{\circ}$\,S--$10^{\circ}$\,N}, \mbox{$140^{\circ}$\,E--$80^{\circ}$\,W}. Both components
are scaled to unit variance and signed so that PC1 correlates positively with Ni\~no~3.4 and PC2 with Ni\~no~4
minus Ni\~no~3. The indices are these two components in axes rotated by $45^{\circ}$,
$E=(\mathrm{PC1}-\mathrm{PC2})/\sqrt{2}$ and $C=(\mathrm{PC1}+\mathrm{PC2})/\sqrt{2}$, where the factor
$1/\sqrt{2}$ keeps their variance at one. $E$ is therefore large when the warming is strongest in the east, and $C$
when it is strongest in the center.

The second pair is the $N_{\mathrm{CT}}$ (cold-tongue) and $N_{\mathrm{WP}}$ (warm-pool) indices of
\citet{RenJin2011}. They are built from the canonical Ni\~no~3 and Ni\~no~4 indices, $N_3$ and $N_4$, which are the SST
anomalies averaged over \mbox{$5^{\circ}$\,S--$5^{\circ}$\,N}, \mbox{$150^{\circ}$--$90^{\circ}$\,W} and over
\mbox{$5^{\circ}$\,S--$5^{\circ}$\,N}, \mbox{$160^{\circ}$\,E--$150^{\circ}$\,W}, respectively. Because $N_3$ and $N_4$ are
strongly correlated, each new index subtracts a fraction $\kappa$ of the other, which separates the two types of
event: $N_{\mathrm{CT}}=N_3-\kappa N_4$ and $N_{\mathrm{WP}}=N_4-\kappa N_3$. The fraction is $\kappa=0.4$ if
$N_3\cdot N_4>0$ and $\kappa=0$ otherwise.

\section{Model description}
\label{sec:model}

\subsection{Overview}
\label{sec:overview}

\begin{figure*}[tp]
\includegraphics[width=\textwidth,height=0.73\textheight,keepaspectratio]{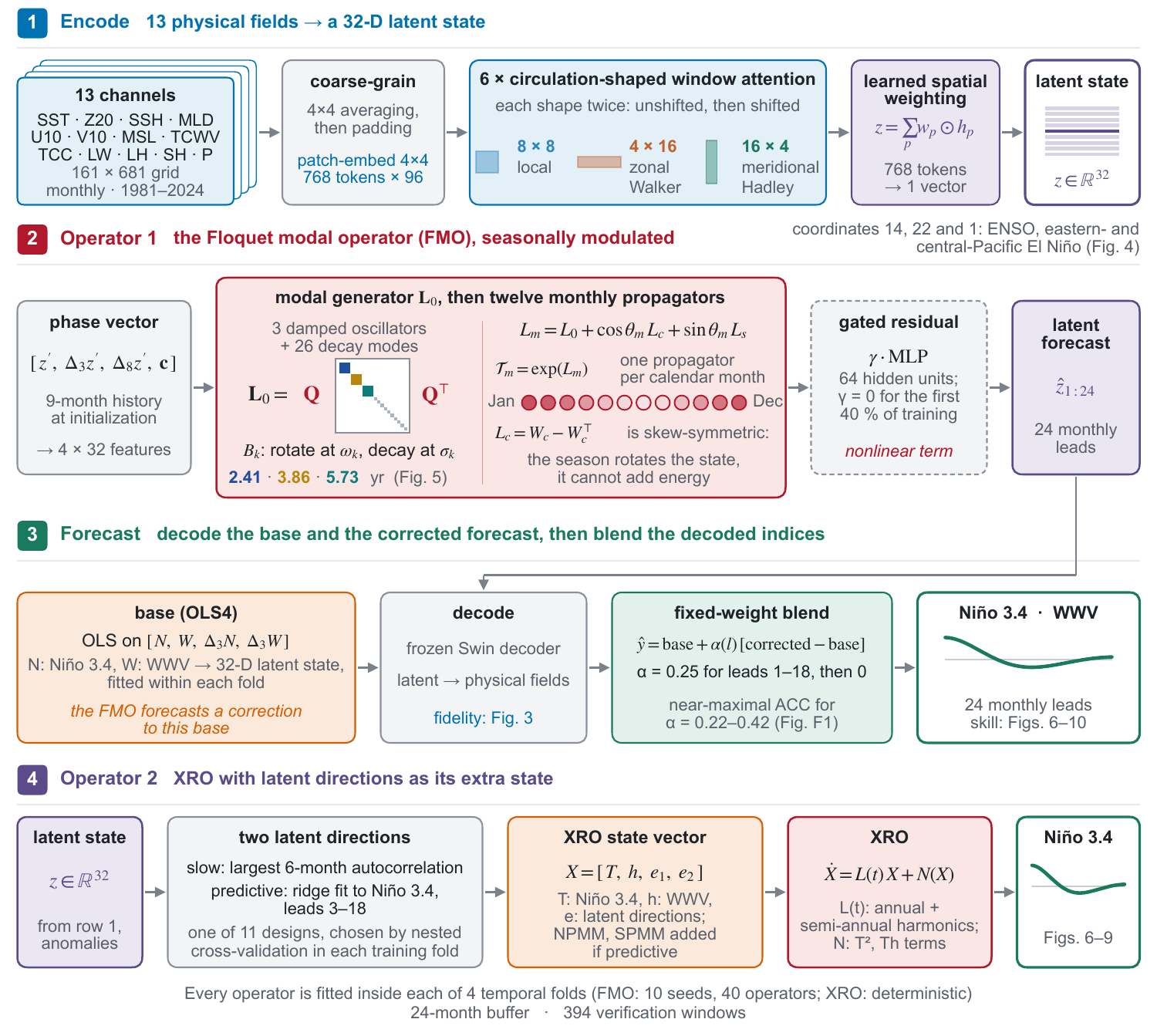}
\caption{The model at a glance. (1) Encode: thirteen monthly fields on a $161\times681$ tropical-Pacific grid
are coarse-grained and patch-embedded into 768 tokens, passed through six window-attention blocks whose windows
follow the tropical circulation, and reduced by a learned spatial weighting to the 32-dimensional state $\vec{z}$.
(2) Operator 1: a phase vector built from a 9-month history drives a modal generator with three damped
oscillators and 26 decay modes; a seasonal (Floquet) modulation gives one propagator per calendar month, and a
gated residual adds a nonlinear term. Three further terms present in every trained operator (a correction to
the initial state, a state-dependent gain and a low-rank direct term) are omitted for legibility
(Sect.~\ref{sec:operator}). (3) Forecast: the operator corrects a four-predictor least-squares base; base and
corrected forecast are decoded, and the reported forecast blends them with weight $\alpha=0.25$ to lead 18.
(4) Operator 2: XRO with two latent directions in its state, chosen by nested cross-validation inside each
training fold. Channels: LW, top-of-atmosphere net long-wave radiation; LH and SH, surface latent and sensible
heat fluxes; P, precipitation.}
\label{fig:workflow}
\end{figure*}

LatENSO has a learned state and two operators that use it (Fig.~\ref{fig:workflow}). An encoder compresses each
monthly 13-channel field into a state $\vec{z}\in\mathbb{R}^{32}$, the latent state, and a decoder maps that state back to the fields. The two are trained together as an autoencoder and then frozen, so that their weights no longer change (Sects.~\ref{sec:encoder}--\ref{sec:ae-loss}). The first operator, the FMO, advances the state through twelve monthly propagators (matrices that carry the state forward by one month) built from three damped oscillators and decay modes. Its output is a correction to a four-predictor least-squares forecast, which we call the base, and the reported forecast blends the two
(Sects.~\ref{sec:operator}--\ref{sec:training}). The second operator is the nonlinear operator XRO \citep{Zhao2024} with two directions
of the latent state as its extra state variables (Sect.~\ref{sec:xrolat}). The autoencoder is trained once; the
two operators are fitted separately within each cross-validation fold. Symbols are listed in
Appendix~\ref{app:symbols} (Table~\ref{tab:symbols}), and Table~\ref{tab:config} collects the configuration.

\subsection{Encoder}
\label{sec:encoder}

The encoder is a windowed transformer of the Swin family \citep{LiuEtAl2021Swin}, modified in the geometry of its
attention windows (Algorithm~\ref{alg:encoder}). Each monthly field is first coarse-grained by $4\times4$ average
pooling, from the native $0.25^{\circ}$ grid to $40\times170$ cells of $1^{\circ}$ (the northernmost row and the easternmost
column of the $161\times681$ grid do not enter the pooling and are restored by the final resampling). This discards sub-degree structure that a state of 32 numbers could not carry in any case and reduces the number of tokens sixteenfold (a token is the feature vector of one patch of the field). The
pooled field is padded by replication to $64\times192$ cells, the smallest extent divisible by the patch size and
by every window dimension used below. It is then embedded by a convolution with a stride of $4$, followed by a layer normalization, into $16\times48=768$ tokens of dimension 96, each token covering a $4^{\circ}\times4^{\circ}$ block. The replicated margin is removed
after decoding and never enters the loss.

Six window-attention blocks follow. The original architecture partitions the token grid into square windows. Here three geometries are used, chosen to match the anisotropy of the tropical circulation rather than for
computational convenience: $8\times8$ tokens ($32^{\circ}\times32^{\circ}$), a locally isotropic window;
$4\times16$ tokens ($16^{\circ}$ of latitude by $64^{\circ}$ of longitude), zonally elongated after the Walker
circulation; and $16\times4$ tokens ($16^{\circ}$ of longitude by the full depth of the token grid), meridionally
elongated after the Hadley circulation. Each geometry partitions the $16\times48$ token grid into exactly twelve
windows of 64 tokens, so the three differ in shape at identical cost. The meridional window is the limiting
case: the token grid is sixteen rows deep, so one $16\times4$ window covers every row, and each token in a
$16^{\circ}$ sector of longitude attends to the whole $20^{\circ}$\,S--$20^{\circ}$\,N band at once.

Each geometry appears twice, once with the window grid aligned to the domain and once displaced by half a
window, by $(4,4)$, $(2,8)$ and $(8,2)$ tokens respectively. The displacement is how windowed attention carries information across window boundaries. It is implemented as a cyclic roll with the accompanying attention mask,
so that tokens carried from one edge of the domain to the other cannot attend to one another: the domain is a bounded sector of the Pacific, not a periodic channel. In its shifted form the meridional window is therefore split at the equator by this mask, so attention across the whole band occurs in one of the two meridional blocks. Attention has four heads at dimension 96, and each block
is a pre-norm residual attention followed by a pre-norm two-layer perceptron (MLP) with expansion factor four and a GELU activation.
The attention blocks have no relative position bias and no additive positional encoding. Position enters through the per-token pooling weights of Eq.~(\ref{eq:pool}), the per-token basis and offset of Eq.~(\ref{eq:decoder}) and the fields themselves (their mean state and coastline), while the windows decide which tokens exchange information. The choice of window geometry is therefore a physical modeling assumption about the circulation, not merely an implementation detail.

The 768 token features are reduced to the state by a learned spatial weighting, rather than by global average
pooling or a summary token,
\begin{equation}
\vec{z}=\sum_{p=1}^{768}\vec{w}_p\odot\vec{h}_p,\qquad \vec{w}_p\in\mathbb{R}^{32},
\label{eq:pool}
\end{equation}
where $\vec{h}_p$ is the feature of token $p$ after a linear projection from 96 to 32 dimensions and $\odot$ is
the element-wise product. Thus, each of the 32 components of the latent state, which we call the latent coordinates, has its own 768-element weight map over the domain,
24\,576 parameters in all. The maps are initialized uniformly at $1/768$, so that at the start of training
$\vec{z}$ is the domain average of the token features, and each coordinate acquires a distinct weight map only as training proceeds.

\begin{algorithm}[t]
\caption{Encoding one monthly field into the latent state. MSA: multi-head self-attention; LN: layer normalization; MLP: two-layer perceptron.}
\label{alg:encoder}
\begin{algorithmic}[1]
\REQUIRE standardized field $\mathbf{X}\in\mathbb{R}^{13\times161\times681}$ (land set to 0)
\ENSURE latent state $\vec{z}\in\mathbb{R}^{32}$
\STATE $\mathbf{X}_{\mathrm{p}}\leftarrow$ AvgPool$_{4\times4}(\mathbf{X})$ \hfill $13\times40\times170$
\STATE $\mathbf{X}_{\mathrm{p}}\leftarrow$ replicate-pad $\mathbf{X}_{\mathrm{p}}$ to $13\times64\times192$
\STATE $\mathbf{H}\leftarrow$ Conv$_{4\times4,\ \mathrm{stride}\ 4}(\mathbf{X}_{\mathrm{p}})$ \hfill 768 tokens $\times$ 96
\FOR{window size $(n_1,n_2)$ in $(8,8),(8,8),(4,16),(4,16),(16,4),(16,4)$, alternating shift $0$ and $(n_1/2,n_2/2)$}
\STATE roll $\mathbf{H}$ by the shift; partition into twelve $n_1\times n_2$ windows; mask pairs of tokens that the
roll brought from opposite edges
\STATE $\mathbf{H}\leftarrow\mathbf{H}+\mathrm{MSA}_{4\ \mathrm{heads}}(\mathrm{LN}(\mathbf{H}))$ within each
window
\STATE $\mathbf{H}\leftarrow\mathbf{H}+\mathrm{MLP}_{96\to384\to96}(\mathrm{LN}(\mathbf{H}))$; roll back
\ENDFOR
\STATE $\vec{h}_p\leftarrow\mathbf{W}_{\mathrm{proj}}\,\mathbf{H}_p$ for every token $p$ \hfill $96\to32$
\STATE $\vec{z}\leftarrow\sum_{p}\vec{w}_p\odot\vec{h}_p$ \hfill Eq.~(\ref{eq:pool})
\RETURN $\vec{z}$
\end{algorithmic}
\end{algorithm}

\subsection{Decoder}
\label{sec:decoder}

The decoder maps the 32-dimensional state back to the thirteen fields, approximately inverting the encoder, in a
form that is transparent at its first stage. A state $\vec{z}$ is expanded
against a learned per-token basis $\mathbf{\Phi}\in\mathbb{R}^{768\times32}$ and a learned per-token offset $\mathbf{\Psi}\in\mathbb{R}^{768\times32}$, and mapped back to the 96-dimensional token space by a linear layer with weight $\mathbf{W}\in\mathbb{R}^{96\times32}$ and bias $\vec{b}_0$:
\begin{equation}
\tilde{\vec{h}}_p=\mathbf{W}\big(\vec{z}\odot\vec{\Phi}_p+\vec{\Psi}_p\big)+\vec{b}_0
=\sum_{d=0}^{31}z_d\,\Phi_{pd}\,\mathbf{W}_{\cdot d}+\big(\mathbf{W}\vec{\Psi}_p+\vec{b}_0\big).
\label{eq:decoder}
\end{equation}
The token field produced by this first stage is therefore affine in the state (linear plus a constant): 32 fixed spatial patterns, one per coordinate,
superposed on a fixed learned background field. This structure is similar to that of an EOF expansion, with patterns learned
jointly with the encoder instead of eigenvectors of a covariance matrix. It is what makes it meaningful to speak of the token-level footprint of coordinate $d$, the column $\vec{\Phi}_{\cdot d}$ of $\mathbf{\Phi}$. The affine stage is followed by six
window-attention blocks in the mirror order of the encoder, a transposed $4\times4$ convolution back to thirteen
channels on the $64\times192$ grid, removal of the replicated margin and bilinear resampling to
$161\times681$. These later stages are not linear in $\vec{z}$, so the complete map from state to field is
nonlinear. One consequence is used throughout: the blend of Sect.~\ref{sec:blend} is applied to decoded
indices, not to latent states, because decoding does not commute with a linear combination of states.

\subsection{Representation objective and resolution}
\label{sec:ae-loss}

The autoencoder is trained on reconstruction with two penalties on the latent state $\vec{z}$:
\begin{equation}
\mathcal{L}_{\mathrm{AE}}=\big\lVert\mathbf{X}-\tilde{\mathbf{X}}\big\rVert^2_{\mathrm{MSE}}
+\lambda_{\mathrm{dec}}\sum_{i\neq j}\big[\mathrm{Corr}(\vec{z})\big]_{ij}^{2}
+\frac{\lambda_{2}}{32}\,\mathbb{E}\big[\lVert\vec{z}\rVert^2\big],
\label{eq:aeloss}
\end{equation}
with $\lambda_{\mathrm{dec}}=10^{-4}$ and $\lambda_2=10^{-3}$. The reconstruction term is a mean squared error (MSE) over all thirteen channels and grid points in standardized units. In the second term, $\mathrm{Corr}(\vec{z})$ is the $32\times32$ correlation matrix of the latent state across the minibatch (the months drawn for one optimization step), and the term penalizes its off-diagonal entries. Because the minibatch holds 32 samples and the latent state has 32 dimensions, that matrix is a noisy estimate, and the penalty acts as a stochastic regularizer accumulated over training rather than a constraint enforced at any one step. The decorrelation achieved is therefore measured after training (Sect.~\ref{sec:res-coords}). The third term, in which $\mathbb{E}$ is the mean over the minibatch, bounds the size of the latent state, so the network cannot inflate latent variance to carry structure the decoder
cannot use. Together the penalties remove the incentive to concentrate variance in a few leading directions,
which is why the latent space is close to isotropic, with nearly the same variance in every direction.

Thirty-two numbers cannot hold a $13\times161\times681$ field, and the architecture sets the character of what is
lost. Pooling by four followed by four-cell patches means that no structure finer than a $4^{\circ}$ token is
represented; the final bilinear resampling restores the grid but not the information. The representation is
therefore a low-pass filter with a known cut-off, and results obtained through it are statements about the
large-scale field. Its fidelity as a function of scale is measured in Sect.~\ref{sec:res-fidelity}.

The autoencoder is trained on the full 1981--2024 record. This is deliberate: it uses all the available data. The representation is unsupervised:
it never pairs a state with a later state, it is scored against the raw reanalysis rather than against any
forecast target, and every anomaly used by the operators is formed from a climatology of the training months of
each fold. As a control, Sect.~\ref{sec:res-xro} compares the latent state with 32 EOFs of the same fields
computed both ways, from the full record and from the training months of each fold.

\subsection{The Floquet modal operator (FMO)}
\label{sec:operator}

\subsubsection{Input}

The FMO advances the latent anomaly $\vec{z}'(t)=\vec{z}(t)-\bar{\vec{z}}_{m(t)}$, where $t$ is the month index of the record, $m(t)\in\{0,\dots,11\}$ its calendar month (January $=0$), and $\bar{\vec{z}}_m$ is the mean latent state of calendar month $m$ over the training months of the fold (Sect.~\ref{sec:cv}). A forecast is initialized at month $t$, the last month of its history. The input of the FMO is a phase vector
assembled from the nine-month history $\vec{z}'(t-8),\dots,\vec{z}'(t)$: the current anomaly, its three-month and eight-month differences $\Delta_j\vec{z}'=\vec{z}'(t)-\vec{z}'(t-j)$, and a convolutional summary of the history,
\begin{equation}
\vec{\phi}=\big[\,\vec{z}'(t),\ \Delta_3\vec{z}',\ \Delta_8\vec{z}',\ \vec{c}\,\big]\in\mathbb{R}^{128},
\label{eq:phase}
\end{equation}
where $\vec{c}\in\mathbb{R}^{32}$ is the output of a one-dimensional convolution of kernel length nine over the
nine-month history of the 32 latent coordinates, so that it mixes over lag and over coordinates at once. The state handed to the propagator is a learned correction
of the current anomaly, $\vec{z}_0=\vec{z}'(t)+\mathrm{MLP}_{\phi}(\vec{\phi})$, where $\mathrm{MLP}_{\phi}$ is a two-layer network ($128\to32\to32$, SiLU activation) whose output layer is initialized at zero so that $\vec{z}_0=\vec{z}'(t)$ at the start of training.

\subsubsection{Modal generator}

The time-invariant part of the operator is a modal generator, the matrix that gives the rate of change of the state before the seasonal terms are added,
\begin{equation}
\mathbf{L}_0=\mathbf{Q}\begin{pmatrix}\mathbf{B}_1 & & & \\ & \mathbf{B}_2 & & \\ & & \mathbf{B}_3 & \\
& & & \mathbf{D}\end{pmatrix}\mathbf{Q}^{\top},
\qquad
\mathbf{B}_k=\begin{pmatrix}-\sigma_k & \omega_k\\ -\omega_k & -\sigma_k\end{pmatrix},
\label{eq:L0}
\end{equation}
where $\mathbf{Q}=\exp(\mathbf{\Omega}-\mathbf{\Omega}^{\top})$, with $\mathbf{\Omega}$ a learned $32\times32$ matrix, is orthogonal by construction. Each
$2\times2$ block $\mathbf{B}_k$ is a damped oscillator with angular frequency $\omega_k$ and damping rate
$\sigma_k$, and $\mathbf{D}$ is a diagonal block of 26 non-oscillatory decay modes, giving the full 32
dimensions. Frequencies and damping rates are bounded by their parameterization rather than learned freely. The damping rate is $\sigma_k=0.005+0.045\,\mathrm{sigmoid}(\cdot)$ per month, with a free parameter as the argument, and the 26 decay rates $d_j$ (the diagonal of $-\mathbf{D}$) are bounded in the same way, so that every mode is damped and stability is imposed rather than discovered. The frequency $\omega_k$ (in radians per month) is confined in the same way to periods $2\pi/\omega_k$ between one and ten years.
The oscillators are initialized at periods of 2.5, 4.0 and 6.0 years and at an identical damping of
$0.0275$\,month$^{-1}$, the midpoint of the bound, so that any ordering of the damping rates after training is
acquired rather than assumed.

\subsubsection{Seasonal modulation}

The generator is modulated by the calendar to represent ENSO's phase locking to the seasonal cycle, giving a
linear but time-periodic (Floquet) system,
\begin{equation}
\mathbf{L}_m=\mathbf{L}_0+\cos\theta_m\,\mathbf{L}_c+\sin\theta_m\,\mathbf{L}_s,\qquad
\theta_m=\frac{2\pi m}{12},\qquad \mathcal{T}_m=\exp(\mathbf{L}_m),
\label{eq:floquet}
\end{equation}
so that twelve monthly propagators $\mathcal{T}_m$, $m=0,\dots,11$ (January $=0$), follow from one shared
generator.
The seasonal terms $\mathbf{L}_c=\mathbf{W}_c-\mathbf{W}_c^{\top}$ and $\mathbf{L}_s=\mathbf{W}_s-\mathbf{W}_s^{\top}$, with $\mathbf{W}_c$ and $\mathbf{W}_s$ learned $32\times32$ matrices, are skew-symmetric by construction. This restricts the seasonal modulation to a rotation of the state: it can move energy
between latent directions but cannot add or remove it, so all growth and decay sits in $\mathbf{L}_0$.

\subsubsection{Forecast}

A forecast initialized in calendar month $m=m(t)$ is the corrected initial state $\vec{z}_0$, the latent anomaly encoded from the reanalysis fields plus its learned correction, advanced by the ordered product of the propagators of the months traversed, scaled by a bounded, state-dependent gain, plus two additive terms that depend only on the phase vector at the initialization month:
\begin{equation}
\vec{r}(l)=\underbrace{g(\vec{\phi})\;\vec{z}_0^{\top}\,\mathcal{T}_{m+1}\mathcal{T}_{m+2}\cdots\mathcal{T}_{m+l}}
_{\text{Floquet propagation}}
+\underbrace{\gamma\,\mathrm{MLP}(\vec{\phi},\theta_m)\big|_l+\mathbf{W}_{\mathrm{out}}(\mathbf{W}_{\mathrm{in}}\vec{\phi})\big|_l}
_{\text{additive terms}},\qquad l=1,\dots,24.
\label{eq:fmo}
\end{equation}
The gain $g(\vec{\phi})=1+0.5\tanh(\cdot)\in[0.5,1.5]$ has as its argument a linear function of $\vec{\phi}$, batch-normalized without affine parameters. It therefore averages one across a minibatch and redistributes amplitude between forecasts rather than inflating all of them. Being one positive factor on the whole trajectory, it changes amplitude but not sign. In the gated residual $\gamma\,\mathrm{MLP}(\vec{\phi},\theta_m)$, $\gamma$ is a learned scalar gate. The MLP is a two-layer network of 64 hidden units with layer
normalization, a SiLU activation and dropout of 0.2. It takes the phase vector and seasonal encoding $(\sin\theta_m,\cos\theta_m)$
of the initialization month as input, and predicts the full $24\times32$ output at once. The low-rank direct term projects $\vec{\phi}$ to rank four ($\mathbf{W}_{\mathrm{in}}\in\mathbb{R}^{4\times128}$) and expands it ($\mathbf{W}_{\mathrm{out}}\in\mathbb{R}^{768\times4}$) to the same $24\times32$ output. The two additive terms capture what a seasonally modulated linear propagator cannot. The gated residual allows a nonlinear, sign-asymmetric dependence on the initial state. The direct term gives each lead its own linear path from the initial state, in addition to the path through the Floquet propagation. The subscript $|_l$ selects the block of lead $l$ (in months), and the month subscripts are taken modulo 12. The state is propagated as a row vector; written for a column vector, one month of propagation is $\vec{v}\leftarrow\mathcal{T}^{\top}\vec{v}$, as in Algorithm~\ref{alg:fmo}. The periods and damping rates of $\mathbf{L}_0$ are the same in both forms.

The FMO's output $\vec{r}(l)$ is a correction to the base forecast $\vec{b}(l)$ of Sect.~\ref{sec:blend}: the
corrected latent forecast is $\hat{\vec{z}}(l)=\vec{b}(l)+\vec{r}(l)$, and this sum is what the training loss
scores (Sect.~\ref{sec:training}).

The interpretation of the operator rests on two properties of Eq.~(\ref{eq:fmo}). First, both additive terms are
functions of the initial phase vector alone and are not applied recursively; the only part of the forecast that
compounds with lead is the ordered product of propagators. Section~\ref{sec:res-modes} reports the periods and
damping rates of its time-invariant generator $\mathbf{L}_0$. Second, every learned term beyond the propagation of $\vec{z}'(t)$ is zero at the start of training: $\mathbf{W}_{\mathrm{out}}$, the
output layers of the two MLPs and the linear layer of the gain are initialized at zero, and $\gamma$ is held at
zero for the first 40\,\% of training. Training therefore starts from a pure Floquet propagation of the observed
state, and each of these terms is acquired only insofar as it lowers the loss. Algorithm~\ref{alg:fmo} details the complete forecasting workflow.

\begin{algorithm}[t]
\caption{One FMO forecast, from an initialization month to the reported Ni\~no 3.4 forecast.}
\label{alg:fmo}
\begin{algorithmic}[1]
\REQUIRE latent states $\vec{z}(t-8),\dots,\vec{z}(t)$; fold climatologies $\bar{\vec{z}}_m$ (latent state) and $\bar{N}_m$ (SST averaged over the Ni\~no 3.4 box); frozen decoder $\mathrm{Dec}$; base coefficients $\boldsymbol{\beta}_l$; ten trained operators; $\alpha(l)$
\ENSURE Ni\~no 3.4 forecast $\hat{y}(l)$, $l=1,\dots,24$
\STATE $\vec{z}'(i)\leftarrow\vec{z}(i)-\bar{\vec{z}}_{m(i)}$ for $i=t-8,\dots,t$
\STATE $\vec{\phi}\leftarrow[\vec{z}'(t),\Delta_3\vec{z}',\Delta_8\vec{z}',\vec{c}]$
\hfill Eq.~(\ref{eq:phase})
\STATE $\vec{b}(l)\leftarrow\boldsymbol{\beta}_l^{\top}[N(t),\,W(t),\,\Delta_3N,\,\Delta_3W,\,1]$ for $l=1,\dots,24$ \hfill base, Eq.~(\ref{eq:base})
\FOR{each of the ten operators}
\STATE $\vec{z}_0\leftarrow\vec{z}'(t)+\mathrm{MLP}_{\phi}(\vec{\phi})$; $\vec{v}\leftarrow\vec{z}_0$
\FOR{$l=1$ \TO $24$}
\STATE $\vec{v}\leftarrow\mathcal{T}_{m(t)+l}^{\top}\vec{v}$ \hfill one month of Floquet propagation
\STATE $\vec{r}(l)\leftarrow g(\vec{\phi})\,\vec{v}+\gamma\,\mathrm{MLP}(\vec{\phi},\theta_{m(t)})|_l
+\mathbf{W}_{\mathrm{out}}\mathbf{W}_{\mathrm{in}}\vec{\phi}|_l$
\ENDFOR
\ENDFOR
\STATE $\langle\vec{r}\rangle(l)\leftarrow$ mean of $\vec{r}(l)$ over the ten operators
\STATE $y_{\mathrm{B}}(l)\leftarrow$ Ni\~no~3.4 box mean of $\mathrm{Dec}\big(\vec{b}(l)+\bar{\vec{z}}_{m(t)+l}\big)$, minus $\bar{N}_{m(t)+l}$
\STATE $y_{\mathrm{F}}(l)\leftarrow$ Ni\~no~3.4 box mean of $\mathrm{Dec}\big(\vec{b}(l)+\langle\vec{r}\rangle(l)+\bar{\vec{z}}_{m(t)+l}\big)$, minus $\bar{N}_{m(t)+l}$
\STATE $\hat{y}(l)\leftarrow y_{\mathrm{B}}(l)+\alpha(l)\,\big[y_{\mathrm{F}}(l)-y_{\mathrm{B}}(l)\big]$
\hfill Eq.~(\ref{eq:blend})
\RETURN $\hat{y}(1{:}24)$
\end{algorithmic}
\end{algorithm}

\subsection{Base and blend}
\label{sec:blend}

The base is an ordinary least-squares (OLS) forecast of the latent anomaly from four observed predictors at
initialization: Ni\~no 3.4, WWV and their three-month changes. For each lead $l$ and each fold,
\begin{equation}
\vec{b}(l)=\boldsymbol{\beta}_l^{\top}\big[N(t),\ W(t),\ N(t)-N(t-3),\ W(t)-W(t-3),\ 1\big],
\label{eq:base}
\end{equation}
with $\boldsymbol{\beta}_l\in\mathbb{R}^{5\times32}$ fitted on the training windows of the fold (a window is a nine-month history followed by 24 monthly leads; Sect.~\ref{sec:cv}) and $N$ and $W$ the
Ni\~no 3.4 and WWV anomalies, respectively. We call this system OLS4. Its coefficients do not depend on the initialization
month. After performing the main comparison (Sect.~\ref{sec:res-skill}), this choice was tested against two alternatives with seasonal coefficients, each judged on training data only (Sect.~\ref{sec:sens-base}); neither met its acceptance criterion, and OLS4 is kept as the base of LatENSO v1.0.

The reported forecast is a fixed-weight blend of the decoded base and the decoded corrected forecast,
\begin{equation}
\hat{y}(l)=y_{\mathrm{B}}(l)+\alpha(l)\,\big[y_{\mathrm{F}}(l)-y_{\mathrm{B}}(l)\big],\qquad
\alpha(l)=\begin{cases}0.25, & l\le18,\\ 0, & l>18,\end{cases}
\label{eq:blend}
\end{equation}
applied to decoded indices for the reason given in Sect.~\ref{sec:decoder}. The blend is needed because the correction does not carry over equally to every fold. Unblended, the corrected forecast has a higher anomaly correlation coefficient (ACC; Sect.~\ref{sec:metrics}) than its base in the first and last folds and a lower one in the other two. Over leads 1--18 its ACC falls below that of the base (0.531 against 0.565; Table~\ref{tab:skill} and Sect.~\ref{sec:res-base-gain}). We call the blended system OLS4+FMO and the unblended one ($\alpha=1$) OLS4+FMO (unblended). The weight and the cut-off at lead 18 were set
on the cross-validated forecasts: $\alpha=0.25$ lies close to the value that maximizes the ACC over leads 1--18 (0.31), and ACC stays within 0.005 of its maximum for $\alpha$ between 0.22 and
0.42 (Appendix~\ref{app:alpha}). Beyond lead 18 the reported forecast is the base alone. The correlation, across the verification windows, between the correction $y_{\mathrm{F}}-y_{\mathrm{B}}$ and the error of the base (the observed anomaly minus $y_{\mathrm{B}}$) fades with lead, from $+0.26$ [$+0.02$, $+0.46$] at lead 17 to values whose interval includes zero from lead 18 on, turning negative from lead 22. Here and below, brackets give the 95\,\% interval of the paired bootstrap, and the evidence is graded as in Sect.~\ref{sec:stats}. A weight of 0.25 kept beyond lead 18 would lower the ACC below that of the base from lead 20 onward ($-0.113$ [$-0.213$, $-0.008$] at lead 24, clear
evidence). Neither the weight nor the cut-off was re-tuned afterward, and every comparison in this paper uses the
same weight. Because Eq.~(\ref{eq:blend}) is linear in $\alpha$, every score can be recomputed for any weight from the
stored forecasts (see Appendix~\ref{app:alpha}).

\subsection{Training}
\label{sec:training}

\subsubsection{Autoencoder}

The autoencoder is trained once, for 2500 epochs at batch size 32 over the 522-month record (about 42\,500
optimization steps), with CA-AdamW, a curvature-aware variant of AdamW \citep{AnEtAl2026}, at learning rate $2\times10^{-3}$, weight decay $10^{-4}$, curvature scale 0.1, a warm-up of 50 steps followed by cosine decay to $5\times10^{-4}$,
gradient clipping at unit norm and mixed precision. It has 1\,462\,317 parameters and takes 1.9\,h on one
NVIDIA RTX 4090 (2.8\,s per epoch). It is frozen thereafter.

\subsubsection{Operator}

Within each fold, the operator is trained on the training windows to predict the latent anomaly trajectory as a
correction to the base (Algorithm~\ref{alg:train}). The objective combines terms computed in latent space
with terms computed on the two indices after decoding:
\begin{equation}
\mathcal{L}=\mathcal{L}^{\mathrm{MSE}}_{\mathrm{lat}}+\mathcal{L}^{\mathrm{ACC}}_{\mathrm{lat}}
+\frac{1}{|S|}\sum_{l\in S}\Big[2\,\mathcal{L}^{\mathrm{MSE}}_{N}(l)+\mathcal{L}^{\mathrm{ACC}}_{N}(l)
+2\,\mathcal{L}^{\mathrm{MSE}}_{W}(l)+\mathcal{L}^{\mathrm{ACC}}_{W}(l)\Big].
\label{eq:fmoloss}
\end{equation}
Here $\mathcal{L}^{\mathrm{ACC}}=1-r$, with $r$ the Pearson correlation across the minibatch (for the latent term, averaged over the 24 leads and 32 coordinates), so that
correlation is optimized directly rather than only through the squared error. The latent squared error is weighted
linearly by lead from 0.5 at lead 1 to 3.0 at lead 24, so that long leads are not dominated by short ones. The
index terms need decoding, which is too expensive at all 24 leads for every minibatch: $S$ is a random subset of
six leads drawn per minibatch without replacement, with probability increasing linearly with lead over the same 0.5--3.0 range. The WWV terms are divided by the training-period standard deviation of the WWV anomaly.
Because the decoder is frozen, gradients from the index terms pass through it into the operator without
changing the representation. Evaluating a loss on decoded fields and differentiating it through a fixed,
pre-trained decoder or generator is an established procedure, used for example in latent-space inversion of geophysical
data, in variational data assimilation in the latent space of an autoencoder and in physics-informed solvers for
heterogeneous media \citep[e.g.][]{Laloy2019,Mosser2020,MelincZaplotnik2024,Panahi2026}.

Optimization uses CA-AdamW at learning rate $10^{-3}$, weight decay $10^{-4}$, curvature scale 0.1 and a
cosine schedule to $10^{-4}$, with gradients clipped at unit norm, for 301 epochs at batch size 32. The gate
$\gamma$ is held at zero, with its gradient disabled, for the first 120 epochs and is then set to 0.05 and
trained, so that the linear operator is fitted before the gated residual is available. Ten operators are
trained per fold from ten fixed seeds, 40 in all, each with 79\,682 parameters; one takes on average 8.6\,min on
the same GPU and the forty take 5.7\,h. The reported forecast uses the mean of the ten corrections
(Algorithm~\ref{alg:fmo}); no seed is selected.

\begin{algorithm}[t]
\caption{Training the ten FMO operators of one fold.}
\label{alg:train}
\begin{algorithmic}[1]
\REQUIRE frozen autoencoder; training windows of the fold (targets outside the held-out period and its 24-month buffers)
\ENSURE ten trained operators for the fold
\STATE compute the monthly climatologies $\bar{\vec{z}}_m$, $\bar{N}_m$, $\bar{W}_m$ (latent state, Ni\~no 3.4 and WWV box means) from the training months; form anomalies
\STATE fit the base $\boldsymbol{\beta}_l$ by least squares on the training windows, $l=1,\dots,24$
\FOR{seed $i=1$ \TO $10$}
\STATE initialize the operator; zero the output layers, the gain layer and $\mathbf{W}_{\mathrm{out}}$;
periods $2.5,4,6$\,yr; $\sigma_k=0.0275$\,month$^{-1}$
\FOR{epoch $j=1$ \TO $301$}
\IF{$j\le120$}
\STATE $\gamma\leftarrow0$, gradient off
\ELSIF{$j=121$}
\STATE $\gamma\leftarrow0.05$, gradient on
\ENDIF
\FOR{each minibatch of 32 windows}
\STATE $\hat{\vec{z}}(l)\leftarrow\vec{b}(l)+\vec{r}(l)$ for $l=1,\dots,24$ \hfill Eq.~(\ref{eq:fmo})
\STATE draw six leads $S$ with probability $\propto$ lead weight; decode $\hat{\vec{z}}(l)+\bar{\vec{z}}_{m(t)+l}$ for $l\in S$
\STATE evaluate $\mathcal{L}$ (Eq.~\ref{eq:fmoloss}); back-propagate through the frozen decoder; clip; step
\ENDFOR
\STATE cosine learning-rate step
\ENDFOR
\ENDFOR
\RETURN the ten operators
\end{algorithmic}
\end{algorithm}

\subsection{XRO with the latent state}
\label{sec:xrolat}

The second operator is the extended nonlinear recharge oscillator \citep{Zhao2024},
\begin{equation}
\frac{\mathrm{d}\vec{X}}{\mathrm{d}t}=\mathbf{L}(t)\,\vec{X}+\vec{N}(\vec{X})+\boldsymbol{\xi},\qquad
\mathbf{L}(t)=\mathbf{A}_0+\sum_{j=1}^{2}\big[\mathbf{A}^{c}_j\cos(j\omega_a t)+\mathbf{A}^{s}_j\sin(j\omega_a t)\big],
\label{eq:xro}
\end{equation}
where $\omega_a=2\pi/(12\ \mathrm{months})$ is the annual frequency and $\mathbf{A}_0$, $\mathbf{A}^c_j$ and $\mathbf{A}^s_j$ are the five fitted coefficient matrices (unrelated to the generator $\mathbf{L}_0$ of Eq.~\ref{eq:L0}). The first two variables of $\vec{X}$ are the ENSO temperature $T$ (the Ni\~no 3.4 anomaly, $N$ in Eq.~\ref{eq:base}) and the recharge $h$ (the WWV anomaly, $W$ in Eq.~\ref{eq:base}; we keep the XRO notation here). The vector $\vec{N}$ collects the quadratic terms $T^2$ and $Th$ in the $T$ equation, and
$\boldsymbol{\xi}$ is noise. The operator is fitted by least squares to monthly tendencies with the XRO code, with
its seasonal phase anchored to January, and forecasts are deterministic (the noise is switched off). In the published XRO the
remaining state variables are indices of other modes, namely the two meridional modes and the six indices of other basins listed in
Sect.~\ref{sec:indices}. Here they are replaced by directions of the latent state.

The 32 latent coordinates cannot all be added to the XRO state. Its seasonal operator carries five coefficient matrices (a mean plus annual and semi-annual harmonics), so a 34-variable XRO ($T$, $h$ and the 32 latent coordinates) would need about $34^2 \times 5 \approx 5800$ coefficients, to be fitted to roughly 30 years of monthly data. Two kinds of compact directions (linear combinations of the latent coordinates) are therefore built from the standardized latent anomalies of the training months (Algorithm~\ref{alg:xrolat}). \emph{Slow} directions maximize the lag-6 autocorrelation, that is, the persistence of the latent state. After removing a linear trend, they are the leading solutions of the generalized eigenproblem
$\mathbf{C}_6\vec{x}=\mu\,\mathbf{C}_0\vec{x}$, where $\mathbf{C}_0$ is the covariance and $\mathbf{C}_6$ the symmetrized six-month lag covariance of the latent anomalies, and $\mu$ is the lag-6 autocorrelation of the corresponding direction.
\emph{Predictive} directions carry information on the future Ni\~no~3.4 anomaly that the indices already in the state do not. They are obtained in three steps. First, at each lead $l = 3, 6, \ldots, 18$ months, both the latent anomalies and the Ni\~no~3.4 anomaly at lead $l$ are regressed on these indices, and the residuals are retained. Second, a ridge regression (penalty $n\,\mathbf{I}$, with $n$ the number of training pairs) of the residual target on the residual latent anomalies gives one weight vector per lead. Third, the leading left singular vectors of the matrix formed by stacking the six weight vectors define the directions.

The projections of the standardized latent anomalies onto the $K$ directions, $e_1,\ldots,e_K$, are then added to the XRO state as extra variables. In this latent-augmented state each variable, $T$ and $h$ included, is divided by its standard deviation over the training months, and the forecast is rescaled afterward.

We compare eleven designs for the latent contribution to the XRO state. The first group uses the predictive directions in place of all other indices, $\vec{X}=[T,h,e_1,\dots,e_K]$, with $K=2$, 4 and 8, where $K$ is capped at the six directions that exist. The second group uses the slow directions in the same way, with $K=2$ and 4. In the third group, the predictive directions are added to the two in-domain meridional modes, giving $\vec{X}=[T,h,\mathrm{NPMM},\mathrm{SPMM},e_1,\dots,e_K]$ with $K=2$ and 4. The last group keeps all 32 latent coordinates and controls their number effectively through a ridge penalty on their coefficients, applied at four strengths (0.3, 0.1, 0.03 and 0.01 times the mean diagonal entry of the normal-equation matrix for those coefficients).

Within each training fold, the design is selected by an inner cross-validation. We hold out one 48-month block of training months at a time, separated from the remaining data by an 18-month buffer, and score each design by its mean ACC over leads 1--18. Any design whose Ni\~no 3.4 forecast exceeds ten standard deviations of the training Ni\~no 3.4 anomaly at any lead in any inner block is excluded as unstable. Among the remaining designs, we keep the simplest one that scores within 0.005 of the best (the designs are ranked from the simplest in the order of Table~\ref{tab:designs}), which favors parsimony when differences are negligible. Every design adds at least two latent variables, so XRO-dom itself is not a candidate (Sect.~\ref{sec:res-xro} reports the selection with it). The selected design is then refitted on all training months and used to forecast the held-out period. We call the result XRO-lat.

To separate the contribution of the latent state from that of the pipeline itself, we run two controls through the identical procedure, replacing the latent coordinates with a different set of candidate variables. The first uses 32 EOFs of the same thirteen fields (XRO-EOF; the principal components of the standardized fields after the average pooling of the encoder, with EOFs and climatology from the full record), which tests whether the autoencoder adds anything beyond a linear compression of the same size. The second uses the eight remote-basin indices of the published XRO (XRO-remote), and serves as a positive control.

\begin{algorithm}[t]
\caption{XRO with the latent state, for one outer fold.}
\label{alg:xrolat}
\begin{algorithmic}[1]
\REQUIRE latent anomalies $\vec{z}'$; indices $T$, $h$, NPMM, SPMM; training months $\mathcal{M}$ of the fold; the eleven designs
\ENSURE XRO-lat forecasts for the held-out windows of the fold
\STATE split $\mathcal{M}$ into blocks of 48 months
\FOR{each design $i$}
\FOR{each block $j$}
\STATE $\mathcal{M}_j\leftarrow\mathcal{M}$ without block $j$ and 18 months on either side
\STATE standardize $\vec{z}'$ on $\mathcal{M}_j$; build the directions of design $i$ on $\mathcal{M}_j$
\STATE fit XRO (Eq.~\ref{eq:xro}) on $\mathcal{M}_j$, January phase; forecast every month of block $j$
\ENDFOR
\STATE score$(i)\leftarrow$ mean ACC over leads 1--18 of the pooled block forecasts; $-\infty$ if any
forecast exceeds 10 standard deviations
\ENDFOR
\STATE $i^{*}\leftarrow$ first design, in order of simplicity, with score $\ge\max_i\mathrm{score}(i)-0.005$
\STATE rebuild the directions of design $i^{*}$ on $\mathcal{M}$; fit XRO on $\mathcal{M}$; forecast the held-out
windows
\RETURN the forecasts
\end{algorithmic}
\end{algorithm}

\begin{table*}[t]
\caption{Configuration of LatENSO v1.0. Window sizes are in tokens; one token is a $4^{\circ}\times4^{\circ}$
block of the pooled grid.}
\label{tab:config}
\begin{tabular}{ll}
\tophline
Setting & Value \\
\middlehline
\multicolumn{2}{l}{\textit{Autoencoder}} \\
\quad Input & 13 channels, $161\times681$ points, $0.25^{\circ}$, January 1981--June 2024 \\
\quad Pooling, patches, tokens & $4\times4$ mean; $4\times4$ patches; $16\times48=768$ tokens of dimension 96 \\
\quad Window geometries & $8\times8$, $4\times16$, $16\times4$ tokens; each unshifted and shifted \\
\quad Attention & 4 heads; MLP expansion 4; no positional encoding \\
\quad Latent state & 32 dimensions; learned spatial weighting (24\,576 weights) \\
\quad Loss & MSE $+\,10^{-4}\sum_{i\neq j}[\mathrm{Corr}(\vec{z})]^2_{ij}+10^{-3}\,\mathbb{E}[\lVert\vec{z}\rVert^2]/32$ \\
\quad Training & 2500 epochs, batch 32, CA-AdamW, 1.9\,h on one GPU \\
\quad Parameters & 1\,462\,317 \\
\middlehline
\multicolumn{2}{l}{\textit{Floquet modal operator}} \\
\quad Phase vector & $[\vec{z}',\Delta_3\vec{z}',\Delta_8\vec{z}',\vec{c}]$ from a 9-month history, dimension 128 \\
\quad Oscillators & 3, initialized at 2.5, 4.0, 6.0\,yr; periods bounded to 1--10\,yr \\
\quad Damping & $0.005$--$0.05$\,month$^{-1}$ on all 32 modes, imposed \\
\quad Seasonality & 12 monthly propagators from one generator; skew-symmetric seasonal terms \\
\quad Gain and additive terms & gain in $[0.5,1.5]$; gated MLP (64 hidden); rank-4 direct term \\
\quad Base and blend & OLS on 4 observed predictors; $\alpha=0.25$ to lead 18, then 0 \\
\quad Training & 301 epochs, batch 32, $\gamma=0$ for 120 epochs; 10 seeds per fold; 8.6\,min each \\
\quad Parameters & 79\,682 per operator \\
\middlehline
\multicolumn{2}{l}{\textit{XRO with the latent state}} \\
\quad Operator & XRO, annual and semi-annual harmonics; $T^2$ and $Th$ terms; deterministic forecasts \\
\quad Extra state & 2--8 latent directions (slow or predictive), or all 32 with ridge; 11 designs \\
\quad Selection & nested cross-validation, 48-month inner blocks, 18-month buffer, mean ACC over leads 1--18 \\
\quad Cost & about 1\,s per fit on a CPU; nested selection about 5\,min per fold \\
\bottomhline
\end{tabular}
\end{table*}

\section{Evaluation protocol}
\label{sec:evaluation}

\subsection{Cross-validation}
\label{sec:cv}

Skill must be measured on years that played no part in the fitting. The difficulty lies in the length of the
record: a single test period would verify about one decade in four. Instead, we use temporal cross-validation over
four folds (Table~\ref{tab:folds}). Each fold holds out one period of about eleven years for verification,
1981--1991, 1991--2002, 2002--2013 or 2013--2024, and is trained on the remaining years, about 30. Around each
held-out period a 24-month buffer is removed from training, so that no verification target overlaps, even partly,
with a training window.

\begin{table*}[t]
\caption{The four folds of the cross-validation. In each fold, every model is fitted on the training months and
forecasts only the held-out period; the buffer months are removed from that fold's training. Only the autoencoder
is fitted on the full record (Sect.~\ref{sec:ae-loss}).}
\label{tab:folds}
\scriptsize
\begin{tabular}{llll}
\tophline
Fold & Held out for verification & Training months & Buffer, removed from training \\
\middlehline
1 & Jan 1981--Nov 1991 & Dec 1993--Jun 2024 (30.6 years) & Dec 1991--Nov 1993 \\
2 & Dec 1991--Sep 2002 & Jan 1981--Nov 1989 and Oct 2004--Jun 2024 (28.7 years) & Dec 1989--Nov 1991 and Oct 2002--Sep 2004 \\
3 & Oct 2002--Jul 2013 & Jan 1981--Sep 2000 and Aug 2015--Jun 2024 (28.7 years) & Oct 2000--Sep 2002 and Aug 2013--Jul 2015 \\
4 & Aug 2013--Jun 2024 & Jan 1981--Jul 2011 (30.6 years) & Aug 2011--Jul 2013 \\
\bottomhline
\end{tabular}
\end{table*}

By a forecast window we mean a nine-month history followed by 24 monthly leads. A window is held out only if all
its 33 months lie in the held-out period. This rule has a cost: there are no initializations in December 1989--July
1992, October 2000--May 2003, August 2011--March 2014 or after June 2022. The held-out periods thus contain 394
initializations (the verification windows), from 36 initialization years.

For the test to be fair, every quantity that depends on the data is recomputed inside each fold from its training
months: the latent, Ni\~no 3.4 and WWV climatologies, the base coefficients, the operators, the standardization and
directions of Sect.~\ref{sec:xrolat}, and the reference models of Sect.~\ref{sec:references}. The verifying
observation is the Ni\~no 3.4 anomaly with respect to the same training climatology. The one exception is the
autoencoder, which is fitted on the full record (Sect.~\ref{sec:ae-loss}).

\subsection{Reference models on identical data}
\label{sec:references}

Published skill curves cannot be compared with ours: they are verified over other years, with three-month means,
on other data and after training on other periods. We therefore refit every reference model on our data, our
folds and our 394 windows, and use the published curves only to check that each model was run correctly
(Appendix~\ref{app:references}).

\begin{itemize}
\item \textbf{nRO}: the nonlinear recharge oscillator, XRO restricted to $T$ and $h$, with its $T^2$ and $Th$
terms \citep{Zhao2024}. It is the recharge oscillator of \citet{Jin1997} with seasonal coefficients.
\item \textbf{XRO-dom}: XRO with only the indices that our domain supports, $[T,h,\mathrm{NPMM},\mathrm{SPMM}]$,
the meridional modes truncated as in Sect.~\ref{sec:indices}. Like our model, it uses only in-domain information.
\item \textbf{XRO} (also called the three-basin XRO below, for the Pacific, Indian and Atlantic Oceans): the published XRO model \citep{Zhao2024}, with the eight remote-basin indices: the six indices that describe climate modes in other basins and the meridional-mode indices calculated over the full boxes (Sect.~\ref{sec:indices}). It therefore uses information from outside the domain.
\item \textbf{DESN-dom} and \textbf{DESN}: the deep echo-state network of \citet{Zhang2026}, run with the
authors' code and settings (20 one-layer and 20 two-layer reservoirs of 20\,000 and 20\,000 + 12\,000 units,
ridge readout, averaged over the 40 members), on the XRO-dom and full XRO indices respectively.
\end{itemize}

XRO and nRO are fitted with the authors' code and options, the seasonal phase anchored to January in every fold.
XRO and nRO are deterministic least-squares fits, and DESN is run with its authors' seeds and settings; none was
tuned on our verification years. The base OLS4 (Sect.~\ref{sec:blend}) is reported too, but not as a reference model: its
coefficients do not vary with the season, so the like-for-like linear benchmark is the recharge oscillator, nRO.

\subsection{Metrics}
\label{sec:metrics}

The primary score is the ACC of monthly Ni\~no 3.4, that is, the Pearson correlation between the forecast and the observed anomaly. It is computed per lead over the 394 windows and averaged over leads 1--18 and over two lead ranges within them, leads 3--9 and 10--18 (Table~\ref{tab:skill}). Event
discrimination is measured by the area under the receiver operating characteristic (ROC) curve (AUC, the
Mann--Whitney statistic), for El Ni\~no (observed anomaly above $+1.0\,^{\circ}$C) and La Ni\~na (below $-1.0\,^{\circ}$C). The forecast anomaly is the score, negated for La Ni\~na, and the AUC is the probability that a window with an event has a higher score than a window without one. With this threshold the observed record has 11 runs of consecutive months above $+1.0\,^{\circ}$C and 14 below
$-1.0\,^{\circ}$C (some events split into several runs). At lead 12, 48 of the 394 windows verify on El Ni\~no
conditions (19, 12, 5 and 12 in the four folds) and 59 on La Ni\~na conditions (16, 18, 21 and 4). The threshold, stricter than the usual $\pm0.5\,^{\circ}$C, restricts the score to events of at least moderate
strength rather than weak anomalies. Appendix~\ref{app:blocks} reports thresholds from $+0.5$ to $+1.5\,^{\circ}$C.

Calibration is described by two regression slopes at each lead, the amplitude slope
$a=\mathrm{cov}(f,o)/\mathrm{var}(o)=r\,\sigma_f/\sigma_o$ and the reliability
$\rho=\mathrm{cov}(f,o)/\mathrm{var}(f)=r\,\sigma_o/\sigma_f$, with $f$ the forecast, $o$ the observation, $r$ their correlation, and $\sigma_f$ and $\sigma_o$ their standard deviations over the windows. Their product is $r^2$. A calibrated forecast, $o=f+\epsilon$ with $\epsilon$ uncorrelated
with $f$, has $\rho=1$ and $\sigma_f=r\,\sigma_o$, hence an amplitude slope of $r^2$. A calibrated forecast of imperfect skill therefore
has an amplitude slope below 1. Reliability below 1 means that the forecast is overconfident. The mean-square skill
score against the sample climatology, $\mathrm{MSSS}=1-\mathrm{MSE}/\mathrm{var}(o)$ \citep{Murphy1988}, combines the two. For unbiased anomalies $\mathrm{MSSS}=2rs-s^2$, with the amplitude ratio $s=\sigma_f/\sigma_o$; it reaches its maximum at $s=r$ (where it
equals $r^2$) and is equal to $2r-1$ at full amplitude.

\subsection{Statistics}
\label{sec:stats}

How much of a difference between two scores is due to the sample of years? This can be estimated by resampling. One
natural approach would be to resample the forecast windows one by one. However, overlapping 24-month windows
initialized in the same year are not independent, so this would understate the uncertainty. Instead, we use a
paired block bootstrap with the initialization year as the block: each bootstrap sample draws 36 initialization
years with replacement and keeps all windows of a drawn year together. Every score is recomputed on each sample,
with all systems scored on the same samples so that differences are paired, and each 95\,\% interval is the 2.5th
to 97.5th percentile of the resampled values. All intervals are 95\,\% confidence intervals computed from 10\,000
paired year-block bootstrap draws (seed 0). A 95\,\% interval that excludes zero is the strongest grade on the
evidence scale used below, called clear evidence.

Of course, a calendar year is not a perfect block. ENSO events and the corresponding 18-month forecast targets can
extend across calendar-year boundaries, which means that one-year bootstrap blocks may not fully preserve the
temporal dependence between forecast windows. As a result, uncertainty could be underestimated. To assess the
sensitivity of the results to block length, Appendix~\ref{app:blocks} repeats the main comparisons using blocks of
two, three and four consecutive years.

What can a record of this length resolve? With 36 initialization years, the smallest difference in mean ACC that
can be resolved (the half-width of the 95\,\% interval) ranges from about 0.02 to 0.10 depending on the pair of
systems (Table~\ref{tab:diff}), and many comparisons fall short of it. To report the strength of the evidence
rather than only its passage through the 95\,\% threshold, we also give the share of bootstrap resamples in which a
difference is positive, and we grade it. The evidence is clear when the share is at least 0.975 (or at most 0.025;
this is equivalent to a 95\,\% interval that excludes zero), strong between 0.95 and 0.975, moderate between 0.90
and 0.95, weak between 0.75 and 0.90, and inconclusive otherwise. For a negative difference, the same bounds apply
to the share of resamples in which it is negative. Note, however, that the share is a resampling frequency, and not
the probability that the effect is real. The evidence grades are intended to complement the intervals, not to
replace them.

Not all comparisons in this paper carry the same weight. Four comparisons were prespecified and evaluated using
acceptance criteria fixed in advance: (i) whether the latent state improves existing models when used as an
additive corrector (Appendix~\ref{app:designs}), (ii) whether including the latent state in XRO improves forecast
skill (Sect.~\ref{sec:xrolat}), and (iii--iv) whether using a seasonal or a lead-dependent base improves the
forecast operator (Sect.~\ref{sec:sens-base}). The retraining test of Sect.~\ref{sec:res-ablations} was likewise
read with criteria fixed before it was run. All other comparisons are exploratory and descriptive, and no
adjustment for multiple comparisons is applied to them. This means that, among so many comparisons, some will look
convincing by chance, and an isolated clear difference should not be overemphasized.

\subsection{Verification of the implementation}
\label{sec:verification}

Two checks were used to verify the results. First, the full training pipeline was rerun from scratch for one operator (one fold and one seed). The rerun reproduced the stored forecasts to within 0.006\,$^{\circ}$C (root mean square 0.001\,$^{\circ}$C), matched the stored ACC to four decimal places, and reproduced the latent base forecast to within $10^{-4}$. Second, XRO-lat and its two control configurations were independently re-implemented using only a written description of the protocol. The independent implementation reproduced every forecast to within $10^{-12}\,^{\circ}$C, made the same design choices in every fold, and yielded identical confidence intervals.

\subsection{Sensitivity to the base}
\label{sec:sens-base}

The FMO corrects a base forecast, so its skill inherits the strengths and weaknesses of that base. One weakness of
OLS4 is that it uses the same coefficients for every initialization month, although ENSO predictability depends
strongly on the season. This matters, because the operator falls short of the reference models at leads 3--9
(Sect.~\ref{sec:res-skill}), and the base alone, refitted with seasonal coefficients, gains $+0.069$ [$+0.023$,
$+0.108$] at leads 3--9 on the verification windows. After the results of Sect.~\ref{sec:results} were known, we
therefore tested whether a base that varies with the season removes this deficit, and whether the operator still
adds skill over it.

The difficulty lies in the timing: with the results on the verification windows already known, a base chosen on
those windows would be tuned to them. Instead, to keep the choice of base independent of the held-out periods, each
alternative was compared with OLS4 by cross-validation within the training data of each fold. It would have
replaced OLS4 only if it scored higher in at least three of the four folds, a criterion set before each comparison.

The first alternative is a seasonal base. It uses the same four predictors, with coefficients that vary with the
initialization month through annual and semi-annual harmonics. It scored higher in only one of the four folds
(Appendix~\ref{app:base}). The reason is a trade-off between leads: the seasonal base improved leads 3--9 in every
fold but overfitted leads 10--18 in every fold. OLS4 was therefore kept, and no operator was trained on this base.

The second alternative follows from this trade-off. It combines the two into a composite base, OLS4c: the seasonal
base up to a lead $L^{*}$, where seasonality helps, and OLS4 beyond it, where it overfits. The lead $L^{*}$ was
chosen in each fold from the training data (11, 8, 9 and 9 months), and all 40 operators were retrained on OLS4c
with the same seeds and settings. Of course, scoring a base on the data that chose this lead would favor it. With
$L^{*}$ chosen anew in each inner split, OLS4c scored higher in two of the four folds, so OLS4 remains the base of
the system described in this paper. The retrained system, OLS4c+FMO, serves in Sect.~\ref{sec:res-base} to measure
how much of the operator's deficits comes from its base, and Table~\ref{tab:skill} reports it, with OLS4c alone,
alongside all other systems.

\section{Results}
\label{sec:results}

Table~\ref{tab:skill} reports the performance of every forecasting system on each verification metric, and
Table~\ref{tab:diff} presents the paired differences between the two LatENSO operators and the reference models (and, for OLS4+FMO, its base). We
first assess the learned representation and then the predictive operators.

\subsection{Fidelity and compactness of the latent state}
\label{sec:res-fidelity}

\begin{figure*}[tp]
\includegraphics[width=\textwidth,height=0.73\textheight,keepaspectratio]{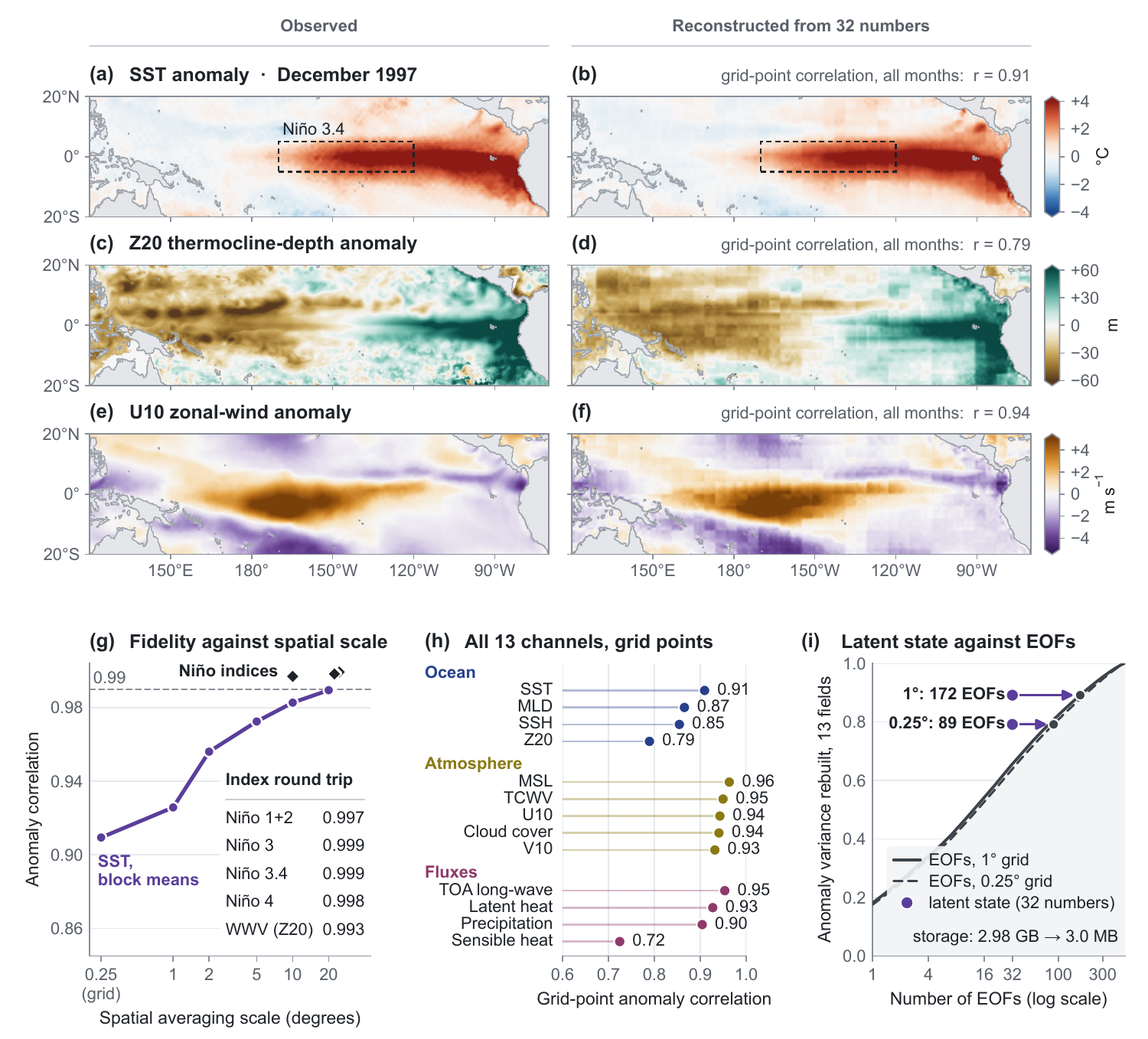}
\caption{Fidelity and compactness of the latent state. (a)--(f) Observed and reconstructed anomalies at the
December 1997 El Ni\~no peak, for SST, Z20 and U10. (g) Anomaly correlation of reconstructed SST against the
scale of spatial averaging (means over square blocks; blocks that are less than half ocean are left out), with the standard Ni\~no indices placed at the scale of their own boxes (three of the
four marks nearly coincide); the table summarizes the round-trip correlation of each index. (h) Grid-point anomaly correlation for all thirteen channels, grouped (and colored) into ocean, atmosphere and fluxes (TOA: top of the atmosphere). (i) Share of the anomaly variance of all thirteen fields
rebuilt by the leading EOFs (curves, at $1^{\circ}$ and $0.25^{\circ}$) and by the 32 latent numbers (dots); the
arrows mark the number of EOFs needed to match the latent state, and the last line gives the storage size of the
1981--2024 record of the thirteen fields and of its latent form (latent trajectories and decoder). The number above (b), (d) and (f) is the grid-point
anomaly correlation over all months. Color limits are $\pm4\,^{\circ}$C, $\pm60$\,m and $\pm5$\,m\,s$^{-1}$; values
beyond them take the end color. All statistics are calculated on ocean points only.}
\label{fig:fidelity}
\end{figure*}

The autoencoder discards structure finer than its $4^{\circ}$ tokens (Sect.~\ref{sec:ae-loss}). How much of the
large-scale field survives? And how does that compare with a classical basis of the same size?
Figure~\ref{fig:fidelity} answers both questions.

The fidelity of the reconstruction depends on the spatial scale. In this section and the next, anomalies are
departures from the monthly climatology of the full record, and grid-point statistics pool all months and all ocean
points. The anomaly correlation between the reconstructed and the observed SST is 0.909 on the native
$0.25^{\circ}$ grid. It rises when both fields are first averaged over square blocks, to 0.926 for blocks of
$1^{\circ}$, 0.972 for $5^{\circ}$ and 0.989 for $20^{\circ}$ (Fig.~\ref{fig:fidelity}g). The bottleneck is thus a
large-scale filter: it loses sub-degree detail and keeps the basin-scale field. In particular, both indices on
which the forecast is scored, recomputed from the reconstructed fields, survive the round trip with correlations
above 0.99: 0.999 for Ni\~no 3.4 (root-mean-square error 0.04\,$^{\circ}$C) and 0.993 for WWV. Their amplitude
ratios are 0.990 and 0.966, where by amplitude ratio we mean the standard deviation of the reconstructed anomaly
divided by that of the observed one.

Across the thirteen channels, grid-point correlations range from 0.725 for the surface sensible heat flux, which is
noisy and rich in small-scale detail, to 0.963 for sea-level pressure, and lie between 0.79 and 0.91 for the four
ocean fields (Fig.~\ref{fig:fidelity}h). The reconstructed amplitudes are close to the observed ones: eight of the
thirteen channels have amplitude ratios within 5\,\% of one. The subsurface fields are damped most (Z20: 0.89, MLD:
0.92, SSH: 0.95), whereas the sensible heat flux is slightly inflated (1.11).

Turning to the second question, the same 32 numbers are a compact description of the coupled fields
(Fig.~\ref{fig:fidelity}i). We measure compactness by the share of the anomaly variance of the thirteen
standardized fields that is rebuilt: one minus the sum of the squared errors divided by the sum of the squared
anomalies. For the leading EOFs of the same anomalies, this is their cumulative share of variance. At $1^{\circ}$
(the fields averaged over blocks of four by four grid points), the latent state rebuilds 0.891 of the anomaly
variance of all thirteen fields, whereas 32 EOFs of the same fields rebuild 0.656, and 172 EOFs are needed to match
it. At the native $0.25^{\circ}$, where the latent state rebuilds 0.791 of the anomaly variance, 89 EOFs are
needed. Note that both the EOFs and the autoencoder are fitted on, and evaluated on, the full record, so these are
in-sample fidelities for both. Per field, at the native resolution, the number of these EOFs needed has a median of
101 and ranges from 15 to 183. The one field for which 32 EOFs do better than the latent state is the surface
sensible heat flux (0.528 against 0.380 of its anomaly variance).

The latent representation also saves storage. As a lossy compression, the 2.98\,GB of the 1981--2024 record reduce
to 3.0\,MB for the latent trajectories and the decoder (in single precision, like the record), three orders of
magnitude less. An EOF basis of equal fidelity at $0.25^{\circ}$ holds $89\times1\,293\,706\approx115$ million
numbers (one value per EOF, field and ocean point), 79 times the 1.46 million parameters of the autoencoder.
High-fidelity decoding from a 32-dimensional state is what allows the operators of Sect.~\ref{sec:model} to remain
small. We return to its uses in Sect.~\ref{sec:discussion}.

\subsection{What the latent coordinates are}
\label{sec:res-coords}

\begin{figure*}[tp]
\includegraphics[width=\textwidth]{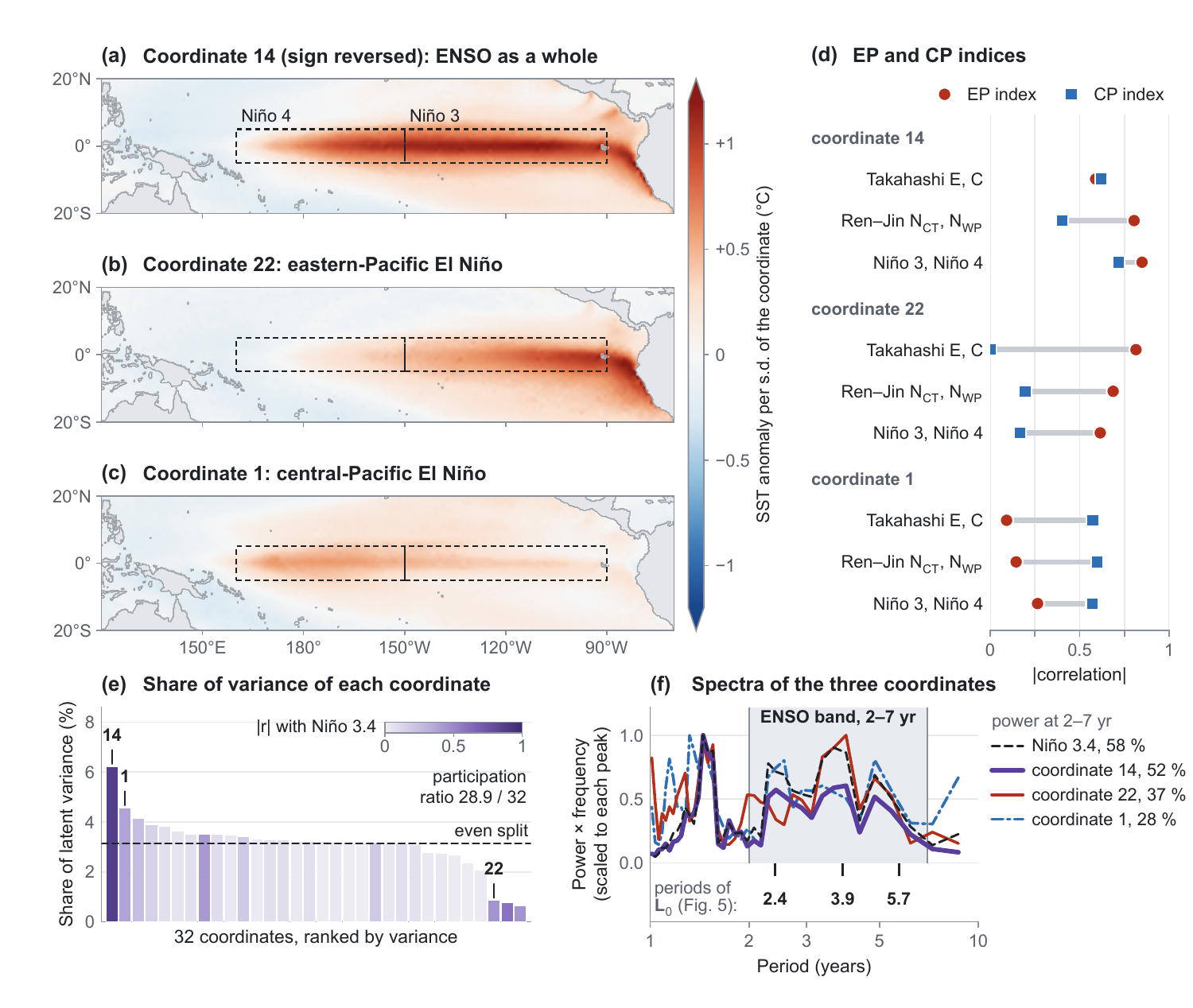}
\caption{The latent coordinates (numbered 0 to 31). (a)--(c) SST anomaly regressed on coordinates 14 (sign
reversed), 22 and 1, per standard deviation of the coordinate, with the Ni\~no 3 and Ni\~no 4 boxes marked.
(d) Absolute correlation of the three coordinates with three pairs of eastern-Pacific (EP, circles) and
central-Pacific (CP, squares) indices: $E$ and $C$ \citep{Takahashi2011}, $N_{\mathrm{CT}}$ and $N_{\mathrm{WP}}$
\citep{RenJin2011}, and Ni\~no 3 and Ni\~no 4. (e) Share of latent variance of all 32 coordinates, ranked, shaded by
the absolute correlation with Ni\~no 3.4, with the three coordinates marked. (f) Variance-preserving multitaper
spectra (time-bandwidth product 2, three tapers, as in Fig.~\ref{fig:modes}a) of the anomalies of the three coordinates and of Ni\~no 3.4 (not detrended), each scaled to its
own peak, with the periods of the generator $\mathbf{L}_0$ of the Floquet operator (Sect.~\ref{sec:res-modes}); the
legend gives the share of each spectrum's power, integrated over all frequencies above zero, in the 2--7-year band. The
peak near 1.45 years, present in Ni\~no 3.4 itself, is the combination tone of ENSO with the annual cycle
\citep{Stuecker2013}.}
\label{fig:coords}
\end{figure*}

What do the individual latent coordinates represent? Ranking them by variance alone would pick near-arbitrary
directions, because the latent space is close to isotropic. Its participation ratio, $1/\sum_d v_d^2$ with $v_d$
the variance of the latent anomaly in coordinate $d$ divided by the sum over all coordinates, counts how many
coordinates effectively share the variance. It is 28.9 of 32, and the coordinate with the largest share holds
6.2\,\% of the variance, against 3.1\,\% for an even split (Fig.~\ref{fig:coords}e). This is the intended effect of
the decorrelation and size penalties of Eq.~(\ref{eq:aeloss}), and not an emergent property. We therefore
characterize the coordinates by their association with ENSO instead.

ENSO as a whole is concentrated in one coordinate. Coordinate 14 (coordinates are numbered 0 to 31) has the largest
share of variance (6.2\,\%) and the strongest correlation with Ni\~no 3.4 ($r=-0.875$); only coordinates 14 and 2
exceed $|r|=0.5$. This coordinate also holds 52\,\% of its spectral power in the 2--7-year band, close to the
58\,\% of Ni\~no 3.4 itself (Fig.~\ref{fig:coords}f). Its footprint, the SST anomaly regressed on it, is the
canonical cold-tongue warming, strongest east of $150^{\circ}$\,W (Fig.~\ref{fig:coords}a). On the two box-based
pairs of indices (Sect.~\ref{sec:indices}) it leans toward the eastern Pacific, with $N_{\mathrm{CT}}$ ($-0.80$)
against $N_{\mathrm{WP}}$ ($-0.40$) and Ni\~no 3 ($-0.85$) against Ni\~no 4 ($-0.72$). However, it correlates
equally with $E$ and $C$ ($-0.59$, $-0.62$). It therefore tracks the amplitude of ENSO rather than its flavor
(eastern-Pacific or central-Pacific).

The flavor sits in two other coordinates (Fig.~\ref{fig:coords}b--d). Coordinate 22 correlates with the
eastern-Pacific index $E$ ($r=+0.81$) and not with $C$ ($0.00$), and with $N_{\mathrm{CT}}$ ($+0.69$) and Ni\~no 3
($+0.62$) rather than $N_{\mathrm{WP}}$ ($-0.20$) and Ni\~no 4 ($+0.17$). Its footprint is confined to the eastern
Pacific. Coordinate 1 correlates with every central-Pacific index and with none of the eastern-Pacific ones
($C=+0.57$ against $E=-0.09$, $N_{\mathrm{WP}}=+0.60$ against $N_{\mathrm{CT}}=+0.15$, Ni\~no 4 $=+0.57$ against
Ni\~no 3 $=+0.27$). Its footprint is centered near the date line. Unlike coordinate 14, coordinates 22 and 1
correlate only moderately with Ni\~no 3.4 ($+0.44$ and $+0.38$) and hold less of their power in the 2--7-year band
(37 and 28\,\%). Coordinate 1 carries the second-largest share of variance (4.6\,\%; Fig.~\ref{fig:coords}e, f).

The autoencoder thus separates a coordinate for ENSO as a whole from one coordinate for each flavor of El Ni\~no
\citep{Kug2009,Capotondi2015}, without any prior information on either. The eastern-Pacific coordinate carries little
variance (0.9\,\%, rank 30 of 32), so the flavor enters as a small correction to the common mode. No other
coordinate shows a comparable preference. For example, coordinate 24 correlates with all six indices at $-0.24$ to
$-0.40$. Of course, this classification is descriptive. It was made only after training, on the full record, and it
belongs to this training run: a retrained autoencoder would have a different basis.

\subsection{The modal spectrum}
\label{sec:res-modes}

\begin{figure*}[tp]
\includegraphics[width=0.92\textwidth]{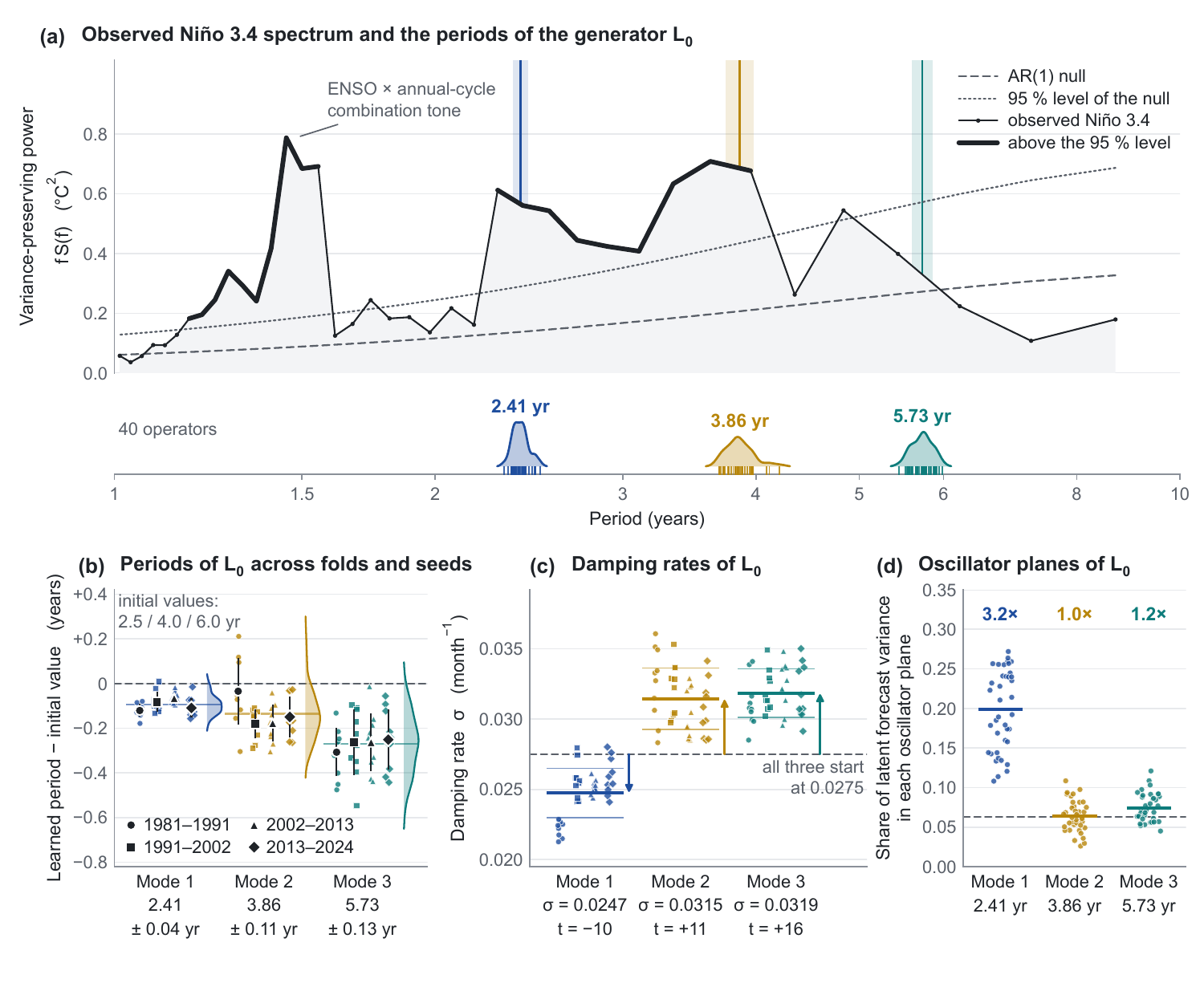}
\caption{Periods and damping rates of the time-invariant generator $\mathbf{L}_0$ of the Floquet operator,
across all 40 trained operators. (a) Multitaper spectrum
\citep{Thomson1982} of the observed Ni\~no 3.4 anomaly against a first-order autoregressive, AR(1), null, with the periods of $\mathbf{L}_0$;
dots mark the Fourier frequencies and the heavy line the part above the 95\,\% level of the null; vertical lines
and bands: mean and one standard deviation of each period over the 40 operators, whose distribution is shown in
the strip below. (b) Learned period of each oscillator minus its initial value; small markers: the 40 operators,
by fold; black markers and bars: mean and standard deviation of each fold. (c) Learned damping rate; all three
oscillators start at the same value (dashed); thick and thin lines: mean and one standard deviation; $t$:
one-sample $t$ statistic of the mean learned rate against the initial value (40 operators). (d) Share $\zeta_k$ of the
latent forecast variance in each oscillator plane of $\mathbf{L}_0$ (Eq.~\ref{eq:share}), against the value for an isotropic state
(dashed, 2/32); the numbers give the ratio of the mean share to that value. In (a), the anomaly is the Ni\~no 3.4 index of
the full record with its monthly climatology and linear trend removed. The spectrum $S(f)$, with $f$ in cycles per
month, uses three tapers with a time-bandwidth product of 2 and is drawn in variance-preserving form, $fS(f)$. The
null is $S_0(f)\propto(1-r_1^2)/(1-2r_1\cos2\pi f+r_1^2)$, with $r_1$ the lag-1 autocorrelation of the anomaly, scaled to
the variance of the anomaly, and its 95\,\% level is $S_0(f)$ times the 0.95 quantile of $\chi^2_{6}/6$, with two degrees of freedom for each of the three
tapers.}
\label{fig:modes}
\end{figure*}

In all 40 operators, the three oscillators of the time-invariant generator $\mathbf{L}_0$ of the Floquet operator
converge on the same periods $2\pi/\omega_k$, $2.405\pm0.041$, $3.864\pm0.115$ and $5.729\pm0.127$ years (mean
$\pm$ standard deviation over the 40 operators), with damping rates $\sigma_k$ of $0.0247\pm0.0018$,
$0.0315\pm0.0022$ and $0.0319\pm0.0017$\,month$^{-1}$ (Fig.~\ref{fig:modes}b, c). Note, however, that these are
parameters of $\mathbf{L}_0$, and not eigenperiods of the full seasonal system. The seasonal terms are larger than
$\mathbf{L}_0$ (Frobenius norms, averaged over the operators, of about 1.7 for $\mathbf{L}_c$ and $\mathbf{L}_s$
against 0.43 for $\mathbf{L}_0$), so the twelve-month propagator, the product of the twelve monthly propagators
$\mathcal{T}_m$ of Eq.~(\ref{eq:floquet}), mixes the modes of $\mathbf{L}_0$. In every operator it has 15 or 16
pairs of complex-conjugate eigenvalues (oscillatory pairs), among which the three oscillators cannot be singled
out.

Further properties bound how the $\mathbf{L}_0$ parameters can be read. The periods start in the ENSO band, at 2.5,
4.0 and 6.0 years, so they are displaced from their initial values within a range that the parameterization allows,
and not discovered from scratch. The displacement is consistent across the 40 operators of the four folds
(one-sample $t$ test, over the 40 operators, of the learned period against its initial value $x_0$; the statistic
is $\sqrt{40}\,(\bar{x}-x_0)/\mathrm{sd}(x)$, with $\bar{x}$ and $\mathrm{sd}(x)$ the mean and standard deviation of the learned periods;
$t_{39}=-14.5$, $-7.5$ and $-13.4$), and no period sits at a bound. Stability is imposed, so the fact that every
mode is damped carries no information. What is learned is the ordering of the damping rates. The three rates start
at the same value and separate during training (the same test of the learned damping rate against its initial
value, $t_{39}=-9.9$, $+11.4$, $+15.9$), and the 2.4-year mode becomes the least damped.

We next ask how much of the forecast variance each oscillator carries. The columns $\vec{q}_1,\dots,\vec{q}_{32}$
of the orthogonal matrix $\mathbf{Q}$ in Eq.~(\ref{eq:L0}) form an orthonormal basis of the latent space. In the
coordinates $\vec{u}=\mathbf{Q}^{\top}\vec{z}'$ of the latent anomaly, the generator $\mathbf{L}_0$ is
block-diagonal, so under $\mathbf{L}_0$ the pair $(u_{2k-1},u_{2k})$ of oscillator $k$ evolves on its own. It
rotates with angular frequency $\omega_k$ and decays at rate $\sigma_k$, because the eigenvalues of $\mathbf{B}_k$
are $-\sigma_k\pm\mathrm{i}\omega_k$. This is the same idea as the oscillatory normal modes of a linear inverse
model \citep{PenlandSardeshmukh1995}. We call $P_k=\mathrm{span}(\vec{q}_{2k-1},\vec{q}_{2k})$ the plane of
oscillator $k$. The 26 decay modes act on the remaining coordinates, each of which only decays, without rotation:
$u_j\propto\mathrm{e}^{-d_j\tau}$ after $\tau$ months, with $d_j$ its decay rate.

The share of the forecast variance that lies in $P_k$ is
\begin{equation}
\zeta_k=\frac{\vec{q}_{2k-1}^{\top}\hat{\mathbf{C}}\,\vec{q}_{2k-1}+\vec{q}_{2k}^{\top}\hat{\mathbf{C}}\,\vec{q}_{2k}}{\mathrm{tr}\,\hat{\mathbf{C}}},
\label{eq:share}
\end{equation}
where $\hat{\mathbf{C}}$ is the sample covariance matrix of the corrected latent forecasts $\hat{\vec{z}}(l)$ of
the held-out period of a fold, with every initialization and all 24 leads pooled. In words, $\zeta_k$ is the variance
along the two directions of $P_k$ divided by the total variance. We compute $\zeta_k$ for each of the 40 operators,
with its own $\mathbf{Q}$ and the forecasts of its fold. These forecasts use the mean correction of the ten
operators of the fold (Algorithm~\ref{alg:fmo}), so only $\mathbf{Q}$ differs between the ten operators of a fold.
If the forecasts were isotropic, that is, if $\hat{\mathbf{C}}=c\,\mathbf{I}$ with the same variance $c$ in every
direction, then $\vec{q}^{\top}\hat{\mathbf{C}}\,\vec{q}=c$ for every unit vector $\vec{q}$, and $\zeta_k=2/32=0.0625$
for any $\mathbf{Q}$.

Of the three oscillator planes of $\mathbf{L}_0$, only the leading one stands out in the forecasts. The 2.4-year
plane ($k=1$) carries $0.199\pm0.051$ of the forecast variance (mean $\pm$ standard deviation of $\zeta_1$ over the 40
operators), which is 3.2 times the isotropic value. The 3.9- and 5.7-year planes carry only 1.0 and 1.2 times that
value (Fig.~\ref{fig:modes}d). The 5.7-year mode also lies outside the band in which the observed spectrum exceeds
its 95\,\% AR(1) level (2.29--3.95 years; Fig.~\ref{fig:modes}a). Further, its period does not depend on the fold
in which it is trained. In a one-way analysis of variance of the learned period across the four folds, ten
operators each, the ratio of the between-fold to the within-fold mean square is $F_{3,36}=0.36$, against 4.80 and
4.64 for the other two. Overall, the generator holds one prominent quasi-biennial oscillator. In these diagnostics,
the 3.9- and 5.7-year oscillators are not distinguishable from the 26 decay modes, which fade without cycling.
Finally, $\mathbf{L}_c$ and $\mathbf{L}_s$ are skew-symmetric and rotate the state between directions through the
year, so $P_k$ is a plane of $\mathbf{L}_0$ alone. The share $\zeta_k$ therefore measures the forecast variance in the
fixed basis of $\mathbf{L}_0$, and not in a subspace that the seasonal propagators keep invariant.

The periods of $\mathbf{L}_0$ are a property of the operator, and not of its base (the least-squares forecast that
it corrects). When the operators are retrained on the lead-dependent base of Sect.~\ref{sec:sens-base}, with the
same seeds, the mean periods of each fold change by at most 0.05 years (for example 2.38, 3.97 and 5.69 years
against 2.38, 3.94 and 5.71 in the first fold), which indicates stability under a change of training target. When
they are retrained without the two additive terms of Eq.~(\ref{eq:fmo}), the periods shorten by 0.03, 0.12 and 0.09
years on average and by at most 0.18 years in the mean of a fold, less than two standard deviations of the 40
operators (Sect.~\ref{sec:res-ablations}). This stability supports reading the spectrum as learned dynamics, and
not as an independent discovery of the periods.

\subsection{Skill on identical data}
\label{sec:res-skill}

\begin{figure*}[tp]
\includegraphics[width=\textwidth]{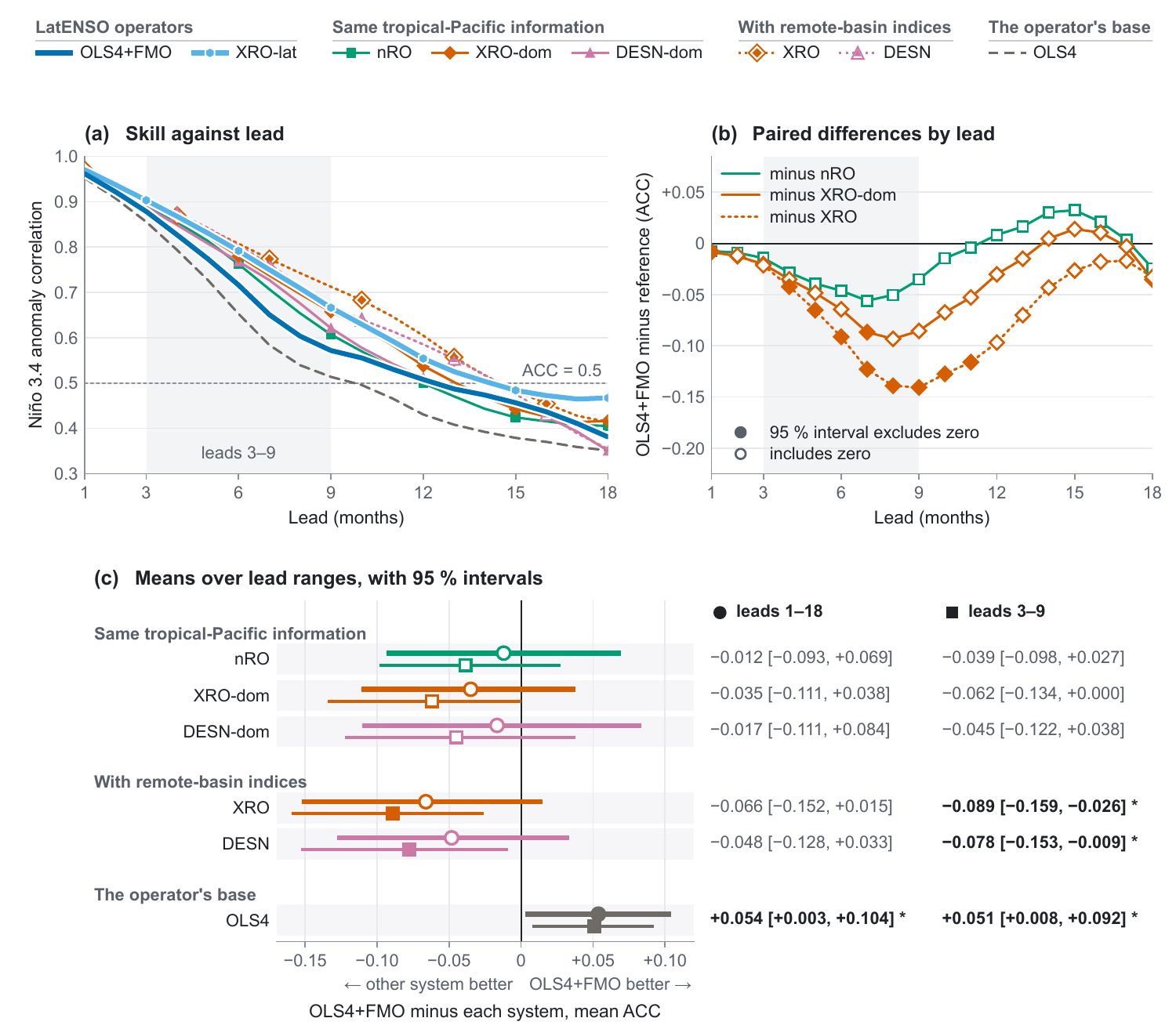}
\caption{Skill of every system on the same 394 windows. (a) Ni\~no 3.4 ACC against lead. (b) Paired differences
of the Floquet operator (OLS4+FMO) from nRO, XRO-dom and XRO at each lead; filled markers: the 95\,\% interval
excludes zero. (c) Paired differences, mean over leads 1--18 (circles) and 3--9 (squares), with 95\,\% intervals,
for the Floquet operator against each reference model and its base, OLS4 (Table~\ref{tab:diff} gives the same for XRO-lat); filled
markers: the interval excludes zero, and the asterisk marks the same in the numbers at the right. In Figs.~\ref{fig:skill}--\ref{fig:season} and
\ref{fig:e2}, dotted lines and ringed markers mark XRO and DESN, the reference models with remote-basin indices.}
\label{fig:skill}
\end{figure*}

\begin{table*}[t]
\caption{Skill of every system on the same 394 windows. ACC: anomaly correlation coefficient of monthly Ni\~no 3.4, mean over
the leads given. AUC: area under the ROC curve for El Ni\~no ($>+1\,^{\circ}$C) and La Ni\~na ($<-1\,^{\circ}$C)
events, mean over leads 10--18. Amplitude: slope $\mathrm{cov}(f,o)/\mathrm{var}(o)$; reliability: slope
$\mathrm{cov}(f,o)/\mathrm{var}(f)$ (1 is calibrated, below 1 overconfident). JJA: ACC for June--August targets
at lead 18. MSSS: mean-square skill score against the sample climatology, mean over leads 1--18. Persistence:
the observed Ni\~no 3.4 anomaly at initialization, held at every lead.}
\label{tab:skill}
\scriptsize
\begin{tabular}{lccccccccccc}
\tophline
 & \multicolumn{3}{c}{ACC, leads} & \multicolumn{2}{c}{AUC, leads 10--18} & Amplitude & \multicolumn{3}{c}{Reliability, lead} & JJA & MSSS \\
System & 1--18 & 3--9 & 10--18 & El Ni\~no & La Ni\~na & lead 6 & 6 & 12 & 18 & lead 18 & 1--18 \\
\middlehline
OLS4+FMO (LatENSO, operator 1) & \textbf{0.619} & \textbf{0.717} & \textbf{0.471} & \textbf{0.819} & \textbf{0.693} & \textbf{0.519} & \textbf{0.987} & \textbf{0.863} & \textbf{0.690} & \textbf{0.218} & \textbf{0.408} \\
XRO-lat (LatENSO, operator 2) & \textbf{0.673} & \textbf{0.788} & \textbf{0.521} & \textbf{0.765} & \textbf{0.777} & \textbf{0.709} & \textbf{0.884} & \textbf{0.947} & \textbf{1.082} & \textbf{0.436} & \textbf{0.469} \\
OLS4+FMO, unblended ($\alpha=1$) & 0.531 & 0.636 & 0.368 & 0.777 & 0.629 & 0.729 & 0.559 & 0.385 & 0.239 & 0.008 & $-$0.089 \\
\middlehline
OLS4c+FMO (lead-dependent base) & 0.631 & 0.749 & 0.470 & 0.817 & 0.694 & 0.682 & 0.862 & 0.871 & 0.692 & 0.215 & 0.417 \\
OLS4 (base) & 0.565 & 0.667 & 0.406 & 0.733 & 0.682 & 0.449 & 0.948 & 0.756 & 0.635 & 0.283 & 0.342 \\
OLS4c (lead-dependent base) & 0.592 & 0.732 & 0.406 & 0.733 & 0.683 & 0.644 & 0.863 & 0.756 & 0.635 & 0.283 & 0.372 \\
\middlehline
nRO & 0.631 & 0.756 & 0.463 & 0.752 & 0.718 & 0.663 & 0.877 & 0.849 & 0.799 & 0.381 & 0.422 \\
XRO-dom & 0.654 & 0.779 & 0.491 & 0.760 & 0.735 & 0.704 & 0.865 & 0.912 & 1.001 & 0.418 & 0.447 \\
DESN-dom & 0.636 & 0.762 & 0.468 & 0.767 & 0.687 & 0.695 & 0.848 & 0.505 & 0.298 & 0.322 & 0.239 \\
XRO (remote basins) & 0.685 & 0.806 & 0.532 & 0.761 & 0.795 & 0.745 & 0.875 & 0.950 & 1.007 & 0.272 & 0.472 \\
DESN (remote basins) & 0.667 & 0.795 & 0.507 & 0.763 & 0.733 & 0.780 & 0.826 & 0.601 & 0.346 & 0.149 & 0.347 \\
\middlehline
Persistence & 0.198 & 0.407 & $-$0.124 & 0.471 & 0.327 & 0.392 & 0.401 & $-$0.082 & $-$0.217 & $-$0.338 & $-$0.618 \\
\bottomhline
\end{tabular}
\end{table*}

\begin{table*}[t]
\caption{Paired differences between the two LatENSO operators and each reference model, and between OLS4+FMO and its
base, OLS4 (last row of the first block; Sect.~\ref{sec:res-base-gain}), with 95\,\% intervals
from the paired year-block bootstrap (36 initialization years, 10\,000 draws). Superscripts give the strength of
evidence for a difference of the sign shown (Sect.~\ref{sec:stats}): C, clear (the 95\,\% interval excludes zero);
S, strong (the 90\,\% interval excludes zero); M, moderate; W, weak; no mark, inconclusive. ACC differences are means
over the leads given; AUC differences are for El Ni\~no events, mean over leads 10--18. Appendix~\ref{app:blocks}
repeats the main comparisons with longer blocks.}
\label{tab:diff}
\scriptsize
\begin{tabular}{lcccc}
\tophline
 & \multicolumn{3}{c}{ACC difference, leads} & El Ni\~no AUC \\
Comparison & 1--18 & 3--9 & 10--18 & leads 10--18 \\
\middlehline
OLS4+FMO $-$ nRO & $-$0.012 [$-$0.093, $+$0.069] & $-$0.039 [$-$0.098, $+$0.027]$^{\mathrm{W}}$ & $+$0.008 [$-$0.121, $+$0.132] & $+$0.067 [$-$0.020, $+$0.150]$^{\mathrm{M}}$ \\
OLS4+FMO $-$ XRO-dom & $-$0.035 [$-$0.111, $+$0.038]$^{\mathrm{W}}$ & $-$0.062 [$-$0.134, $+$0.000]$^{\mathrm{S}}$ & $-$0.019 [$-$0.140, $+$0.096] & $+$0.059 [$-$0.013, $+$0.134]$^{\mathrm{M}}$ \\
OLS4+FMO $-$ DESN-dom & $-$0.017 [$-$0.111, $+$0.084] & $-$0.045 [$-$0.122, $+$0.038]$^{\mathrm{W}}$ & $+$0.003 [$-$0.139, $+$0.160] & $+$0.052 [$-$0.053, $+$0.174]$^{\mathrm{W}}$ \\
OLS4+FMO $-$ XRO & $-$0.066 [$-$0.152, $+$0.015]$^{\mathrm{M}}$ & $-$0.089 [$-$0.159, $-$0.026]$^{\mathrm{C}}$ & $-$0.061 [$-$0.199, $+$0.072]$^{\mathrm{W}}$ & $+$0.058 [$-$0.019, $+$0.133]$^{\mathrm{M}}$ \\
OLS4+FMO $-$ DESN & $-$0.048 [$-$0.128, $+$0.033]$^{\mathrm{W}}$ & $-$0.078 [$-$0.153, $-$0.009]$^{\mathrm{C}}$ & $-$0.036 [$-$0.161, $+$0.095] & $+$0.056 [$-$0.035, $+$0.161]$^{\mathrm{W}}$ \\
OLS4+FMO $-$ OLS4 & $+$0.054 [$+$0.003, $+$0.104]$^{\mathrm{C}}$ & $+$0.051 [$+$0.008, $+$0.092]$^{\mathrm{C}}$ & $+$0.065 [$-$0.014, $+$0.146]$^{\mathrm{M}}$ & $+$0.086 [$+$0.020, $+$0.164]$^{\mathrm{C}}$ \\
\middlehline
XRO-lat $-$ nRO & $+$0.042 [$-$0.003, $+$0.092]$^{\mathrm{S}}$ & $+$0.032 [$-$0.008, $+$0.096]$^{\mathrm{M}}$ & $+$0.058 [$-$0.020, $+$0.134]$^{\mathrm{M}}$ & $+$0.013 [$-$0.077, $+$0.081] \\
XRO-lat $-$ XRO-dom & $+$0.019 [$-$0.002, $+$0.040]$^{\mathrm{S}}$ & $+$0.009 [$-$0.007, $+$0.025]$^{\mathrm{W}}$ & $+$0.031 [$-$0.012, $+$0.074]$^{\mathrm{M}}$ & $+$0.005 [$-$0.045, $+$0.049] \\
XRO-lat $-$ DESN-dom & $+$0.037 [$-$0.036, $+$0.119]$^{\mathrm{W}}$ & $+$0.026 [$-$0.028, $+$0.087]$^{\mathrm{W}}$ & $+$0.053 [$-$0.066, $+$0.191]$^{\mathrm{W}}$ & $-$0.002 [$-$0.113, $+$0.115] \\
XRO-lat $-$ XRO & $-$0.012 [$-$0.049, $+$0.021]$^{\mathrm{W}}$ & $-$0.018 [$-$0.048, $+$0.009]$^{\mathrm{M}}$ & $-$0.010 [$-$0.073, $+$0.048] & $+$0.004 [$-$0.040, $+$0.038] \\
XRO-lat $-$ DESN & $+$0.006 [$-$0.061, $+$0.068] & $-$0.007 [$-$0.059, $+$0.038] & $+$0.015 [$-$0.095, $+$0.121] & $+$0.002 [$-$0.073, $+$0.088] \\
\bottomhline
\end{tabular}
\end{table*}

We now turn to forecast skill. By the Floquet operator we mean, from here on, the blended system OLS4+FMO. Over
leads 1--18 it reaches an ACC of 0.619 and cannot be distinguished from any reference model built from the same
domain (Fig.~\ref{fig:skill}; Table~\ref{tab:diff}). The evidence for a difference is inconclusive against nRO and
DESN-dom and weak against XRO-dom. Over the same leads, XRO and DESN with their eight remote-basin indices are
ahead by margins that 36 initialization years do not resolve. The evidence of a deficit for OLS4+FMO is moderate
against XRO and weak against DESN. Every system is far above persistence (Table~\ref{tab:skill}); OLS4+FMO exceeds
it by $+0.421$ [$+0.291$, $+0.535$].

These differences depend on lead. At leads 3--9 the operator trails XRO, DESN and XRO-dom, with clear evidence of
a deficit against the first two and strong evidence against the third (Table~\ref{tab:diff}). The gap to XRO is
largest at lead 9 ($-0.141$), while the gap to XRO-dom peaks at lead 8 and closes by lead 14
(Fig.~\ref{fig:skill}b). At leads 10--18 the operator is level with every in-domain reference model.
Section~\ref{sec:res-base-gain} shows what the operator adds to its base, and Sect.~\ref{sec:res-base} traces the
mid-lead deficit to that base.

Three further points put these numbers in context. First, skill varies more between decades than between systems.
Across the four folds the operator scores between 0.340 (2002--2013) and 0.793 (1981--1991), and XRO between 0.456
and 0.783 in the same two folds. In 2002--2013 the operator lowers the skill of its own base, which reaches 0.449.
Every system except DESN-dom finds 2002--2013 hardest. Second, all reference models in this work were refitted on folds
of about 30 years, which is shorter than the 1958--1999 training of their published versions. On the 197
verification windows of 2002--2023, with the published definition of skill, XRO scores lower at leads 12 and 15
when trained on our folds than when trained on 1958--1999 (moderate evidence over leads 10--18;
Appendix~\ref{app:references}). The comparison is on equal terms, but every absolute number here is lower
than a published one for the same model. Third, the reported forecast is the mean of ten operators, and not a
selected seed. Single seeds reach 0.597--0.620 (mean 0.612), and XRO and nRO are deterministic fits with no seed to
choose.

The operator also forecasts WWV, the second index on which it is trained, with the same blend weights. Over leads
1--18 its WWV ACC is 0.617, against 0.688--0.709 for nRO, XRO-dom and XRO and 0.253 for persistence. At
leads 3--9, where its Ni\~no 3.4 skill trails XRO and XRO-dom, its WWV skill is level with that of the oscillator
models, and the operator adds $+0.051$ [$+0.027$, $+0.079$] to its base (clear
evidence). At leads 10--18 it falls behind them (0.394 against 0.535--0.558), with clear evidence against XRO-dom
($-0.164$ [$-0.331$, $-0.019$]) and strong evidence against nRO. This deficit is already in the base, which reaches
0.399. Over leads 1--18 the evidence of a WWV deficit is therefore clear against XRO-dom ($-0.092$ [$-0.172$,
$-0.021$]) and strong against nRO and XRO.

\subsection{What the Floquet operator adds to its base}
\label{sec:res-base-gain}

The comparisons of this section are with the operator's own base. They show where its skill comes from, and not how
good it is. Over leads 1--18 the operator adds $+0.054$ to the ACC of its base, OLS4 (clear evidence;
Table~\ref{tab:diff}), and the Floquet propagation of Eq.~(\ref{eq:fmo}) carries this gain. With the two additive
terms removed from the trained operators, the propagation alone reaches 0.631 at the same blend weight, which is
$+0.066$ over the base (Table~\ref{tab:blocks}) and $+0.012$ [$-0.0002$, $+0.026$] over the full operator (strong
evidence). The gain of the propagation over the base is clear with year blocks of one to four years, and that of
the full operator only with one-year blocks (Appendix~\ref{app:blocks}). The additive terms add no skill. The gated
residual changes nothing (0.632), and the direct term lowers the ACC to 0.619, pulling the unblended operator
below its base (Table~\ref{tab:skill}); without the direct term it reaches 0.589 and stays above the base. Trained
without the two terms from the start, the operator reaches the same ACC over leads 1--18 as the full one, although
unblended it also falls below its base (Sect.~\ref{sec:res-ablations}).

Nor does the gain rest on the blend. Unblended, with the gain factor $g$ set to one and without the learned
correction of the initial state, the linear propagation of the observed state reaches 0.614, which is $+0.049$
[$+0.005$, $+0.094$] over the base, and 0.744 at leads 3--9, above the full operator (Table~\ref{tab:skill}). At
the blend weight, this plain propagation adds $+0.032$ [$+0.017$, $+0.048$] to the base, with clear evidence at every block
length and also for the operators trained on the lead-dependent base of Sect.~\ref{sec:sens-base}. About 40\,\% of
the full operator's gain can be reproduced by letting the base coefficients vary with the season. The numerical
output of these tests is archived with the data (see Code and data availability).

The gain over the base does not rest on the fold used during development (2013--2024;
Sect.~\ref{sec:res-ablations}). It keeps its sign whichever fold is left out, although it depends most on
1981--1991. With one fold left out, the gain is $+0.071$ [$+0.012$, $+0.126$] without 1991--2002 and $+0.058$
[$+0.012$, $+0.112$] without 2002--2013 (clear evidence), and $+0.058$ [$-0.003$, $+0.121$] without 2013--2024
(strong evidence). Without 1981--1991, the fold in which it is largest, it is $+0.020$ [$-0.047$, $+0.088$]
(inconclusive).

Part of the gain is a correction of bias. Anomalies are taken with respect to the climatology of each fold's
training months (Sect.~\ref{sec:cv}), so the means of forecasts and observations differ between folds, and a
correlation pooled over the four folds also rewards a forecast that follows these means. The operator lowers the
mean absolute bias of its base over the four folds from 0.150 to 0.101\,$^{\circ}$C, the level of the least biased
reference models, DESN and XRO-dom (Table~\ref{tab:bias}). With the mean of each fold removed from forecasts and observations, the
gain is $+0.041$ (strong evidence), about three quarters of the pooled value, and the evidence for the gain of the
Floquet propagation alone stays clear. Most other comparisons barely change (Appendix~\ref{app:blocks}).

\subsection{The latent state in XRO}
\label{sec:res-xro}

\begin{figure*}[tp]
\includegraphics[width=\textwidth]{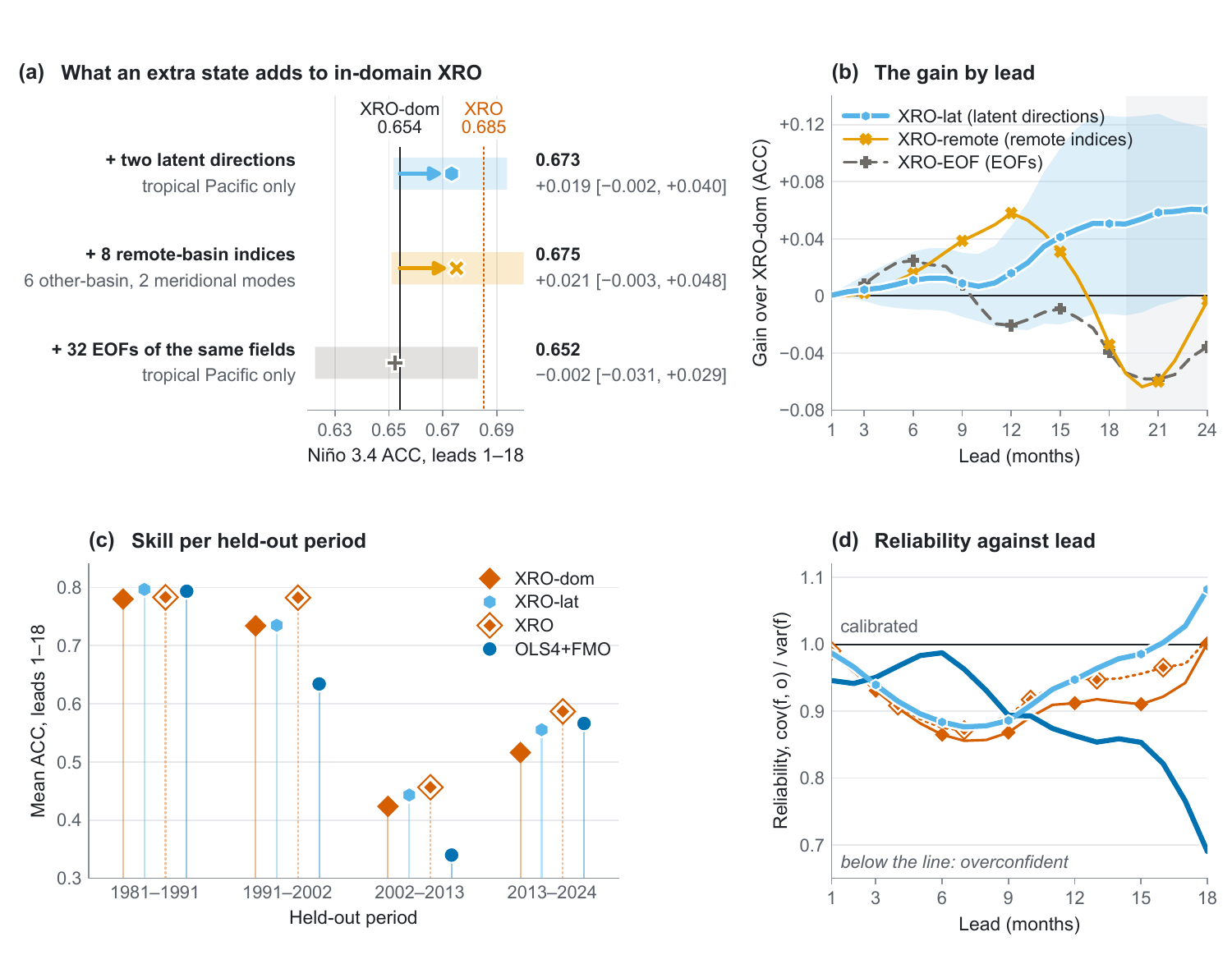}
\caption{The latent state as part of the state of XRO. (a) Mean ACC over leads 1--18 of XRO-dom and of XRO whose extra
state is chosen by the same selection pipeline from the latent directions, the eight remote-basin indices or 32 EOFs
(with or without the in-domain meridional modes), with the three-basin XRO for reference; bands: 95\,\% intervals.
(b) Gain over XRO-dom at each lead of the three systems in (a), with the 95\,\% interval of XRO-lat; shading:
leads 19--24, beyond the main evaluation window. (c) Mean ACC over leads 1--18 in
each fold, scored on that fold's forecasts only. (d) Reliability against lead (1 is calibrated); colors as in (c).}
\label{fig:latent}
\end{figure*}

\begin{table*}[t]
\caption{The latent state as part of the state of XRO: paired differences in mean ACC over leads 1--18 with 95\,\%
and 90\,\% intervals, the share of bootstrap resamples in which the difference is positive, and the grade of
Sect.~\ref{sec:stats}. XRO-lat, XRO-remote and XRO-EOF use the identical selection pipeline (Sect.~\ref{sec:xrolat}).}
\label{tab:latent}
\scriptsize
\begin{tabular}{llccccl}
\tophline
Comparison & Meaning & Effect & 95\,\% interval & 90\,\% interval & Share $>0$ & Grade \\
\middlehline
XRO-lat $-$ XRO-dom & latent directions selected into the state & $+$0.019 & [$-$0.002, $+$0.040] & [$+$0.001, $+$0.036] & 0.960 & strong $+$ \\
XRO-remote $-$ XRO-dom & eight remote-basin indices selected (positive control) & $+$0.021 & [$-$0.003, $+$0.048] & [$+$0.001, $+$0.043] & 0.956 & strong $+$ \\
XRO-EOF $-$ XRO-dom & 32 EOFs selected (control) & $-$0.002 & [$-$0.031, $+$0.029] & [$-$0.027, $+$0.023] & 0.432 & inconclusive \\
XRO-lat $-$ XRO-remote & latent directions against remote-basin indices & $-$0.002 & [$-$0.030, $+$0.022] & [$-$0.025, $+$0.019] & 0.430 & inconclusive \\
XRO-lat $-$ XRO-EOF & latent directions against EOFs & $+$0.021 & [$-$0.012, $+$0.053] & [$-$0.005, $+$0.047] & 0.911 & moderate $+$ \\
XRO-lat $-$ XRO & latent directions against the three-basin XRO & $-$0.012 & [$-$0.049, $+$0.021] & [$-$0.043, $+$0.016] & 0.239 & weak $-$ \\
XRO $-$ XRO-dom & three-basin XRO against XRO-dom & $+$0.031 & [$-$0.010, $+$0.072] & [$-$0.002, $+$0.065] & 0.939 & moderate $+$ \\
\bottomhline
\end{tabular}
\end{table*}

The second operator gives the latent state its clearest test: does it make a good model better? XRO with two latent
directions in its state, XRO-lat, reaches 0.673 over leads 1--18, the highest of all systems that use only
tropical-Pacific information (Table~\ref{tab:skill}). Over XRO-dom, which has the in-domain indices but no latent
directions, the gain is $+0.019$ (strong evidence). Through the identical pipeline, the eight remote-basin indices
of the published XRO add $+0.021$ (strong evidence), and 32 EOFs of the same thirteen fields add no net skill
(inconclusive; Fig.~\ref{fig:latent}a; Table~\ref{tab:latent}), a gain at leads up to 9 that is offset by a loss
beyond (Fig.~\ref{fig:latent}b). These EOFs, like the autoencoder, are fitted on the full record
(Sect.~\ref{sec:ae-loss}). Computed from the training months of each fold instead, they add nothing (0.000;
[$-0.025$, $+0.024$]), and the latent state is ahead of them by $+0.019$ [$-0.009$, $+0.047$] (moderate evidence).

Within what 36 initialization years can resolve, the latent state therefore adds about as much skill to XRO as the
remote-basin indices do through the same pipeline, and more than the EOFs selected in the same way, whichever
record they are computed from. It adds that skill at longer leads, where the remote-basin indices no longer help.
Two qualifications apply. First, the three-basin XRO, fitted without the selection pipeline, gains more over
XRO-dom, with moderate evidence (Table~\ref{tab:latent}). Second, the EOF result holds for the selected
systems only: used in every fold without the selection, four EOF designs also beat XRO-dom
(Appendix~\ref{app:designs}).

XRO-lat is indistinguishable in skill from the three-basin XRO (weak evidence of a difference) and from DESN (inconclusive evidence of a
difference), and it is ahead of the recharge oscillator nRO with strong evidence (Table~\ref{tab:diff}). For the
comparison with XRO-dom, the criterion set in advance was a 95\,\% interval that excludes zero, and the interval
misses it by 0.002. The 90\,\% interval excludes zero, and the gain is positive in 96\,\% of bootstrap resamples,
which we grade as strong evidence (Table~\ref{tab:latent}; Sect.~\ref{sec:stats}).

The gain grows with lead, from near zero at lead 1 to $+0.051$ at leads 17 and 18 (Fig.~\ref{fig:latent}b). It is
positive in all four folds ($+0.001$ to $+0.039$), but over leads 1--18 XRO-lat stays behind
the three-basin XRO in three of the four (Fig.~\ref{fig:latent}c). Beyond lead 14 the order reverses. From lead 15
to lead 24 XRO-lat is the most skillful of all systems at every lead (Fig.~\ref{fig:latent}b), with 0.469 over
leads 19--24 against 0.375 for XRO ($+0.094$ [$-0.014$, $+0.219$], strong evidence), and it is ahead of XRO there
in all four folds. Note, however, that these leads lie beyond the main evaluation window.

In every fold the nested selection chose two latent directions: slow directions in the second and third folds, and
predictive directions added to the meridional modes in the first and fourth. In the second and third folds the
latent directions therefore replace the meridional modes rather than adding to them. Every candidate design
contains latent directions, so XRO-dom itself is not a candidate. With XRO-dom added to the candidates, the same
selection keeps latent directions in two folds of four, the remote-basin indices in all four and the EOFs in none,
and the gain of XRO-lat over XRO-dom becomes $+0.007$ [$-0.010$, $+0.024$] (weak evidence).

The gain depends on the design being compact. With each of the eleven designs used in every fold, without the
selection, only four latent designs beat XRO-dom (the slow and the predictive-added ones with two or four
directions), whereas the remote-basin indices beat it in all eleven (Appendix~\ref{app:designs}). Averaged over the
eleven designs, the latent state and the EOFs are level (Table~\ref{tab:designs}; Appendix~\ref{app:designs}), and four designs of each EOF
basis beat XRO-dom. The EOFs of the training months score higher than those of the full record in eight of the eleven designs, so fitting a
linear basis of these fields to the full record gives it no advantage. Adding all 32 latent coordinates at once
does not help.

What the latent state contributes is in-domain, low-dimensional and slow. Added to the four in-domain indices, it
improves the out-of-sample reconstruction of the North Pacific meridional mode, but it carries no information on
the Indian Ocean or Atlantic indices, whose held-out $R^2$ falls when it is added (Table~\ref{tab:remote};
Appendix~\ref{app:designs}). The latent state does not replace the information of the remote basins; it
adds different information. The gain from the remote-basin indices peaks at lead 12 and turns negative from lead
17, whereas that of the latent state stays near its maximum, about $+0.06$, to lead 24 (Fig.~\ref{fig:latent}b), and the two increments over
XRO-dom are only weakly correlated across forecasts ($r=0.30$). However, added to the three-basin XRO, two slow
latent directions leave its skill unchanged ($-0.004$ [$-0.021$, $+0.013$] over leads 1--18; 0.377 against 0.375
over leads 19--24), so the forecasts do not show that the two sources of skill add up. XRO-lat is also among the
best-calibrated systems at long leads, with reliability close to one at leads 12--18
(Fig.~\ref{fig:latent}d). At lead 18 it is the most skillful of all systems for winter, summer and autumn targets
(0.497, 0.436 and 0.485; Fig.~\ref{fig:season}b; Sect.~\ref{sec:res-season}).

\subsection{Event detection}
\label{sec:res-events}

\begin{figure*}[tp]
\includegraphics[width=\textwidth]{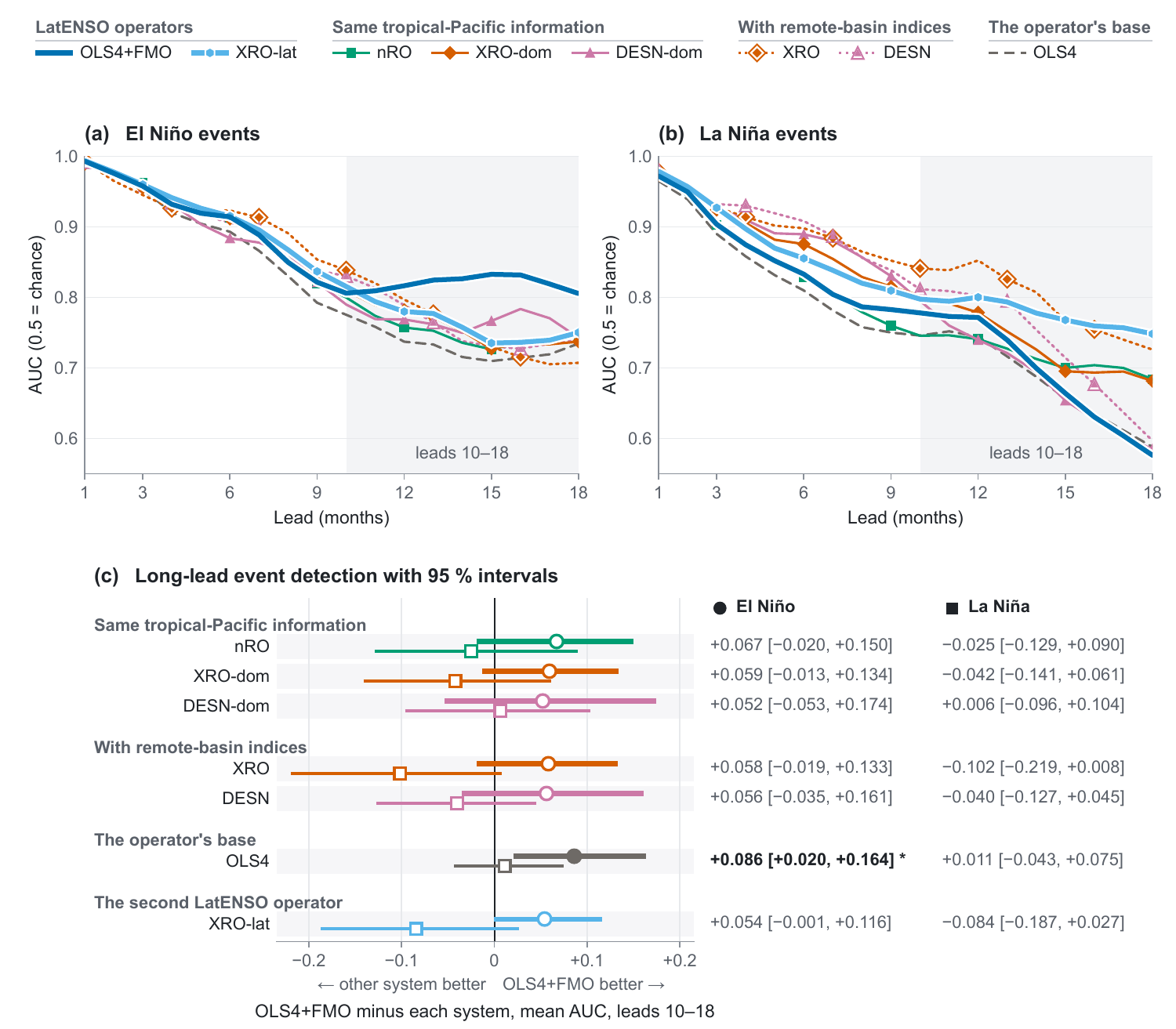}
\caption{Event detection. (a) El Ni\~no AUC (observed anomaly above $+1\,^{\circ}$C) against lead for every system.
(b) The same for La Ni\~na (below $-1\,^{\circ}$C). (c) Paired differences of the Floquet operator from each of the
other systems in AUC averaged over leads 10--18, El Ni\~no and La Ni\~na, with 95\,\% intervals; filled markers: the
interval excludes zero, and the asterisk marks the same in the numbers at the right.}
\label{fig:events}
\end{figure*}

At every lead from 12 to 18, the El Ni\~no AUC of the Floquet operator is higher than that of every reference
model (Fig.~\ref{fig:events}a). Averaged over leads 10--18 it reaches 0.819, against 0.752--0.767 for the reference
models (Table~\ref{tab:skill}). At shorter leads (6--11) XRO detects El Ni\~no best. The evidence for the long-lead
advantage over the reference models is moderate against nRO, XRO-dom and XRO and weak against the two DESN models,
and no interval excludes zero (Table~\ref{tab:diff}). It is strong against XRO-lat and clear against OLS4
(Fig.~\ref{fig:events}c). The advantage comes from the first and the last fold; in the second the systems are
level, and the third holds only five El Ni\~no events at lead 12. The advantage belongs to substantial events. At
the conventional threshold of $+0.5\,^{\circ}$C it nearly vanishes. From $+0.75\,^{\circ}$C upward it is positive
against every reference model at every threshold up to $+1.5\,^{\circ}$C, but never with an interval that excludes
zero (Table~\ref{tab:thresh}).

The advantage is specific to El Ni\~no events. For La Ni\~na the operator is below XRO, XRO-lat, XRO-dom, DESN and
nRO at leads 10--18 and above only DESN-dom (Fig.~\ref{fig:events}b; Table~\ref{tab:skill}), and its base already
reaches nearly the same AUC. The El Ni\~no advantage comes from the learned
correction of the initial state (Sect.~\ref{sec:operator}). This correction is a nonlinear function of the phase
vector, so it can respond differently to El Ni\~no and La Ni\~na conditions, which the linear propagation cannot.
Without it, the propagation of the observed state reaches an El Ni\~no AUC of 0.758 at the same blend weight, like
the reference models. The correction raises this AUC by $+0.054$ [$+0.012$, $+0.100$] (clear evidence), against
$+0.010$ for La Ni\~na, and the gain factor $g$ adds a further $+0.011$ [$+0.004$, $+0.021$] (clear).

\subsection{Seasonality and the boreal-summer deficit}
\label{sec:res-season}

\begin{figure*}[tp]
\includegraphics[width=\textwidth]{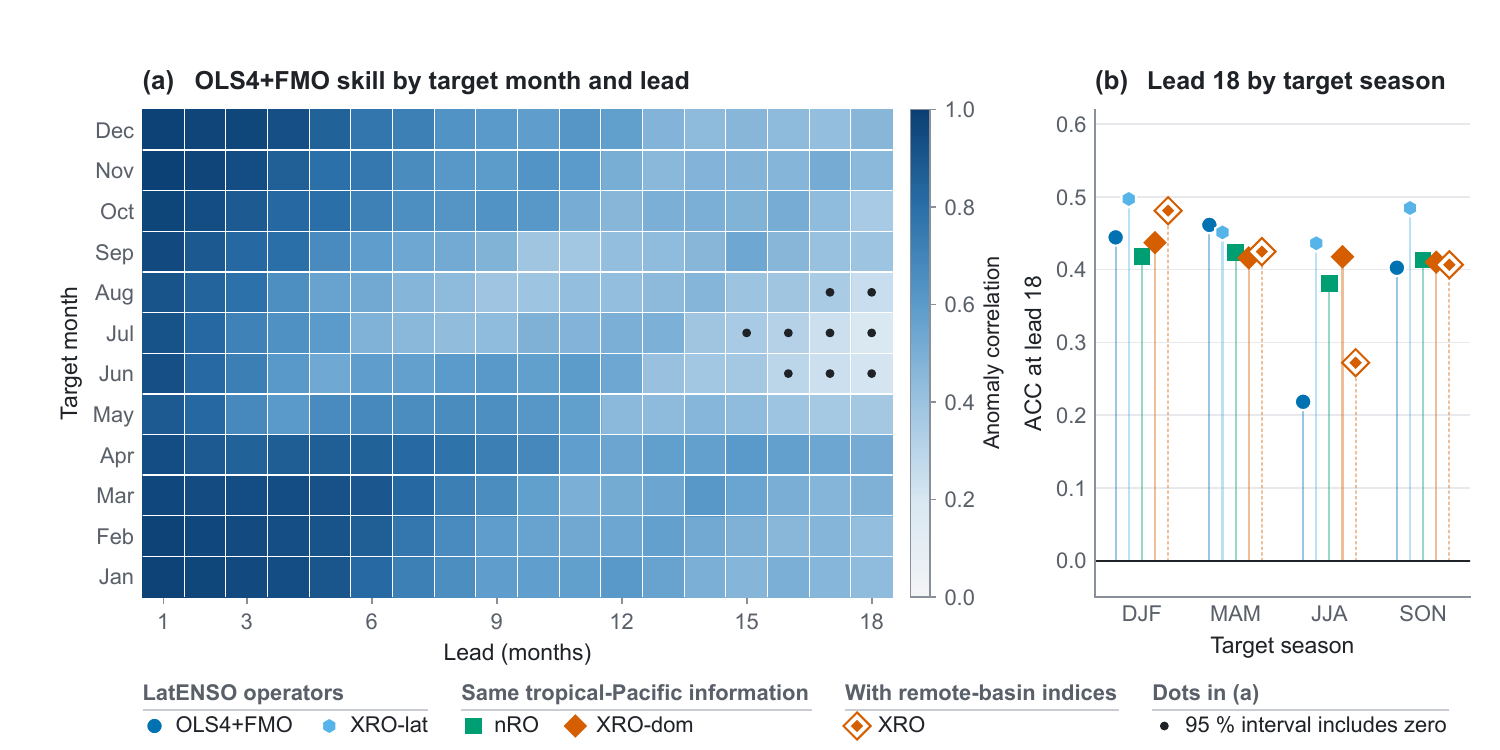}
\caption{Seasonality of skill. (a) ACC of the Floquet operator by target month and lead; dots mark the cells whose
95\,\% interval includes zero. (b) ACC at lead 18 by target season for both LatENSO operators, nRO, XRO-dom and
XRO. DJF, MAM, JJA and SON: December--February, March--May, June--August and September--November.}
\label{fig:season}
\end{figure*}

The skill of the Floquet operator follows the seasonal cycle of ENSO. Averaged over leads 4--9, the ACC is highest
for targets in March (0.829) and lowest for targets in August (0.509). Across the 216 combinations of target month
and lead (12 months $\times$ 18 leads), the 95\,\% interval excludes zero in all but nine, and these nine all occur
for June--August targets at leads 15--18 (Fig.~\ref{fig:season}a). At lead 18 the operator is numerically ahead of
nRO and XRO-dom for winter and spring targets (December--February, DJF, and March--May, MAM). For summer targets,
however, it is well behind them (June--August, JJA, 0.218 against 0.381 and 0.418; Fig.~\ref{fig:season}b). The
evidence for this summer deficit is clear against both, and it stays clear with longer blocks: at every block
length against nRO and at three out of four block lengths against XRO-dom (Table~\ref{tab:blocks}).

Most of the deficit is already in the base, which reaches 0.283 for summer targets at lead 18. The operator, with
its weight of 0.25, lowers this to 0.218 ($-0.065$ [$-0.199$, $+0.048$], weak evidence). The operator's part comes
from the learned terms around the propagation, and not from the propagation itself. Unblended, the plain
propagation of the observed state reaches 0.276, close to the base, whereas the learned correction of the initial
state and the gain lower it to 0.103, and the direct term to near zero (Table~\ref{tab:skill}). A weight that
depends on the target season would remove the operator's part of the deficit. However, it would have to be chosen
inside the training folds, and it is not part of LatENSO v1.0. XRO-lat has no such deficit: it is the most skillful
system for summer targets at lead 18 (Table~\ref{tab:skill}).

\subsection{Where the deficits come from}
\label{sec:res-base}

\begin{figure*}[tp]
\includegraphics[width=\textwidth]{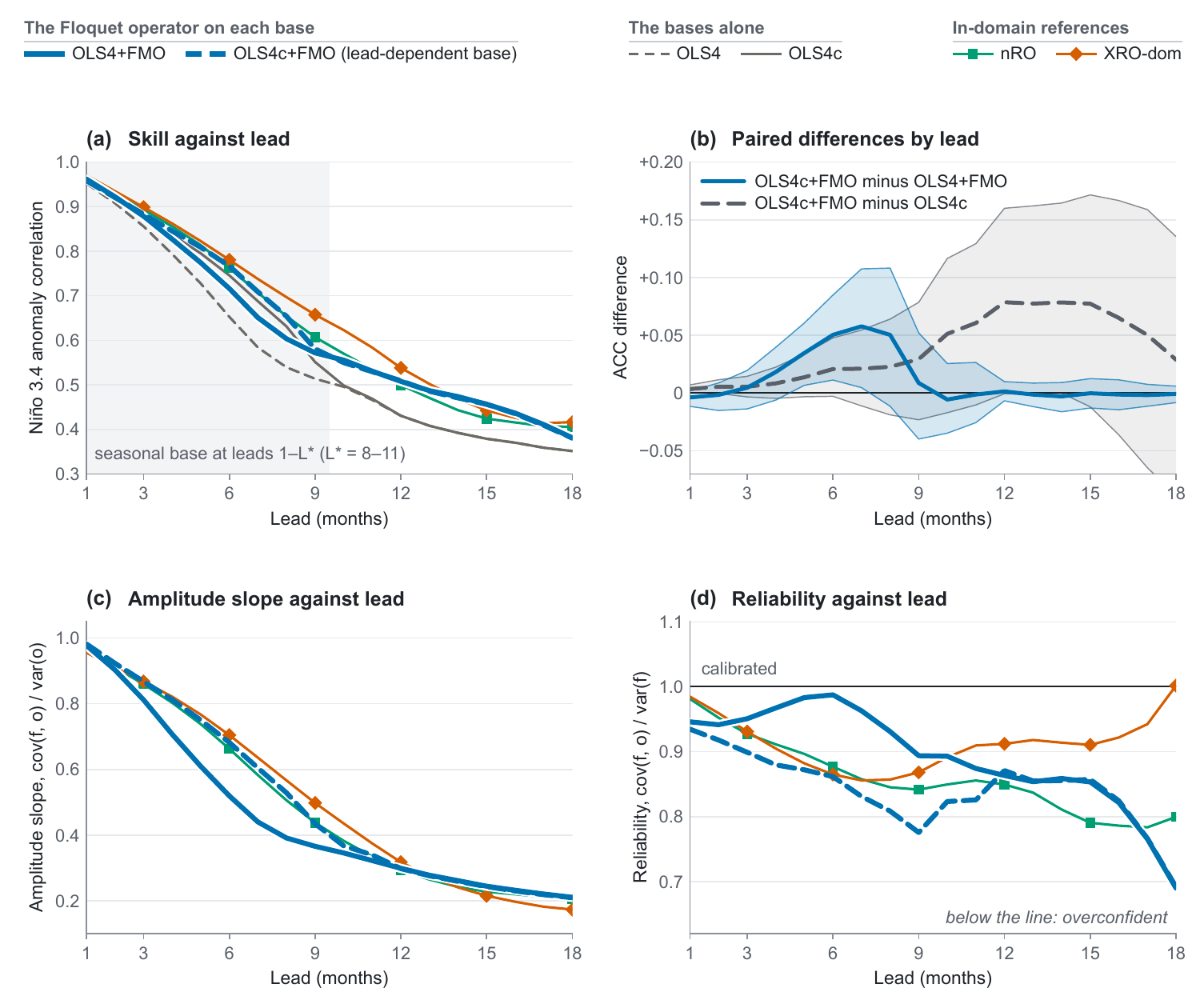}
\caption{The base and the deficits. OLS4+FMO against the same operator retrained on the lead-dependent base OLS4c (OLS4c+FMO; Sect.~\ref{sec:sens-base}; seasonal coefficients up to lead $L^{*}=8$--11, chosen per fold on training data). (a) ACC
against lead, with both bases alone and with nRO and XRO-dom, the in-domain references the text compares against
(all systems are compared in Fig.~\ref{fig:skill}). (b) Paired differences, OLS4c+FMO minus OLS4+FMO and OLS4c+FMO minus OLS4c, with 95\,\% intervals. (c) Amplitude slope against lead.
(d) Reliability against lead.}
\label{fig:base}
\end{figure*}

The operator corrects its base, so a weakness of the base can appear as a weakness of the operator. Retraining the
operator on the lead-dependent base OLS4c (Sect.~\ref{sec:sens-base}) separates the two. With only the base
changed, the mid-lead skill rises. Over leads 3--9 OLS4c+FMO gains $+0.032$ [$-0.003$, $+0.061$] on OLS4+FMO
(strong evidence), with the largest gain at lead 7 (Fig.~\ref{fig:base}a, b). Its gap to XRO-dom at those leads
halves, from $-0.062$ to $-0.030$ [$-0.099$, $+0.030$], and against nRO and DESN-dom the gap at leads 3--9 nearly
vanishes (Table~\ref{tab:skill}). The gap to XRO with remote basins remains, with clear evidence ($-0.057$
[$-0.124$, $-0.001$]). Over leads 1--18 the gain is $+0.011$ [$-0.007$, $+0.026$] (moderate evidence), with an
interval that includes zero. This interval on the verification windows is descriptive. The choice of base was
decided by the fold criterion on training data (Sect.~\ref{sec:sens-base}), which OLS4c met in two of the four
folds, with three required, so OLS4+FMO remains the system described here.

The operator itself is not the source of the deficit. To the lead-dependent base it adds $+0.039$ [$-0.006$,
$+0.083$] over leads 1--18 and $+0.063$ [$-0.013$, $+0.140$] at leads 10--18, where a seasonal regression overfits
(strong evidence in both), and each of the ten seeds adds to the base ($+0.025$ to $+0.042$). The operator's
Floquet propagation alone, with the two additive terms removed, adds $+0.050$ [$+0.012$, $+0.089$] and $+0.082$
[$+0.016$, $+0.154$], with clear evidence in both and, over leads 1--18, at every block length.

The amplitude deficit has the same origin. The amplitude slope of OLS4+FMO at lead 6 is 0.519, below those of nRO
and XRO-dom (Table~\ref{tab:skill}), with clear evidence at every block length (Table~\ref{tab:blocks}). This slope is set by
the base and the weight: $0.449+0.25\,(0.729-0.449)=0.519$, where 0.449 is the slope of the base and 0.729 that of
the unblended forecast. The latter is level with the oscillator models ($+0.066$ [$-0.205$, $+0.333$] against nRO).
Averaging the ten seeds does not damp the amplitude (single seeds 0.496--0.511). With the lead-dependent base the
slope rises to 0.682 (Fig.~\ref{fig:base}c).

Amplitude comes at the expense of calibration (Sect.~\ref{sec:metrics}). At lead 6 OLS4+FMO is calibrated: its
reliability is 0.987 and its amplitude slope matches $r^2$ (Appendix~\ref{app:alpha}). The lead-dependent base
gives the more energetic and less reliable forecast that is typical of the oscillator models
(Fig.~\ref{fig:base}c, d; Table~\ref{tab:skill}). Across leads, OLS4+FMO is the best calibrated of all systems at
leads 4--8, the most overconfident at leads 1--2, and less reliable than XRO and XRO-lat beyond lead 12. Its
mean-square skill score is below those of nRO, XRO-dom and XRO and above those of the two DESN models
(Table~\ref{tab:skill}), whose reliability falls to about one third by lead 18. The evidence for the
difference from DESN-dom, $+0.169$ [$-0.018$, $+0.456$], is strong, and the others are not resolved.

\subsection{Ablations}
\label{sec:res-ablations}

Three design choices of the Floquet operator were tested against alternatives during its development. Each test was
paired over seeds and run on the fold used during development (the fourth, 2013--2024, which is also one of the
four verification folds; Sect.~\ref{sec:discussion}), and the numerical output is archived with the data. The first
test asked whether the operator needs the whole state, since ENSO is concentrated in one latent coordinate
(Sect.~\ref{sec:res-coords}). Restricting the operator to an 8-dimensional subspace of the coordinates most
correlated with Ni\~no 3.4 costs $-0.068$ in mean ACC over leads 1--18 (three seeds), far outside the seed-to-seed
spread of the mean ACC in that fold (standard deviation 0.005), and the 32-dimensional state was kept. ENSO's
output is concentrated in one coordinate, but the state needed to forecast is not.

The second test concerned the number of oscillators. It had been set to three without a test, and the Ni\~no 3.4
spectrum has a peak near 1.45 years (Sect.~\ref{sec:res-coords}) with no oscillator assigned to it. A fourth
oscillator initialized at 1.4 years improved the development fold ($+0.0124$, six of six seeds) but not the
four-fold, ten-seed test ($+0.0030\pm0.0145$, mean $\pm$ standard deviation of the 40 paired differences; 20 of 40
pairs), and the three-oscillator design was kept. The third test added an explicit, bounded amplitude nonlinearity
inside the monthly recursion, because a damped linear recursion can carry an anomaly forward but cannot amplify it.
It did not change mean ACC  (the mean $\pm$ standard deviation of six paired differences being $0.0000\pm0.0008$). Given the
capacity to amplify anomalies, the operator learned additional damping of its second and third oscillators in every
seed, and the term was not adopted.

A fourth test, run on all four folds with the ten seeds, asked whether the Floquet propagation needs the two
additive terms of Eq.~(\ref{eq:fmo}) during training. All 40 operators were retrained with the gated residual and
the direct term switched off from the first step and every other setting unchanged. The criteria for reading the
result were fixed before the test was run. Trained alone, the propagation scores 0.619 over leads 1--18, the same
as the full operator ($-0.001$ [$-0.015$, $+0.014$]), and adds $+0.053$ [$-0.004$, $+0.110$] to its base (strong
evidence). At leads 3--9 it is slightly below the full operator ($-0.011$ [$-0.025$, $+0.002$], strong evidence),
and at leads 10--18 it is level with it ($+0.008$ [$-0.016$, $+0.033$]). Against the propagation of the trained
operators with the terms removed afterward (0.631; Sect.~\ref{sec:res-base-gain}), the difference is $-0.013$
[$-0.031$, $+0.005$] (moderate evidence), which this record does not resolve.

The periods of the retrained operators (2.38, 3.74 and 5.64 years) and their damping rates stay within two standard
deviations of those of the 40 operators of LatENSO v1.0. Without the blend the retrained propagation scores 0.514,
against 0.531 for the full operator and 0.589 for the propagation of the trained operators, and its learned
correction of the initial state is about twice as large in mean square. Its long-lead El Ni\~no AUC is unchanged
(0.818 against 0.819). An operator trained without the two additive terms therefore reaches the same ACC over leads
1--18 as the full operator, with periods and damping rates within the spread of the 40 operators. Whether the terms
help the propagation during training is not resolved by this record. The 40 retrained operators and their forecasts
are archived with the data.

\subsection{Cost, hardware and software}
\label{sec:res-cost}

\begin{table}[t]
\caption{Measured cost. GPU: one NVIDIA RTX 4090; CPU timings on the same workstation. Published timings are those
reported by \citet{Zhang2026}.}
\label{tab:cost}
\scriptsize
\begin{tabular}{ll}
\tophline
Item & Time \\
\middlehline
Autoencoder training, once (2500 epochs) & 1.9\,h (GPU) \\
Encoding all 522 months & 5.4\,s (GPU) \\
FMO training, one operator & 8.6\,min (GPU, mean of 40) \\
FMO training, 40 operators & 5.7\,h (GPU) \\
FMO forecast, 99 initializations (one fold) $\times$ 24 leads, 10 operators & 0.27\,s (GPU) \\
XRO-dom fit, 4 folds & 0.9\,s (CPU) \\
XRO fit, 4 folds & 3.0\,s (CPU) \\
XRO-lat nested selection, per fold & about 5\,min (CPU) \\
Published XRO, echo-state network (ESN), DESN training & 4\,s, 26\,s, 122\,s (CPU) \\
Published 3D-Geoformer \citep{ZhouZhang2023} training & about 12\,h (V100 GPU) \\
\bottomhline
\end{tabular}
\end{table}

LatENSO is cheap to run once trained (Table~\ref{tab:cost}). The autoencoder is trained once, in 1.9\,h on one GPU,
and encoding the whole record takes seconds. A Floquet operator trains in under nine minutes and forecasts in
milliseconds. XRO with the latent state fits in about a second, and its nested selection over eleven designs takes
about five minutes per fold on a CPU. The operators are no cheaper to fit than XRO itself, so the argument for the
latent state is not computational. Instead, it rests on the size of the representation
(Sect.~\ref{sec:res-fidelity}) and on what that representation adds. The code runs in Python 3.11 with PyTorch 2.6
(CUDA 12.4), NumPy, SciPy and xarray. The reference models use XRO (revision d8527a8) and the RC-ENSO code of
\citet{Zhang2026} (revision a506ab2) with ReservoirPy 0.3.12.

\section{Discussion}
\label{sec:discussion}

The central result of LatENSO concerns the learned representation rather than either operator. What does the
learned state add? It was learned solely to reconstruct the fields, so nothing in its training rewards a good
forecast. Even so, its 32 latent coordinates capture information that otherwise requires 89--172 EOF modes, and they
separate eastern- and central-Pacific El Ni\~no events into distinct coordinates. When supplied to XRO as just two
additional state variables, they yield skill gains comparable to those obtained from the eight remote-basin indices
over leads 1--18, and larger ones from lead 15 onward, where the gain from the remote-basin indices fades.

Would a linear compression of the same fields do as well? Through the same selection pipeline, it does not. EOFs
derived from the same fields and passed through that pipeline provide no measurable benefit, whether they are
computed from the full record or from the training months of each fold. Note, however, that this is a statement
about the selection pipeline. Some EOF designs do provide a benefit when used in every fold without the selection
(Appendix~\ref{app:designs}). In fact, averaged over the eleven designs, the latent state and the EOFs score alike
(Sect.~\ref{sec:res-xro}). The latent state is ahead where its slow directions are used: two of them outscore two
slow EOF directions in every fold (0.670 against 0.653 and 0.648; Appendix~\ref{app:designs}).

One feature of the autoencoder may explain why its slow directions carry more than those of a linear basis,
although the present experiments do not test it. Its latent representation is nonlinear in the input fields. This
allows a single latent direction to combine thermocline, wind, and SST anomalies nonlinearly, whereas a
multivariate EOF combines them only linearly. The decoder, in contrast, is affine at the token level in its first
stage, which gives each latent direction a fixed token-level pattern. Its later stages are nonlinear, and for this
reason the physical interpretation in Sect.~\ref{sec:res-coords} rests on regression maps of the observed fields.

The limitations are equally clear. The latent state is learned exclusively from tropical-Pacific fields, which
means that it contains no out-of-sample information about Indian or Atlantic Ocean variability. Accordingly, over
leads 1--18 the three-basin XRO remains ahead of XRO-lat in three of the four folds, although the two are
statistically indistinguishable overall. What the latent representation contributes is therefore information that
is in-domain, low-dimensional, and slowly varying.

Nevertheless, the latent state matches the skill provided by remote-basin indices over leads 1--18 and exceeds it
from lead 15 onward. This suggests that the slow evolution of the Pacific holds predictability of its own. The two
sources also differ in two respects. First, they act at different leads: the gain from the remote-basin indices
peaks at lead 12 and turns negative from lead 17, whereas that of the latent state stays near its maximum, about $+0.06$, to lead 24.
Second, the corrections that they make to individual forecasts are only weakly correlated ($r=0.30$;
Sect.~\ref{sec:res-xro}). These differences point to distinct signals. However, whether the two are truly distinct
or simply different manifestations of the same large-scale dynamics remains unresolved.

Note also that the benefit depends on parsimony. In practice, two to four carefully selected latent directions
improve forecasts, whereas including all 32 coordinates does not. Likewise, only four of the eleven latent designs
outperform XRO-dom when used in every fold without the selection, compared with all eleven designs based on
remote-basin indices. This means that the latent state helps through a few well-chosen directions, and not through
all of its coordinates at once.

It is just as important to say what the evidence does not show. Three conclusions lie beyond its scope and are
therefore not claimed. First, LatENSO cannot be claimed to outperform oscillator models constructed from the same
information. Over leads 1--18, XRO-lat is ahead of nRO and XRO-dom with evidence that is strong at most, and the
Floquet operator is statistically indistinguishable from nRO, XRO-dom, and DESN-dom. At leads 3--9 the Floquet
operator underperforms all three, with strong evidence against XRO-dom and weak evidence against the other two.
Among the systems compared, the Floquet operator's only clear ACC improvement is relative to its own base. For the full operator over leads 1--18, even this improvement appears only under one-year block resampling. One natural reading would be to
count this improvement as skill against a competitor. However, the base is part of the same system, so this
comparison is not a claim against other models. Instead, it measures what the operator adds to its base, and which
part of the operator adds it (see below).

Second, the apparent El Ni\~no advantage at leads 12--18 remains suggestive rather than definitive. In its favor,
the signal is consistent across lead times and across event thresholds from $+0.75$ to $+1.5\,^{\circ}$C. However, the
evidence against any reference model is only moderate, and it is clear at no threshold. Further, the advantage
nearly vanishes under the standard $+0.5\,^{\circ}$C event definition, and it is not mirrored for La Ni\~na.

Third, decoding does not clearly improve field prediction relative to XRO. XRO needs no decoder for this:
regression maps derived from its indices can already forecast pan-tropical SST anomalies \citep{Zhao2024}. Under
cross-validation, our decoded SST scores higher than these maps over the whole domain (0.372 against 0.351 and
0.340), but the decoded fields score lower in the equatorial waveguide and for Z20 and U10
(Appendix~\ref{app:physical}). The significance of the decoded fields therefore lies not primarily in forecast
skill, but in the additional insights they provide (Sect.~\ref{sec:disc-next}).

How much can a record of this length support? An important caveat of this work is the limited size of the available
record. With only 36 initialization years, the smallest difference in mean ACC that a paired bootstrap can resolve
is between about 0.02 and 0.10, depending on the pair of systems (Table~\ref{tab:diff}), while many of the effects
considered here are of similar magnitude. For this reason, we prespecified four primary comparisons, fixed their
acceptance criteria in advance, and treated all remaining analyses as descriptive. One natural way to judge each
comparison would be to ask only whether its 95\,\% confidence interval excludes zero. However, for effects that are
comparable to the resolution of the bootstrap, this would reduce inference to a yes-or-no answer. Instead, we
graded the strength of evidence across comparisons. None of the four primary comparisons met its prespecified
criterion. Two missed it narrowly (the latent state in XRO by 0.002, and the lead-dependent base, preferred in two
of the four folds), and we report all four accordingly.

A second caveat concerns the descriptive analyses. Because no multiple-testing correction was applied to them,
isolated clear differences should not be overemphasized. Instead, the conclusions of this study rest on signals
that recur across lead times, folds, and bootstrap block lengths. Further, more aggressive tuning of the
latent-state representation on a record of this size would likely increase the risk of false positives. Note also
that there is no untouched test period: every development choice, including the blend weight, was made with the
cross-validated forecasts in view. What the Floquet operator adds to its base does not, however, rest on the fold
used during development (2013--2024). With that fold left out, the gain remains positive, with strong evidence
(Sect.~\ref{sec:res-base-gain}).

The limited record also has a cost for the reference models. All systems were refitted on folds of about 30 years.
This is fair, because every system is treated alike, but it costs XRO and DESN skill at long leads. For example,
XRO trained on 1958--1999 reaches 0.54 and 0.38 at leads 12 and 15 on the 197 verification windows of 2002--2023,
and 0.47 and 0.26 when trained on our folds (moderate evidence for XRO and weak for DESN over leads 10--18;
Appendix~\ref{app:references}). The authors of XRO note that a robust fit needs 40--50 years \citep{Zhao2024}. The
autoencoder suffers less from this, since it is fitted once on the whole record without any forecasting target.
However, the operators are fitted on the folds, like the reference models. Two routes would lengthen the training
record: earlier reanalysis years, for which observational data are sparser, and pre-training the operators on
coupled climate model outputs, as \citet{HamEtAl2019} did for a convolutional network. Note, however, that both
require the encoder to map the new data into the same latent space, which is a test of the representation in
itself.

Several aspects of LatENSO's behavior help to clarify both its strengths and its limitations. Where, for example,
do the two main deficits of the Floquet operator come from? Both can be traced to identifiable causes, and neither
is shared by XRO-lat, which is built on the same latent state and is level with XRO-dom at leads 3--9. The first
deficit is the skill gap to the in-domain reference models at leads 3--9, which originates in large part from the
base forecast rather than from the learned dynamics themselves. When the base coefficients are allowed to vary
seasonally at short leads (the lead-dependent base), the gap to XRO-dom is halved and the forecast amplitude
increases. The operator still adds skill over the new base ($+0.039$ against $+0.054$ over leads 1--18), and the
modal spectrum of its generator remains essentially unchanged.

The second deficit, for boreal-summer targets at lead 18, also lies mostly in the base. Alone, the base reaches
only 0.283 for these targets, against 0.381 for nRO and 0.418 for XRO-dom. The operator lowers this score further,
to 0.218, through its learned terms around the propagation, which the fixed blend weight applies in every season. A
season-dependent blend weight would be an obvious remedy for that part. The absence of a comparable deficit in
XRO-lat, which uses neither this base nor a blend and is the most skillful system for these targets, is consistent
with this interpretation.

Neither modification, however, is part of LatENSO v1.0, each for its own reason. The lead-dependent base narrowly
failed the criterion specified in advance: it scored higher than OLS4 on training data in two of the four folds,
where three were required. A season-dependent blend was not tested, because selecting it after inspection of the
verification results would have introduced tuning on the years held out for verification.

The Floquet operator was designed not only to forecast but also to provide a degree of dynamical interpretability,
and the results show that this objective was achieved for the leading mode of its generator. Within the inductive
biases built into the operator (three damped oscillators with bounded damping rates and periods), the learned
time-invariant generator is remarkably stable. Its oscillatory periods remain within the ENSO band, while stability
(every mode is damped) is imposed as part of that inductive bias rather than learned. The generator consistently
holds a prominent quasi-biennial oscillator, the least damped of the three, together with decay modes that fade
without cycling.

Two cautions apply to this reading. First, these are properties of the generator: the seasonal terms are larger and
mix its modes through the year, so they are not eigenperiods of the full seasonal operator. Second, the remaining
oscillators contribute little beyond the isotropic share (the share of the forecast variance that their planes
would carry if that variance were the same in every direction), and adding a fourth oscillator provides no
reproducible benefit. Note, however, that the generator's periods are essentially unchanged when the base is
changed. This indicates that they reflect a property of the learned dynamics rather than an artifact of the
prediction target.

The propagation built on this generator also carries what the operator adds to its base. With the two additive
terms removed, the propagation alone adds $+0.066$ [$+0.022$, $+0.112$] to the ACC of the base over leads 1--18,
with clear evidence at every block length and with the fold means removed, against $+0.054$ for the full operator.
Further, an operator trained without these terms reaches the same ACC over these leads, with generator periods
0.03--0.12 years shorter (Sects.~\ref{sec:res-base-gain} and \ref{sec:res-ablations}; Appendix~\ref{app:blocks}).
The interpretation should nevertheless remain modest. The operator exposes interpretable parameters that a
black-box forecasting system would not provide, but its three-oscillator structure is an inductive bias of the
design rather than an empirical discovery.

Equally informative are the approaches that did not improve performance. As a residual corrector, the latent state
improved nRO by $+0.019$ [$+0.007$, $+0.038$] and XRO-dom by $+0.010$ [$-0.001$, $+0.022$], in every fold, but it
offered no advantage over 32 EOFs derived from the same fields. Restricting the Floquet operator to the latent
coordinates most strongly correlated with Ni\~no 3.4 reduced rather than increased skill. A fourth oscillator
failed to generalize across folds, an explicit amplitude nonlinearity produced no measurable improvement, and a
seasonal base selected within the training data overfitted at leads 10--18. Nor do the two additive terms of the
operator add skill to the forecast (Sects.~\ref{sec:res-base-gain} and \ref{sec:res-ablations}). An operator
without them is therefore a candidate for a simpler next version, although this record does not resolve whether
they help the propagation during training.

Together, these null results place useful constraints on future developments by identifying modifications that
appear plausible but do not survive out-of-sample evaluation. From this perspective, the fourth oscillator is
particularly instructive. On the development fold it improved the forecast in six of six seeds (paired
$t_{5}=5.93$), but in its four-fold test it did so in 20 of 40 pairs (paired $t_{39}=1.31$). This means that, on a
record of this size, a single development fold can produce apparently decisive evidence for an improvement that
does not generalize.

\subsection{Next steps}
\label{sec:disc-next}

The decoder gives the latent state value beyond forecasting. With the decoder, the latent state is not just a
compact predictor. Instead, it provides a two-way mapping between a small number of latent coordinates and the full
set of coupled ocean--atmosphere fields. This creates an interpretable state space in which information can be
compressed, analyzed, modified, and then translated back into physical variables, in the spirit of the learned,
interpretable representations that \citet{Gupta2026} calls for, and of solvers that work in the latent space of a
pre-trained generator \citep{Panahi2026}. For example, a climate model simulation could, in principle, be projected
into the same latent space and compared directly with observations coordinate by coordinate. Ultimately, this would
allow model biases to be diagnosed and potentially corrected in a compact dynamical representation before being
decoded back into physical fields. Of course, whether the encoder learned here transfers to climate-model output
remains an open question. Establishing the extent to which its latent coordinates retain the same physical meaning across models is
therefore the first exploration to be pursued.

A second avenue concerns the spatial domain itself. The present latent state is restricted to the tropical Pacific,
and it therefore cannot explicitly represent variability that originates in other basins. One natural step would be
to extend the autoencoder to the entire tropical belt. The latent representation could then encode Indian Ocean and
Atlantic variability alongside Pacific dynamics within a single, unified state space. In principle, such a state
could capture much of the information that XRO currently obtains through separate remote-basin indices, while
preserving the advantages of a compact and physically interpretable representation. Exploring whether a latent
space that spans the tropics can unify these sources of predictability and improve forecasts is the focus of
ongoing work.

\conclusions
\label{sec:conclusions}

We asked whether a state learned from the coupled fields of the tropical Pacific alone, compressed to a few tens of
numbers and trained without any forecasting objective, can serve as a physically grounded state for seasonal ENSO
operators. LatENSO v1.0 demonstrates that such a compact, learned representation of tropical-Pacific variability
can support ENSO forecasting on a par with leading oscillator and deep-learning systems trained on the same record,
while remaining physically interpretable. Its 32-dimensional state is generated by a Swin-transformer autoencoder
whose attention windows follow the Walker and Hadley circulations and which is trained solely to reconstruct
thirteen coupled ocean--atmosphere fields. Despite its compactness, the representation reproduces these fields as
accurately as 89--172 EOF modes, separates eastern- and central-Pacific El Ni\~no events into distinct coordinates,
and compresses the 1981--2024 record into only 3.0\,MB.

On identical data, folds, and verification windows, the forecasting systems built on this latent state are level
with or ahead of the recharge oscillator and the versions of XRO and of a deep echo-state network that use the same
domain, over leads 1--18. Specifically, the Floquet operator is statistically indistinguishable from all three, and
XRO-lat is ahead of the in-domain XRO with strong evidence. Over the same leads, XRO-lat is also statistically
indistinguishable from the versions of XRO and DESN with remote-basin indices. Note, however, that these versions
remain ahead of the Floquet operator at leads 3--9.

When used as an augmentation of XRO, the latent state contributes about as much forecast skill as the eight
remote-basin indices added through the same selection pipeline ($+0.019$ against $+0.021$ in mean ACC), with a gain
that is positive in every fold and grows with lead. In contrast, EOFs derived from the same fields and selected in
the same way provide no measurable benefit. XRO-lat is therefore the most skillful system considered here that
relies exclusively on tropical-Pacific information, and it achieves an ACC within 0.012 of the full three-basin
XRO. From lead 15 to lead 24 it is the most skillful of all systems considered.

The Floquet operator further identifies El Ni\~no events more effectively than every reference model at leads
12--18 (AUC 0.819 against 0.752--0.767 over leads 10--18), although the evidence for this advantage is moderate at
most. What it adds to its base is carried by its Floquet propagation, which reaches the same ACC over leads 1--18
as the full operator when trained alone. In addition, its time-invariant generator consistently holds a prominent
quasi-biennial oscillator whose period remains stable across model configurations. Among the systems compared, the
Floquet operator is also the best calibrated at leads 4--8, and XRO-lat is among the best calibrated at leads
12--18 (Sects.~\ref{sec:res-base} and \ref{sec:res-xro}). The principal deficiencies of the Floquet operator are
partly attributable to the base forecast and the blending strategy rather than to the latent representation itself.

Of course, these conclusions, and the comparisons between models, are necessarily constrained by the available
record. With only 36 initialization years, many of the differences of interest are comparable to the resolution of
the paired bootstrap, between about 0.02 and 0.10 in mean ACC depending on the pair of systems. This is why we
emphasize graded evidence rather than binary significance tests.

One outcome is intriguing. Although the latent state remains confined to the tropical Pacific and contains no
explicit information from the Indian or Atlantic Oceans, it adds forecast skill at leads where the remote-basin
indices no longer do. Their contribution peaks at lead 12 and turns negative from lead 17, whereas that of the
latent state stays near its maximum, about $+0.06$, to lead 24. This suggests that the slow evolution of the Pacific itself holds long-lead
predictability of its own, but more work is needed to prove it.

More broadly, we believe that the contribution of LatENSO lies not only in competitive forecast skill but, above
all, in the introduction of an interpretable state space that links a small set of coordinates to the full coupled
climate fields. This two-way mapping opens the possibility of comparing observations and climate-model simulations
within a common latent representation, diagnosing model errors, and potentially correcting them before
reconstructing the physical fields. Two questions remain open here: whether the encoder transfers to climate-model
output, and whether a latent state that spans the whole tropical belt can unify the Pacific and remote-basin
sources of predictability. To allow the results to be checked and extended, all code, trained weights and forecasts
are openly available, and every table and figure of this paper can be regenerated from them.

\codedataavailability{The exact version of the LatENSO v1.0 code used to produce the results of this paper (the autoencoder, the Floquet
modal operator and its training, the pipeline that places latent directions in XRO, the reference-model runs on our
folds, the evaluation and the scripts that produce every figure and table) is archived on Zenodo under the MIT
license at \url{https://doi.org/10.5281/zenodo.24374200} \citep{PanahiBracco2026code}. The trained autoencoder, the 40 trained operators, the 40 retrained on the
lead-dependent base (Sect.~\ref{sec:sens-base}) and the 40 retrained without the two additive terms
(Sect.~\ref{sec:res-ablations}), the stored
forecasts of every system on the 394 verification windows, the latent trajectories and the 13-field input on the
model domain (1.3\,GB, compressed) are archived under the Creative Commons Attribution 4.0 license at \url{https://doi.org/10.5281/zenodo.24374299}
\citep{PanahiBracco2026data}. ERA5 monthly averaged data on single levels are available from the Copernicus Climate Change Service
Climate Data Store (\url{https://doi.org/10.24381/cds.f17050d7}; \citealp{Hersbach2020}), and the SODA4.15.2
regridded fields from \url{https://dsrs.atmos.umd.edu/DATA/SODA4.15.2_REG/} \citep{Chepurin2025}. The XRO code is available at
\url{https://github.com/senclimate/XRO} \citep{Zhao2024}; the DESN code and the ORAS5-based climate-mode indices
used here are available at \url{https://github.com/zhangzejing/RC-ENSO} \citep{Zhang2026}. The revisions used are
given in Appendix~\ref{app:references}.
}

\appendix
\section{Reference models: settings and reproduction}
\label{app:references}

\textbf{Settings on our folds.} XRO, XRO-dom and nRO are fitted with the XRO code of \citet{Zhao2024} (revision
d8527a8), with annual and semi-annual harmonics (\texttt{ac\_order}\,=\,2) and the quadratic terms $T^2$ and $Th$
in the ENSO temperature equation. Anomalies are taken with respect to the climatology of the training months of
each fold, and the first training month of every fit is a January. The XRO code tabulates its seasonal operator from the first fitted month, which it takes to be a January, so this keeps the operator in phase with the forecast calendar. Forecasts are integrated without noise from each of the 394 initializations. DESN and
DESN-dom are trained with the GPU code of \citet{Zhang2026} (RC-ENSO, revision a506ab2) and the settings of
their prediction notebook: 20 one-layer reservoirs of 20\,000 units and 20 two-layer reservoirs of 20\,000 and
12\,000 units, spectral radius 1, input connectivity 0.1, a ridge penalty of $5\times10^{-6}$ on the readout, inputs augmented with annual and semi-annual harmonics. The forecast is the mean of the 40 members. The inputs are the same index sets as XRO and
XRO-dom.

\begin{figure*}[tp]
\includegraphics[width=\textwidth]{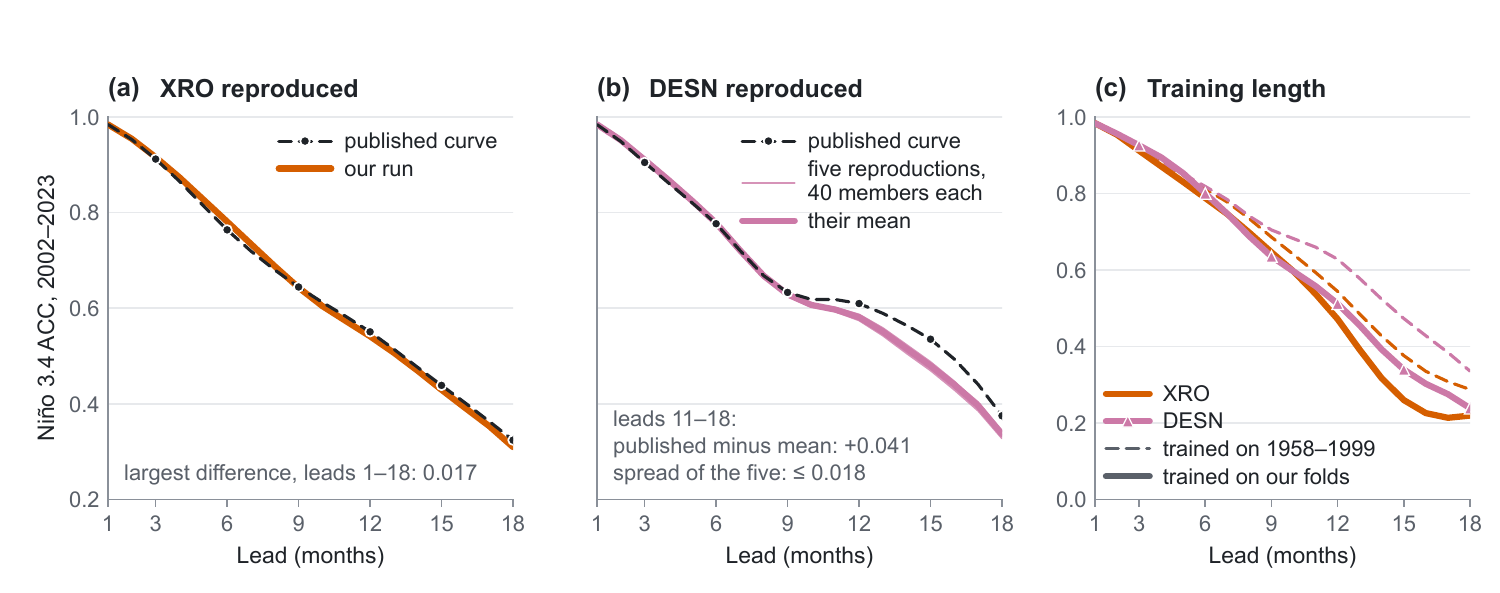}
\caption{The reference models reproduce their published skill. Ni\~no 3.4 ACC for initializations in 2002--2023,
three-month running means, as in the published curves. (a) XRO: published curve and our run of the authors'
setup (ORAS5 training, 1958--1999). (b) DESN: published curve, and five independent 40-member reproductions with
the authors' GPU code and settings, and their mean. (c) XRO and DESN trained
on 1958--1999 (dashed; for DESN one of the ensembles of b) and on our cross-validation folds of about 30 years
(solid), all on the 197 windows of 2002--2023 verified in the cross-validation; (a) and (b) use every
initialization of 2002--2023.}
\label{fig:a1}
\end{figure*}

\textbf{Reproduction of the published curves.} Before any model was run on our folds, each was run in its
authors' published setup and compared with the published skill, verified on 2002--2023 with three-month running
means (Fig.~\ref{fig:a1}). XRO, trained on ORAS5 indices for 1958--1999 with the authors' settings, reproduces the
published curve to within 0.017 at every lead from 1 to 18. DESN, trained with the authors' code and settings,
reproduces the published curve to within 0.02 up to lead 10. From lead 11 to 18, five independent 40-member
reproductions lie on average 0.041 below it, with a spread between reproductions of at most 0.018, so the gap is
systematic rather than member noise. A later version of the DESN repository selects ensemble members using skill
on 2002--2023; we cannot tell whether the published curve used that selection, and our runs do not.

\textbf{Training length.} Our folds leave about 30 years for training, whereas XRO and DESN were published with
training on 1958--1999. To measure what the shorter training costs them, Fig.~\ref{fig:a1}c compares each model
trained on 1958--1999 with the same model trained on our folds of about 30 years, with the published definition of
skill and on the same 197 windows. These are the windows of 2002--2023 that our cross-validation verifies, initialized from June
2003 to July 2011 and from April 2014 to June 2022. Each run is verified against the Ni\~no 3.4 index of the data
it was trained on. For DESN trained on 1958--1999 we use one of the five ensembles of Fig.~\ref{fig:a1}b. At leads
12, 15 and 18, XRO falls from 0.54, 0.38 and 0.29 to 0.47, 0.26 and 0.22, and DESN from 0.63, 0.47 and 0.34 to
0.51, 0.34 and 0.24. Over leads 10--18 the longer training is ahead by $+0.085$ [$-0.020$, $+0.159$] for XRO
(moderate evidence) and by $+0.114$ [$-0.061$, $+0.252$] for DESN (weak evidence), with the 18 initialization
years of these windows as bootstrap blocks. On the same windows and definition the Floquet operator reaches 0.31,
0.29 and 0.25.

\section{The latent state in XRO: designs, controls and remote-basin content}
\label{app:designs}

Table~\ref{tab:designs} gives the skill of each of the eleven designs of Sect.~\ref{sec:xrolat} when the same
design is used in every fold, without the selection, for the latent state, the 32 EOFs computed from the full
record and from the training months of each fold, and the eight remote-basin indices. These numbers are
diagnostic: the reported systems choose their design inside each training fold, and the paired intervals of
Table~\ref{tab:latent} are conditional on those choices. The latent state matches or beats the EOFs of the full record
in eight of the eleven designs (mean 0.648 against 0.640; best 0.670 against 0.664) and those of the training
months in four (mean 0.648 against 0.649; best 0.670 against 0.668). The EOFs of the training months score higher
than those of the full record in eight designs. Four latent designs beat XRO-dom (0.654), as do four designs of
each EOF basis, against all eleven for the remote-basin indices. The latent state is ahead where its slow directions are used:
with two slow directions it outscores the EOFs of the full record by $+0.017$ [$-0.006$, $+0.042$] (moderate
evidence) and those of the training months by $+0.022$ [$-0.002$, $+0.046$] (strong evidence), in every fold. With two predictive directions added to the meridional modes, in contrast, the latent state and the two sets of EOFs are level ($-0.001$
[$-0.012$, $+0.010$] and $-0.001$ [$-0.014$, $+0.012$]). This comparison was not specified in advance. The nested selection chose two latent directions in every fold
(predictive directions added to the meridional modes in the first and fourth folds, slow directions in the second and third). For the EOFs of the full record it chose four different designs, for those of the training months three, and for the remote-basin indices four.

\begin{table}[t]
\caption{Mean ACC over leads 1--18 when one design is used in every fold, without the selection. The EOFs are
computed from the full record or from the training months of each fold. $K$: number of directions, which replace the two meridional modes or are added to them. Predictive designs use at most the six
directions that exist (Sect.~\ref{sec:xrolat}). XRO-dom alone: 0.654.}
\label{tab:designs}
\scriptsize
\begin{tabular}{lcccc}
\tophline
Design & Latent state & EOFs, full record & EOFs, training months & Remote-basin indices \\
\middlehline
predictive, replacing, $K=2$ & 0.642 & 0.640 & 0.645 & 0.691 \\
slow, $K=2$ & \textbf{0.670} & 0.653 & 0.648 & 0.677 \\
predictive, added, $K=2$ & \textbf{0.662} & 0.663 & 0.663 & 0.684 \\
predictive, replacing, $K=4$ & 0.640 & 0.632 & 0.639 & 0.677 \\
slow, $K=4$ & \textbf{0.668} & 0.663 & 0.656 & 0.683 \\
predictive, added, $K=4$ & \textbf{0.663} & 0.664 & 0.668 & 0.683 \\
predictive, replacing, $K=8$ & 0.639 & 0.627 & 0.639 & 0.676 \\
all 32, ridge 0.3 & 0.645 & 0.656 & 0.660 & 0.692 \\
all 32, ridge 0.1 & 0.638 & 0.638 & 0.652 & 0.688 \\
all 32, ridge 0.03 & 0.632 & 0.610 & 0.640 & 0.686 \\
all 32, ridge 0.01 & 0.628 & 0.588 & 0.633 & 0.685 \\
\bottomhline
\end{tabular}
\belowtable{Bold: latent designs that beat XRO-dom.}
\end{table}

\textbf{Remote-basin content of the latent state.} For each remote-basin index we fitted, on the training months of each fold, a lightly regularized (ridge) linear regression on the four in-domain indices alone and on the four indices plus the 32 latent coordinates, at lags 0, 3 and 6 months (the index that many months before the predictors), and scored it on the held-out period (Table~\ref{tab:remote}). The predictors are standardized, and the penalty is one fiftieth of the number of training months. The score is $R^2$, one minus the mean squared error divided by the variance of the index over the four held-out periods together, so it is negative when the regression does worse than the mean of the index. The latent state raises the score of the North Pacific meridional mode, whose box is partly inside the domain (0.82 to 0.93 at lag 0), changes that of the South Pacific mode little at lag 0 and lowers it at longer lags. For the six indices of the Indian Ocean and the Atlantic, adding the latent state lowers the held-out $R^2$ at every lag, with the single exception of the IOD at lag 0
(0.08 to 0.12), and leaves none of them above 0.16. The latent state therefore carries no information on them that survives out of sample.

\begin{table}[t]
\caption{Out-of-sample $R^2$ of each remote-basin index, regressed on the four in-domain indices (dom) and on the same
plus the 32 latent coordinates (+lat), at three lags.}
\label{tab:remote}
\scriptsize
\begin{tabular}{lcccccc}
\tophline
 & \multicolumn{2}{c}{lag 0} & \multicolumn{2}{c}{lag 3} & \multicolumn{2}{c}{lag 6} \\
Index & dom & +lat & dom & +lat & dom & +lat \\
\middlehline
NPMM & 0.82 & 0.93 & 0.48 & 0.56 & 0.34 & 0.38 \\
SPMM & 0.84 & 0.86 & 0.38 & 0.35 & 0.28 & 0.16 \\
IOB & 0.36 & 0.16 & 0.24 & $-0.05$ & 0.17 & $-0.05$ \\
IOD & 0.08 & 0.12 & 0.12 & 0.00 & $-0.01$ & $-0.10$ \\
SIOD & 0.26 & 0.16 & 0.22 & 0.04 & 0.12 & $-0.03$ \\
TNA & $-0.03$ & $-0.16$ & 0.00 & $-0.28$ & 0.01 & $-0.36$ \\
ATL3 & 0.03 & $-0.08$ & $-0.02$ & $-0.10$ & 0.00 & $-0.10$ \\
SASD & 0.16 & 0.00 & 0.18 & $-0.03$ & 0.13 & $-0.01$ \\
\bottomhline
\end{tabular}
\end{table}

\textbf{The latent state as an add-on corrector.} Before testing the latent state inside XRO, we tested it as a corrector of the residuals of nRO, XRO-dom and XRO. The corrector is a per-lead ridge regression of each model's error on the latent state at initialization, with seasonal interactions (each coordinate is also multiplied by the cosine and by the sine of the initialization month). It is fitted on out-of-block residuals of the training months (the errors of forecasts initialized in a four-year block by a model fitted without it), with the penalty chosen at each lead by cross-validation over the blocks. The same regression on 32 EOFs of the training months is the control. The criteria, fixed before the test and
applied to the corrector of XRO-dom over leads 1--18, were a gain of at least $+0.02$, a positive gain in at least
three of the four folds, a gain larger than that of the EOF corrector, and a positive gain for nRO and XRO as well.
The latent corrector improved nRO by $+0.019$
[$+0.007$, $+0.038$], XRO-dom by $+0.010$ [$-0.001$, $+0.022$] and XRO by $+0.003$ [$-0.004$, $+0.012$], positive in
every fold, but the EOF corrector did as well or slightly better ($+0.023$ [$+0.009$, $+0.044$], $+0.012$
[$+0.002$, $+0.024$] and $+0.006$ [$-0.001$, $+0.014$]). The difference between the two on XRO-dom was $-0.002$ [$-0.008$, $+0.002$]. The latent corrector thus met the
second and the fourth criteria but not the first or the third, and it did not outperform EOFs as a corrector.

\section{Block lengths, graded evidence, event thresholds and biases}
\label{app:blocks}

Table~\ref{tab:blocks} repeats the main comparisons with year blocks of one to four years. The deficits stated in
the paper keep their sign and size at every block length. For most of them the evidence stays clear, but the gap to DESN at leads 3--9 becomes borderline with longer blocks (the interval then reaches zero), and the summer deficit against XRO-dom is clear at three of the four block lengths. Over leads 1--18 the evidence for the skill gain of OLS4+FMO over OLS4 is clear only with one-year blocks. At leads 3--9 it is clear at every block length, although about two thirds of this mid-lead gain is reproduced by a base that is seasonal at short leads (at these leads the operator adds $+0.017$ [$-0.008$, $+0.043$] over the lead-dependent base OLS4c). The evidence for the gain of its Floquet propagation alone, with the two additive terms removed (Sect.~\ref{sec:res-base-gain}), is clear at every block length. The gains of the latent state in XRO and of the
Floquet operator in El Ni\~no detection keep their sign and size, with intervals that include zero at every block
length.

\begin{table*}[t]
\caption{Sensitivity to the block length of the paired bootstrap: point estimate and 95\,\% interval with blocks of
1, 2, 3 and 4 years (36, 20, 14 and 11 blocks). The one-year intervals use 10\,000 draws, the others 2000. A block spans that many consecutive calendar years, counted from the first initialization year, and holds the initialization years that fall in it. JJA: June--August targets; amplitude: amplitude slope. An asterisk marks an interval that excludes zero.}
\label{tab:blocks}
\scriptsize
\begin{tabular}{lccccc}
\tophline
Comparison & Effect & 1 year & 2 years & 3 years & 4 years \\
\middlehline
ACC 3--9, OLS4+FMO $-$ XRO & $-0.089$ & [$-0.159$, $-0.026$]$^{*}$ & [$-0.172$, $-0.011$]$^{*}$ & [$-0.170$, $-0.021$]$^{*}$ & [$-0.178$, $-0.019$]$^{*}$ \\
ACC 3--9, OLS4+FMO $-$ DESN & $-0.078$ & [$-0.153$, $-0.009$]$^{*}$ & [$-0.167$, $+0.000$] & [$-0.169$, $+0.002$] & [$-0.188$, $+0.006$] \\
ACC 3--9, OLS4+FMO $-$ XRO-dom & $-0.062$ & [$-0.134$, $+0.000$] & [$-0.136$, $+0.012$] & [$-0.129$, $+0.002$] & [$-0.138$, $+0.004$] \\
ACC 1--18, OLS4+FMO $-$ OLS4 & $+0.054$ & [$+0.003$, $+0.104$]$^{*}$ & [$-0.004$, $+0.118$] & [$-0.008$, $+0.111$] & [$-0.014$, $+0.097$] \\
ACC 3--9, OLS4+FMO $-$ OLS4 & $+0.051$ & [$+0.008$, $+0.092$]$^{*}$ & [$+0.004$, $+0.095$]$^{*}$ & [$+0.001$, $+0.098$]$^{*}$ & [$+0.001$, $+0.087$]$^{*}$ \\
ACC 1--18, Floquet propagation $-$ OLS4 & $+0.066$ & [$+0.022$, $+0.112$]$^{*}$ & [$+0.018$, $+0.127$]$^{*}$ & [$+0.014$, $+0.117$]$^{*}$ & [$+0.011$, $+0.103$]$^{*}$ \\
ACC 1--18, OLS4+FMO $-$ XRO & $-0.066$ & [$-0.152$, $+0.015$] & [$-0.154$, $+0.027$] & [$-0.137$, $+0.009$] & [$-0.156$, $+0.023$] \\
ACC 1--18, XRO-lat $-$ XRO-dom & $+0.019$ & [$-0.002$, $+0.040$] & [$-0.006$, $+0.040$] & [$-0.002$, $+0.035$] & [$-0.003$, $+0.042$] \\
JJA lead 18, OLS4+FMO $-$ nRO & $-0.163$ & [$-0.331$, $-0.011$]$^{*}$ & [$-0.350$, $-0.010$]$^{*}$ & [$-0.402$, $-0.013$]$^{*}$ & [$-0.373$, $-0.044$]$^{*}$ \\
JJA lead 18, OLS4+FMO $-$ XRO-dom & $-0.199$ & [$-0.402$, $-0.012$]$^{*}$ & [$-0.427$, $-0.012$]$^{*}$ & [$-0.475$, $+0.008$] & [$-0.438$, $-0.033$]$^{*}$ \\
Amplitude lead 6, OLS4+FMO $-$ nRO & $-0.143$ & [$-0.269$, $-0.006$]$^{*}$ & [$-0.235$, $-0.040$]$^{*}$ & [$-0.240$, $-0.044$]$^{*}$ & [$-0.252$, $-0.038$]$^{*}$ \\
Amplitude lead 6, OLS4+FMO $-$ XRO-dom & $-0.185$ & [$-0.307$, $-0.051$]$^{*}$ & [$-0.303$, $-0.048$]$^{*}$ & [$-0.302$, $-0.052$]$^{*}$ & [$-0.324$, $-0.039$]$^{*}$ \\
El Ni\~no AUC 10--18, OLS4+FMO $-$ XRO & $+0.058$ & [$-0.019$, $+0.133$] & [$-0.016$, $+0.130$] & [$-0.018$, $+0.130$] & [$-0.021$, $+0.127$] \\
El Ni\~no AUC 10--18, OLS4+FMO $-$ XRO-dom & $+0.059$ & [$-0.013$, $+0.134$] & [$-0.018$, $+0.129$] & [$-0.012$, $+0.130$] & [$-0.033$, $+0.142$] \\
El Ni\~no AUC 10--18, OLS4+FMO $-$ nRO & $+0.067$ & [$-0.020$, $+0.150$] & [$-0.019$, $+0.149$] & [$-0.031$, $+0.148$] & [$-0.057$, $+0.149$] \\
\bottomhline
\end{tabular}
\end{table*}

Table~\ref{tab:thresh} gives the El Ni\~no detection of the Floquet operator against each reference model for event thresholds from $+0.5$ to $+1.5\,^{\circ}$C. At $+0.5\,^{\circ}$C the advantage is small; from $+0.75\,^{\circ}$C upward it is
positive against every reference model at every threshold, with no interval excluding zero. Against its base the evidence is clear from $+0.75\,^{\circ}$C upward. For La Ni\~na the Floquet operator trails XRO at both $-0.5$ and
$-1.0\,^{\circ}$C ($-0.035$ and $-0.102$).

\begin{table*}[t]
\caption{El Ni\~no AUC over leads 10--18, OLS4+FMO minus each system, against the event threshold, with 95\,\%
intervals (2000 draws). In parentheses: verification windows above the threshold at lead 12.}
\label{tab:thresh}
\scriptsize
\setlength{\tabcolsep}{3pt}
\begin{tabular}{lccccc}
\tophline
Threshold & $+0.5\,^{\circ}$C (98) & $+0.75\,^{\circ}$C (70) & $+1.0\,^{\circ}$C (48) & $+1.25\,^{\circ}$C (35) & $+1.5\,^{\circ}$C (28) \\
\middlehline
nRO & $+0.022$ [$-0.049$, $+0.087$] & $+0.066$ [$-0.007$, $+0.132$] & $+0.067$ [$-0.022$, $+0.146$] & $+0.076$ [$-0.030$, $+0.170$] & $+0.068$ [$-0.048$, $+0.172$] \\
XRO-dom & $+0.008$ [$-0.063$, $+0.071$] & $+0.045$ [$-0.019$, $+0.105$] & $+0.059$ [$-0.016$, $+0.130$] & $+0.076$ [$-0.013$, $+0.166$] & $+0.057$ [$-0.030$, $+0.151$] \\
DESN-dom & $+0.031$ [$-0.063$, $+0.125$] & $+0.073$ [$-0.025$, $+0.178$] & $+0.052$ [$-0.053$, $+0.170$] & $+0.044$ [$-0.069$, $+0.192$] & $+0.007$ [$-0.085$, $+0.144$] \\
XRO & $+0.006$ [$-0.079$, $+0.086$] & $+0.023$ [$-0.055$, $+0.091$] & $+0.058$ [$-0.023$, $+0.132$] & $+0.081$ [$-0.004$, $+0.175$] & $+0.062$ [$-0.012$, $+0.147$] \\
DESN & $+0.014$ [$-0.065$, $+0.091$] & $+0.055$ [$-0.027$, $+0.136$] & $+0.056$ [$-0.037$, $+0.158$] & $+0.064$ [$-0.047$, $+0.193$] & $+0.034$ [$-0.060$, $+0.169$] \\
XRO-lat & $+0.001$ [$-0.061$, $+0.062$] & $+0.033$ [$-0.016$, $+0.086$] & $+0.054$ [$-0.002$, $+0.114$] & $+0.067$ [$-0.001$, $+0.141$] & $+0.048$ [$-0.016$, $+0.122$] \\
OLS4 & $+0.048$ [$-0.008$, $+0.109$] & $+0.087$ [$+0.028$, $+0.156$]$^{*}$ & $+0.086$ [$+0.023$, $+0.166$]$^{*}$ & $+0.074$ [$+0.017$, $+0.150$]$^{*}$ & $+0.083$ [$+0.023$, $+0.164$]$^{*}$ \\
\bottomhline
\end{tabular}
\end{table*}

Table~\ref{tab:bias} gives the mean error of the two LatENSO operators, the Floquet operator without the blend, its base
and the five reference models in each fold. Because the anomalies of each fold are
taken with respect to the climatology of its training months, the observed anomaly itself has a mean that differs
between folds, from $-0.249$\,$^{\circ}$C in 2002--2013 to $+0.122$\,$^{\circ}$C in 2013--2024. A correlation pooled over the four folds therefore also rewards a forecast that follows these means. Removing the mean of each fold's
windows from forecasts and observations at every lead, within every bootstrap draw, leaves the main comparisons
nearly unchanged: XRO-lat $-$ XRO-dom is $+0.016$ [$-0.003$, $+0.033$] (strong evidence; $+0.019$
pooled), the remote-basin indices selected in the same way add $+0.022$ [$+0.001$, $+0.048$] (clear; $+0.021$),
XRO $-$ XRO-dom is $+0.031$ [$-0.009$, $+0.070$] (moderate; $+0.031$), and OLS4+FMO $-$ XRO at leads 3--9 is
$-0.087$ [$-0.157$, $-0.024$] (clear; $-0.089$). XRO-lat $-$ nRO falls from $+0.042$ to $+0.026$ [$-0.016$,
$+0.071$] (weak evidence; pooled, it is strong with one-year blocks and moderate with blocks of two to four years),
because removing the fold means raises the mean ACC of nRO and not that of XRO-lat (0.673 and 0.672). The gain of OLS4+FMO over OLS4 falls from $+0.054$ to $+0.041$
[$-0.002$, $+0.090$] (strong), because removing the fold means raises the mean ACC of the base (from 0.565 to
0.587) more than that of the operator (from 0.619 to 0.628), whose fold biases are smaller in magnitude on average (0.101
against 0.150\,$^{\circ}$C). The gain of its Floquet propagation alone falls from $+0.066$ to $+0.050$ [$+0.013$, $+0.095$], and the evidence stays clear.

\begin{table*}[t]
\caption{Mean forecast error (forecast minus observation) of the Ni\~no 3.4 anomaly over leads 1--18, in
$^{\circ}$C, over the windows of each fold, labeled by its held-out period, and over all windows. Last row: mean of
the observed anomaly over the same windows.}
\label{tab:bias}
\scriptsize
\begin{tabular}{lccccc}
\tophline
System & 1981--1991 & 1991--2002 & 2002--2013 & 2013--2024 & All windows \\
\middlehline
OLS4+FMO & $-0.102$ & $-0.087$ & $+0.153$ & $+0.060$ & $+0.006$ \\
XRO-lat & $+0.012$ & $+0.013$ & $+0.150$ & $+0.212$ & $+0.097$ \\
OLS4+FMO, unblended ($\alpha=1$) & $+0.144$ & $-0.233$ & $-0.376$ & $+0.095$ & $-0.091$ \\
OLS4 (base) & $-0.184$ & $-0.038$ & $+0.330$ & $+0.049$ & $+0.038$ \\
nRO & $-0.072$ & $-0.042$ & $+0.304$ & $+0.026$ & $+0.054$ \\
XRO-dom & $-0.002$ & $-0.019$ & $+0.208$ & $+0.184$ & $+0.093$ \\
DESN-dom & $-0.167$ & $-0.156$ & $+0.183$ & $+0.018$ & $-0.031$ \\
XRO & $+0.118$ & $+0.014$ & $+0.189$ & $+0.257$ & $+0.145$ \\
DESN & $+0.085$ & $-0.121$ & $+0.121$ & $+0.062$ & $+0.037$ \\
\middlehline
Observed anomaly & $-0.085$ & $-0.114$ & $-0.249$ & $+0.122$ & $-0.081$ \\
\bottomhline
\end{tabular}
\end{table*}

\section{Base selection}
\label{app:base}

\textbf{Seasonal base.} The seasonal base uses the four predictors and an intercept, each multiplied by
$[1,\allowbreak\cos\theta_m,\allowbreak\sin\theta_m,\allowbreak\cos2\theta_m,\allowbreak\sin2\theta_m]$, with $m$ the initialization month, fitted by least squares. It was compared with OLS4 by an inner cross-validation of the base alone on the training months of each fold (Sect.~\ref{sec:sens-base}), under the criterion, set before the comparison, that it must score higher in at least three of the four folds. It scored higher in one (Table~\ref{tab:base}); it was better at leads 3--9 in every fold and
worse at leads 10--18 in every fold, where its 25 coefficients per latent coordinate overfit. OLS4 was kept, and no operator was trained on the seasonal base.

\textbf{Lead-dependent base.} The lead-dependent base OLS4c uses the seasonal base for leads 1 to $L^{*}$ and OLS4 for longer leads. In each fold $L^{*}$ is the lead that maximizes the inner-cross-validated ACC over leads 1--18 (11, 8, 9 and 9). Scoring OLS4c on the same inner blocks that chose $L^{*}$ would favor it, so the criterion was applied in nested form: $L^{*}$ is chosen without each inner block in turn, and the resulting base is scored on that block. In this nested score OLS4c is higher than OLS4 in two folds of four (Table~\ref{tab:base}), one fold short of the criterion, and OLS4 was kept. As decided before the comparison, all 40 operators were retrained on OLS4c whatever its outcome, with the same seeds and settings, to measure how much of the operator's deficits came from the base (Sect.~\ref{sec:res-base}).

\begin{table}[t]
\caption{Inner cross-validation of the base on training months: mean ACC over leads 1--18 of OLS4 and of the seasonal base, and the lead $L^{*}$ of the lead-dependent base OLS4c with its inner score and its nested score ($L^{*}$ chosen without the block that is scored), which decided the choice.}
\label{tab:base}
\scriptsize
\begin{tabular}{lcccccc}
\tophline
Fold & OLS4 & Seasonal & $L^{*}$ & OLS4c & OLS4c, nested & Higher nested score \\
\middlehline
1981--1991 & 0.508 & 0.501 & 11 & 0.541 & 0.532 & OLS4c \\
1991--2002 & 0.493 & 0.464 & 8 & 0.499 & 0.492 & OLS4 \\
2002--2013 & 0.573 & 0.556 & 9 & 0.583 & 0.577 & OLS4c \\
2013--2024 & 0.639 & 0.646 & 9 & 0.647 & 0.626 & OLS4 \\
\bottomhline
\end{tabular}
\end{table}

\section{Case studies and the forecast in physical space}
\label{app:physical}

\begin{figure*}[tp]
\includegraphics[width=\textwidth]{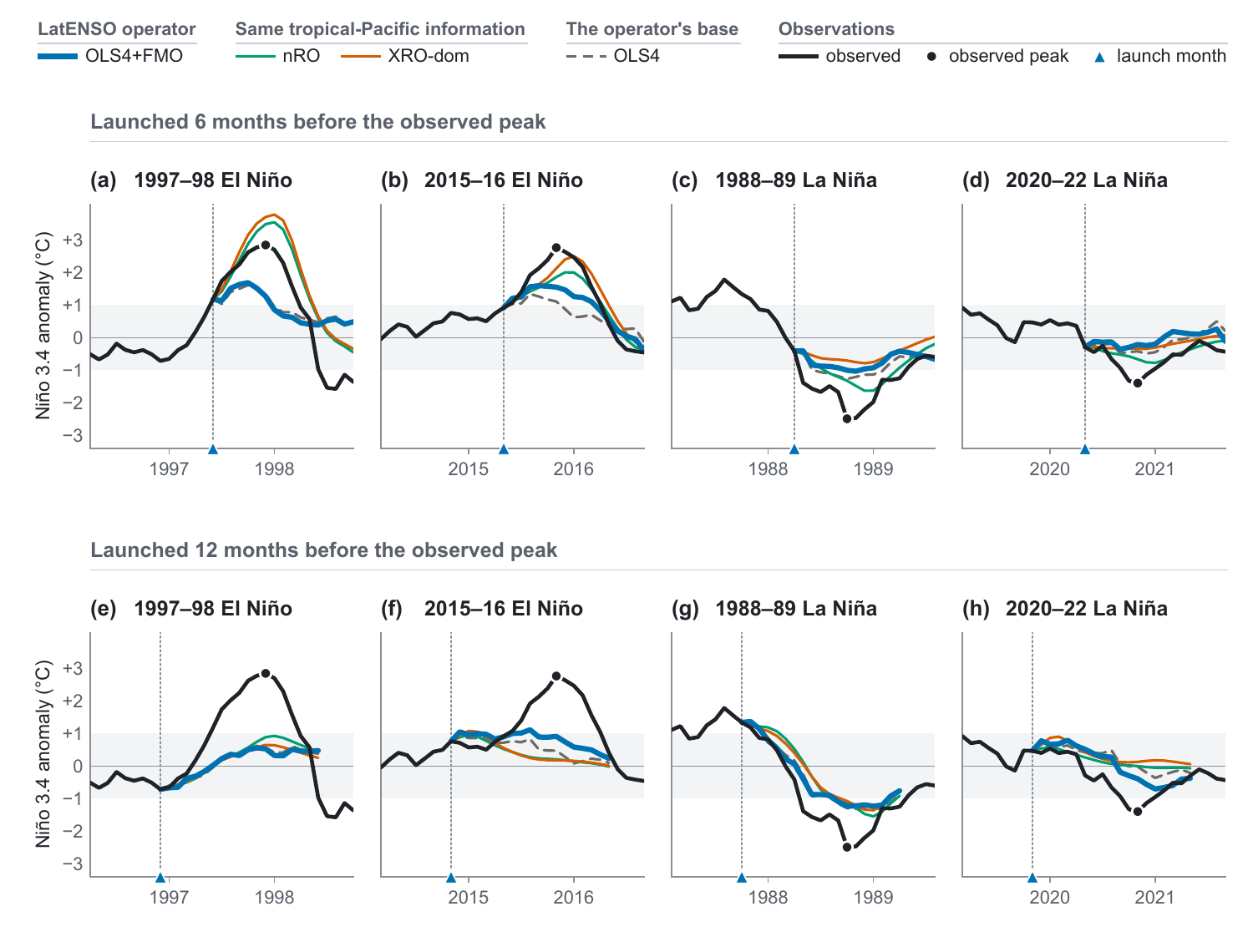}
\caption{Four events, with the Floquet operator, its base and the two in-domain oscillator models (nRO, XRO-dom) launched from the same month, (a)--(d) six and (e)--(h) twelve
months before the observed peak. Observed: Ni\~no 3.4 anomaly with respect to the 1981--2024 climatology; forecasts:
anomalies with respect to the training climatology of their fold, leads 1--18. Shading: anomalies smaller than
$1\,^{\circ}$C in magnitude.}
\label{fig:e1}
\end{figure*}

Figure~\ref{fig:e1} shows four events, with the Floquet operator, its base and the two in-domain oscillator models (nRO, XRO-dom) launched from the same month, six and twelve
months before the observed peak. Launched six months ahead, XRO-dom and nRO overshoot the 1997 peak ($+3.7$ and
$+3.5\,^{\circ}$C against $+2.9\,^{\circ}$C) while the Floquet operator undershoots it ($+1.3\,^{\circ}$C). Launched twelve months ahead, the Floquet operator comes closest to the 2015 peak ($+0.9\,^{\circ}$C against $+0.2\,^{\circ}$C for the two oscillator models) and to the 2020 La Ni\~na. The multi-year La Ni\~na of 2020--2022 is poorly forecast by every system.

\begin{figure*}[tp]
\includegraphics[width=\textwidth,height=0.72\textheight,keepaspectratio]{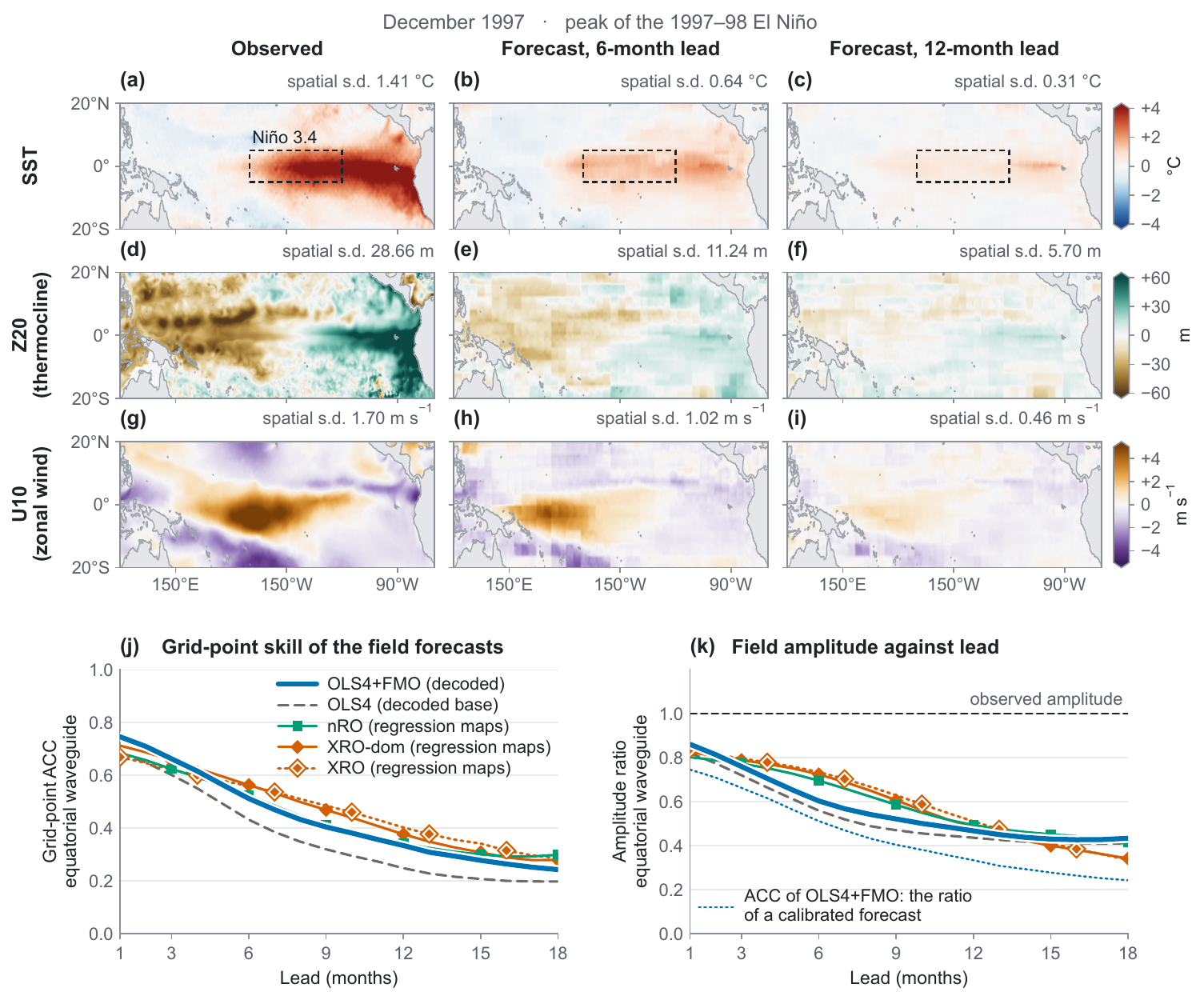}
\caption{The forecast in physical space. (a)--(i) Observed anomalies at the December 1997 peak of SST, Z20 and U10,
and the forecasts of the Floquet operator for that month launched 6 and 12 months earlier, decoded through the frozen
decoder (native grid). (j) Grid-point anomaly correlation over the equatorial waveguide on a $1^{\circ}$ grid, mean of
SST, Z20 and U10, against lead: the decoded forecast and decoded base, and the field forecasts of nRO, XRO-dom and XRO
obtained by regression on their indices. (k) Amplitude ratio over the waveguide, the standard deviation of the forecast anomalies divided by that of the observed anomalies; dotted: the ACC of the decoded forecast, the ratio a calibrated forecast would have. The number above
each map is the spatial standard deviation of the anomaly shown; dashed box: Ni\~no 3.4. Ocean points only.}
\label{fig:e2}
\end{figure*}

Figure~\ref{fig:e2} shows the forecast decoded to fields and compares it with the field forecasts of the reference
models. The reference models forecast indices only. XRO obtains fields by regressing the observed anomalies on its
state with annual and semi-annual harmonics \citep{Zhao2024}, and we apply the same regression, fitted on the
training months of each fold, to nRO, XRO-dom and XRO. All five systems are scored on the same $1^{\circ}$ grid. Over
the equatorial waveguide ($5^{\circ}$\,S--$5^{\circ}$\,N, $160^{\circ}$\,E--$80^{\circ}$\,W), the grid-point anomaly correlation of the decoded forecasts (the correlation across the verification windows at each cell, averaged over the cells with area weights), averaged over SST, Z20 and U10, is 0.75 at lead 1, 0.51 at lead 6 and 0.33 at
lead 12 (Fig.~\ref{fig:e2}j). It is the highest of the five systems at leads 1--3, within 0.02 of nRO at leads 5--14
and behind it at leads 15--18, while XRO-dom or XRO is the most skillful at most leads from lead 4. Over leads 1--18 it averages 0.434, against 0.438 for nRO, 0.466 for XRO-dom and
0.471 for XRO, and 0.362 for the decoded base. For SST alone, the decoded forecast scores 0.522 against 0.549 for
XRO-dom ($-0.027$ [$-0.083$, $+0.033$], inconclusive). Over the whole domain, with every regression fitted on the
training months, the decoded SST forecast scores 0.372 against 0.351 for XRO-dom, 0.340 for XRO and 0.304 for nRO,
while the decoded Z20 and U10 trail (0.327 and 0.178 against 0.366 and 0.213 for XRO-dom). The decoded forecast
therefore scores higher than XRO's regression maps for SST over the whole domain, and lower in the
waveguide and for Z20 and U10. Every system's amplitude ratio (the standard deviation of the forecast anomalies over that of the observed anomalies, across windows and cells) stays above its ACC (Fig.~\ref{fig:e2}k), so all field forecasts are more variable than calibrated ones. At leads 3--13 the two decoded forecasts have a lower amplitude than the three regression-map forecasts.

\section{The blend weight}
\label{app:alpha}

\begin{figure*}[tp]
\includegraphics[width=\textwidth]{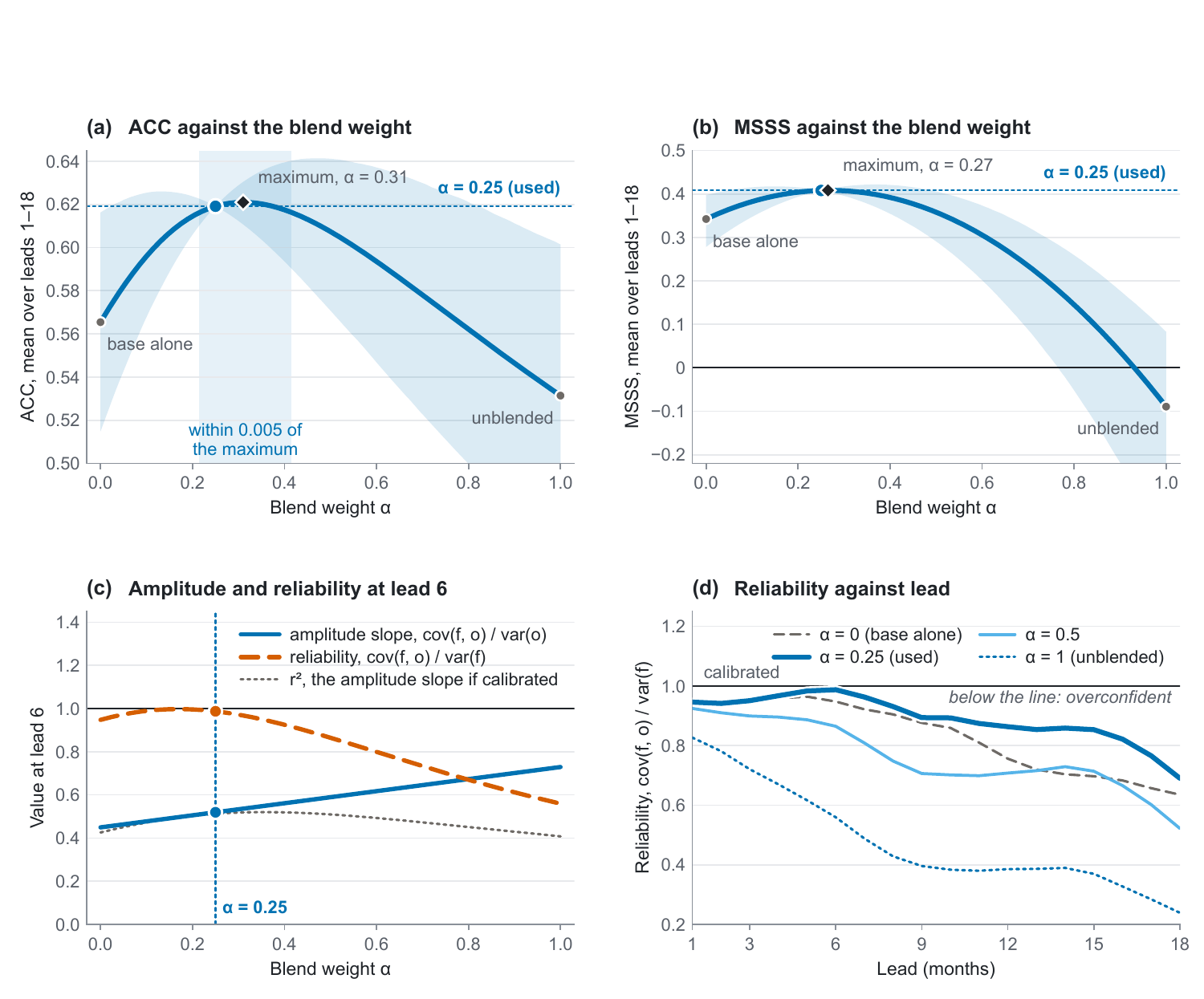}
\caption{The blend weight. (a) ACC over leads 1--18 against $\alpha$; band: 95\,\% interval of the paired difference
from $\alpha=0.25$, so that the evidence that a weight is worse than the one used is clear where the band lies
wholly below the
dashed line. (b) The same for the mean-square skill score. (c) At lead 6: amplitude slope, reliability and $r^2$
against $\alpha$. (d) Reliability against lead for four weights.}
\label{fig:f1}
\end{figure*}

Because the blend is linear in $\alpha$, every score can be written in closed form from five moments per lead (the variances and the covariance of base and correction, and the covariance of each with the observation), and the paired bootstrap can be applied exactly for every weight
(Fig.~\ref{fig:f1}). Over leads 1--18 the ACC is maximal at $\alpha=0.31$ (0.621), within 0.005 of that maximum for
$\alpha$ between 0.22 and 0.42, and 0.619 at the weight used. The difference between $\alpha=0.31$ and $0.25$ is
$+0.002$ [$-0.008$, $+0.011$]. The evidence that ACC is below its value at $\alpha=0.25$ is clear only for
$\alpha<0.02$ and
$\alpha>0.83$, and for the MSSS only for $\alpha<0.05$ and $\alpha>0.49$. The MSSS is maximal at $\alpha=0.27$. At lead 6 the
forecast is calibrated near the weight used (reliability 1 at $\alpha=0.17$, 0.987 at 0.25), and its amplitude slope (0.519) matches $r^2$ (0.513), the slope of a calibrated forecast. The unblended forecast ($\alpha=1$) has the larger amplitude slope (0.729) but a reliability of 0.559 and a negative MSSS ($-0.089$): its extra amplitude is overconfidence, not skill.

\clearpage
\section{List of symbols}
\label{app:symbols}

\begin{table*}[!h]
\caption{Symbols. Symbols that are used only where they are defined (in one sentence, caption or algorithm) are not listed.}
\label{tab:symbols}
\scriptsize
\begin{tabular}{ll}
\tophline
Symbol & Meaning \\
\middlehline
$\mathbf{X}$, $\tilde{\mathbf{X}}$ & Observed and reconstructed 13-channel field, $13\times161\times681$ \\
$\mathbf{X}_{\mathrm{p}}$; $\mathbf{H}$; $\mathbf{W}_{\mathrm{proj}}$ & Pooled and padded field; token features ($768\times96$); projection of the token features to the dimension of the latent state (Algorithm~\ref{alg:encoder}) \\
$p$; $\vec{h}_p$; $\vec{w}_p$ & Token index ($1,\dots,768$; one token per $4^{\circ}\times4^{\circ}$ block); token feature; spatial weight \\
$t$; $m(t)$ & Month index of the record (initialization month of a forecast); its calendar month, $0$ (January) to $11$ \\
$\bar{\vec{z}}_m$, $\bar{N}_m$, $\bar{W}_m$ & Monthly climatologies of the latent state and of the Ni\~no 3.4 and WWV box means over the training months of a fold \\
$\vec{z}\in\mathbb{R}^{32}$, $z_d$; $\vec{z}'$ & Latent state and its coordinate $d$; latent anomaly from the monthly climatology $\bar{\vec{z}}_m$ of the training months \\
$\vec{\Phi}_p$, $\vec{\Phi}_{\cdot d}$; $\mathbf{\Psi}$ & Row $p$ (token) and column $d$ (token-level footprint of coordinate $d$) of the decoder basis $\mathbf{\Phi}$; learned per-token offset \\
$\mathbf{W}$, $\vec{b}_0$ & Weight ($96\times32$, with columns $\mathbf{W}_{\cdot d}$) and bias of the linear layer of the decoder (Eq.~\ref{eq:decoder}) \\
$\lambda_{\mathrm{dec}}$, $\lambda_2$; $\mathrm{Corr}$, $\mathbb{E}$ & Weights of the decorrelation and size penalties; correlation matrix and mean over the minibatch (Eq.~\ref{eq:aeloss}) \\
$\Delta_3$, $\Delta_8$ & Three- and eight-month differences \\
$\vec{\phi}$; $\vec{c}$; $\vec{z}_0$ & Phase vector (Eq.~\ref{eq:phase}); its convolutional block; corrected initial state \\
$\mathrm{MLP}_{\phi}$ & Two-layer network that gives the learned correction of the initial state \\
$\mathbf{L}_0$; $\mathbf{Q}$ & Modal generator; learned orthogonal basis (Eq.~\ref{eq:L0}) \\
$\mathbf{\Omega}$; $\mathbf{W}_c$, $\mathbf{W}_s$ & Learned $32\times32$ matrices that generate $\mathbf{Q}$ and $\mathbf{L}_c$, $\mathbf{L}_s$ \\
$\mathbf{B}_k$; $\omega_k$, $\sigma_k$ & Oscillator block ($k=1,2,3$); angular frequency (radians per month) and damping rate (per month); the period is $2\pi/\omega_k$ \\
$\mathbf{D}$; $d_j$; $\tau$ & Diagonal block of 26 decay modes; their decay rates (per month), the diagonal of $-\mathbf{D}$; time in months \\
$m$, $\theta_m=2\pi m/12$ & Calendar month ($0$ = January, \dots, $11$) and its angle \\
$\mathbf{L}_c$, $\mathbf{L}_s$; $\mathbf{L}_m$; $\mathcal{T}_m$ & Skew-symmetric seasonal terms; month-$m$ generator; month-$m$ propagator (Eq.~\ref{eq:floquet}) \\
$\vec{q}_j$; $\vec{u}$; $P_k$ & Columns of $\mathbf{Q}$; coordinates $\mathbf{Q}^{\top}\vec{z}'$ of the latent anomaly in this basis; plane of oscillator $k$, spanned by $\vec{q}_{2k-1}$ and $\vec{q}_{2k}$ \\
$\hat{\mathbf{C}}$; $\zeta_k$ & Sample covariance matrix of the corrected latent forecasts; share of the forecast variance in the plane of oscillator $k$ (Eq.~\ref{eq:share}) \\
$g$; $\gamma$; $\mathbf{W}_{\mathrm{in}}$, $\mathbf{W}_{\mathrm{out}}$ & State-dependent gain; learned scalar gate of the gated residual; rank-4 direct term \\
$\vec{r}(l)$; $\vec{b}(l)$; $\hat{\vec{z}}(l)$ & Operator correction; base forecast; corrected latent forecast at lead $l$ \\
$\langle\vec{r}\rangle(l)$; $\mathrm{Dec}$ & Mean of the corrections of the ten operators of a fold; the frozen decoder (Algorithm~\ref{alg:fmo}) \\
$l$ & Lead, in months ($1,\dots,24$) \\
$N$, $W$ & Ni\~no 3.4 and WWV anomalies \\
$\boldsymbol{\beta}_l$ & Base coefficients at lead $l$ (Eq.~\ref{eq:base}) \\
$\mathcal{L}$; $\mathcal{L}^{\mathrm{MSE}}$, $\mathcal{L}^{\mathrm{ACC}}$; $S$ & Training loss of the operator; its squared-error and correlation terms; the six leads drawn per minibatch for its index terms (Eq.~\ref{eq:fmoloss}) \\
$y_{\mathrm{B}}$, $y_{\mathrm{F}}$, $\hat{y}$ & Decoded base, decoded corrected forecast, reported forecast (Eq.~\ref{eq:blend}) \\
$\alpha(l)$ & Blend weight: 0.25 for $l\le18$, 0 beyond \\
$\vec{X}$; $T$, $h$; $e_1,\dots,e_K$ & XRO state; ENSO temperature and recharge (Ni\~no 3.4 and WWV anomalies); projections of the latent anomalies onto the $K$ directions (Eq.~\ref{eq:xro}) \\
$\mathbf{L}(t)$; $\vec{N}$; $\boldsymbol{\xi}$; $\omega_a$ & Seasonal linear operator of XRO; its quadratic terms; noise; annual frequency \\
$\mathbf{A}_0$, $\mathbf{A}^c_j$, $\mathbf{A}^s_j$ & The five coefficient matrices of $\mathbf{L}(t)$ (Eq.~\ref{eq:xro}) \\
$\mathbf{C}_0$, $\mathbf{C}_6$; $\mu$, $\vec{x}$ & Covariance and six-month lag covariance of the latent anomalies; their generalized eigenvalue and eigenvector (a slow direction) \\
$\mathcal{M}$, $\mathcal{M}_j$; $n$; $i$, $j$ & Training months of a fold, and without inner block $j$ and its buffer (Algorithm~\ref{alg:xrolat}); number of training pairs; running indices \\
$f$, $o$; $r$; $\sigma_f$, $\sigma_o$ & Forecast and observation; Pearson correlation (in Sect.~\ref{sec:metrics}, of $f$ and $o$); their standard deviations \\
$a$; $\rho$; $s=\sigma_f/\sigma_o$; $\epsilon$ & Amplitude slope; reliability; amplitude ratio; error of a calibrated forecast, $o=f+\epsilon$ \\
ACC; AUC; MSSS & Anomaly correlation coefficient; area under the ROC curve; mean-square skill score \\
$R^2$; $t_{39}$, $t_{5}$; $F_{3,36}$ & Out-of-sample coefficient of determination; $t$ statistics and $F$ ratio, with their degrees of freedom \\
$E$, $C$; $N_{\mathrm{CT}}$, $N_{\mathrm{WP}}$; $\kappa$ & Eastern- and central-Pacific indices; cold-tongue and warm-pool indices, from the Ni\~no 3 and Ni\~no 4 indices $N_3$, $N_4$ and the coefficient $\kappa$ \\
$v_d$ & Share of the variance of the latent anomaly in coordinate $d$ \\
$S(f)$, $S_0(f)$; $r_1$ & Spectrum of the Ni\~no 3.4 anomaly at frequency $f$ (in cycles per month) and its AR(1) null; lag-1 autocorrelation (Fig.~\ref{fig:modes}) \\
$L^{*}$ & Last lead of the seasonal part of the lead-dependent base \\
\bottomhline
\end{tabular}
\end{table*}

\noappendix
\newpage

\authorcontribution{MP designed the model, wrote the code, ran the experiments and the evaluation, and wrote and revised
the manuscript. AB conceived the study with MP, designed the model, defined the benchmarking strategy, and wrote and revised the manuscript.}

\competinginterests{The authors declare that they have no conflict of interest.}

\begin{acknowledgements}
We thank Silvio Gualdi and Antonio Navarra (CMCC) for their guidance and for many helpful discussions. We thank the European Centre for Medium-Range Weather Forecasts for ERA5 and the SODA team for SODA4; Sen Zhao and co-authors for making the XRO code openly available, and Zejing Zhang and co-authors for the RC-ENSO
code, its DESN models and the ORAS5 index data distributed with it. 
The AI assistants Claude Code (Anthropic) and Gemini (Google) were used to help check the verification scripts, to prepare the plotting scripts of the figures, and to organize the code into the released
package. The authors reviewed all code; the results were verified by independent re-implementation and by reproduction from the archived data
(Sect.~\ref{sec:verification}), and the authors take full responsibility for the content of the paper, the code, the results and their interpretation.
\end{acknowledgements}

\financialsupport{This work was supported by the EU under Grant Agreement No. 101294938 (EarthGenerator).
}

\bibliographystyle{copernicus}
\bibliography{references}

\end{document}